\documentclass[12pt,a4paper,final]{article}
\usepackage[utf8]{inputenc}
\usepackage[english]{babel}
\usepackage{amsmath}
\usepackage{amsfonts}
\usepackage{graphicx}
\usepackage{amssymb}
\usepackage[left=2cm,right=2cm,top=2cm,bottom=2cm]{geometry}
\usepackage{tikz-cd}
\usepackage{mathtools}
\newcommand\restr[1]{\raisebox{-.5ex}{$|$}_{#1}}
\usepackage{xcolor, soul}
\sethlcolor{red}

\usepackage[scr=rsfso]{mathalfa}
\usepackage{stmaryrd}
\usepackage{enumitem}
\usepackage{esint}
\newcommand{\Eta}{\text{\rotatebox[origin=c]{90}{$\Xi$}}}
\providecommand\phantomsection{}
\usepackage{amssymb,amsmath}
\usepackage{amsthm}
\usepackage{bbm}
\usepackage{yfonts}
\numberwithin{equation}{section}
\input xy
\xyoption{all} 

\newcommand{\p}{\mbox{\boldmath$\rho$}}

\newcommand{\N}{\mathbb{N}}

\newcommand{\op}[1]{\!\!\mathop{\rm ~#1}\nolimits}

\newtheorem{Theorem}{Theorem}[section]

\newtheorem{Proposition}[Theorem]{Proposition}
 { \theoremstyle{definition}
\newtheorem{Definition}[Theorem]{Definition}

\newtheorem{Example}[Theorem]{Example}
\newtheorem{Remark}[Theorem]{Remark} }

\DeclareMathOperator{\Ber}{Ber}

\usepackage{color}

\font\black=cmbx10 \font\sblack=cmbx7 \font\ssblack=cmbx5 \font\blackital=cmmib10  \skewchar\blackital='177
\font\sblackital=cmmib7 \skewchar\sblackital='177 \font\ssblackital=cmmib5 \skewchar\ssblackital='177
\font\sanss=cmss10 \font\ssanss=cmss8 
\font\sssanss=cmss8 scaled 600 \font\blackboard=msbm10 \font\sblackboard=msbm7 \font\ssblackboard=msbm5
\font\caligr=eusm10 \font\scaligr=eusm7 \font\sscaligr=eusm5  \font\fraktur=eufm10
\font\sfraktur=eufm7 \font\ssfraktur=eufm5 
\font\bsymb=cmsy10 scaled\magstep2
\def\all#1{\setbox0=\hbox{\lower1.5pt\hbox{\bsymb
       \char"38}}\setbox1=\hbox{$_{#1}$} \box0\lower2pt\box1\;}
\def\exi#1{\setbox0=\hbox{\lower1.5pt\hbox{\bsymb \char"39}}
       \setbox1=\hbox{$_{#1}$} \box0\lower2pt\box1\;}

\def\tx#1{{\fam0\relax#1}}

\newfam\bifam
\textfont\bifam=\blackital \scriptfont\bifam=\sblackital \scriptscriptfont\bifam=\ssblackital

\newfam\blfam
\textfont\blfam=\black \scriptfont\blfam=\sblack \scriptscriptfont\blfam=\ssblack

\newfam\bbfam
\textfont\bbfam=\blackboard \scriptfont\bbfam=\sblackboard \scriptscriptfont\bbfam=\ssblackboard

\newfam\ssfam
\textfont\ssfam=\sanss \scriptfont\ssfam=\ssanss \scriptscriptfont\ssfam=\sssanss

\newfam\clfam
\textfont\clfam=\caligr \scriptfont\clfam=\scaligr \scriptscriptfont\clfam=\sscaligr

\newfam\frfam
\textfont\frfam=\fraktur \scriptfont\frfam=\sfraktur \scriptscriptfont\frfam=\ssfraktur

\def\hpb#1{\setbox0=\hbox{${#1}$}
    \copy0 \kern-\wd0 \kern.2pt \box0}
\def\vpb#1{\setbox0=\hbox{${#1}$}
    \copy0 \kern-\wd0 \raise.08pt \box0}

\def\pmb#1{\setbox0\hbox{${#1}$} \copy0 \kern-\wd0 \kern.2pt \box0}
\def\pmbb#1{\setbox0\hbox{${#1}$} \copy0 \kern-\wd0
      \kern.2pt \copy0 \kern-\wd0 \kern.2pt \box0}
\def\pmbbb#1{\setbox0\hbox{${#1}$} \copy0 \kern-\wd0
      \kern.2pt \copy0 \kern-\wd0 \kern.2pt
    \copy0 \kern-\wd0 \kern.2pt \box0}
\def\pmxb#1{\setbox0\hbox{${#1}$} \copy0 \kern-\wd0
      \kern.2pt \copy0 \kern-\wd0 \kern.2pt
      \copy0 \kern-\wd0 \kern.2pt \copy0 \kern-\wd0 \kern.2pt \box0}
\def\pmxbb#1{\setbox0\hbox{${#1}$} \copy0 \kern-\wd0 \kern.2pt
      \copy0 \kern-\wd0 \kern.2pt
      \copy0 \kern-\wd0 \kern.2pt \copy0 \kern-\wd0 \kern.2pt
      \copy0 \kern-\wd0 \kern.2pt \box0}

\mathchardef\za="710B  
\mathchardef\zb="710C  
\mathchardef\zg="710D  
\mathchardef\zd="710E  
\mathchardef\zve="710F 
\mathchardef\zz="7110  
\mathchardef\zh="7111  
\mathchardef\zvy="7112 
\mathchardef\zi="7113  
\mathchardef\zk="7114  
\mathchardef\zl="7115  
\mathchardef\zm="7116  
\mathchardef\zn="7117  
\mathchardef\zx="7118  
\mathchardef\zp="7119  
\mathchardef\zr="711A  
\mathchardef\zs="711B  
\mathchardef\zt="711C  
\mathchardef\zu="711D  
\mathchardef\zvf="711E 
\mathchardef\zq="711F  
\mathchardef\zc="7120  
\mathchardef\zw="7121  
\mathchardef\ze="7122  
\mathchardef\zy="7123  
\mathchardef\zf="7124  
\mathchardef\zvr="7125 
\mathchardef\zvs="7126 
\mathchardef\zf="7127  
\mathchardef\zG="7000  
\mathchardef\zD="7001  
\mathchardef\zY="7002  
\mathchardef\zL="7003  
\mathchardef\zX="7004  
\mathchardef\zP="7005  
\mathchardef\zS="7006  
\mathchardef\zU="7007  
\mathchardef\zF="7008  
\mathchardef\zW="700A  
\mathchardef\zC="7009  

\newcommand{\be}{\begin{equation}}
\newcommand{\ee}{\end{equation}}

\newcommand{\bea}{\begin{eqnarray}}
\newcommand{\eea}{\end{eqnarray}}
\def\*{{\textstyle *}}
\newcommand{\R}{{\mathbb R}}

\newcommand{\Z}{{\mathbb Z}}

\newcommand{\s}{{\textstyle *}}

\newcommand{\h}{{\sf H}}

\def\Hom{\textsf{Hom}}

\def\la{\langle}

\def\xi{\tx{i}}

\def\cM{\cal M}

\def\s*{{\scriptstyle *}}

\def\cO{\mathcal{O}}

\def\cM{\mathcal{M}}

\newcommand{\beas}{\begin{eqnarray*}}
\newcommand{\eeas}{\end{eqnarray*}}

\def\g{\mathfrak{g}}

\newcommand{\0}{\otimes}

\newcommand{\Ci}{\mathcal{C}^\infty}

\begin{document}
\title{The Geometry of Supersymmetry\\ A concise introduction}
\author{Norbert Poncin and Sarah Schouten\footnote{Department of Mathematics, University of Luxembourg, Maison du Nombre, 6 avenue de la Fonte, L-4364 Esch-sur-Alzette, Luxembourg,
{\sf norbert.poncin@uni.lu, sarah.schouten.001@student.uni.lu}}}
\maketitle

\begin{abstract}
This text is a short but comprehensive introduction to the basics of supergeometry and includes some of the recent advances in colored supergeometry. We do not aim for a standard text that states results and proves them more or less rigorously, but all too often offers little insight to the uninformed reader. Instead we opted for a smooth exposition of the successive themes, choosing an order and an approach which are close to the way these pieces of mathematics could have been or were discovered, thereby highlighting the reasons for the various choices and facilitating deeper understanding. We hope that the text will be useful for PhD students and researchers who wish to acquire knowledge in the geometry of supersymmetry.
\end{abstract}

\small{\vspace{2mm}} 
 \medskip


\tableofcontents

\thispagestyle{empty}

\section{Introduction}

The idea of {\it supersymmetry} arose due to insufficiency and incoherence of the so-called standard model of fundamental particles and interactions. The standard model asserts that matter is composed of twelve fundamental particles, which are called fermions and can be further divided into six quarks and six leptons. Moreover, the fundamental interactions between these particles, namely gravitational force, electromagnetic force, weak nuclear interaction and strong nuclear interaction, can also be viewed as particles. The standard model includes the five particles called bosons that correspond to the three last interactions, the photon acting as electromagnetic force, $W_-$, $W_+$ and $Z_0$ acting as weak nuclear interaction and the gluon corresponding to strong nuclear interaction. In order to explain the concept of mass an additional particle called Higgs boson is introduced. The Higgs boson appears in the form of a field the other particles can interact with to obtain mass. However, the standard model does not explain gravity. While gravitational force is mostly negligible when working with subatomic particles it does play an important role in the creation of the universe and in the general theory of relativity. Therefore, it is highly desirable to establish a unified theory that includes all fundamental interactions. One of the theories that might lead to this goal is supersymmetric string theory. String theories are based on the idea that elementary particles originate from vibrating strings, so that the type of vibration determines which of the particles is produced. Supersymmetric means that each of the particles has a corresponding supersymmetric shadow particle. More precisely, with each fermion we associate a boson and conversely each boson is coupled with a fermion.\medskip

Smooth supermanifolds, or $\mathbb{Z}_2$-manifolds, are generalizations of smooth manifolds whose local coordinates consist of standard commuting variables of $\Z_2$-degree $0$ and formal anticommuting parameters of $\Z_2$-degree $1$, so that their function sheaf carries a $\Z_2$-grading. They are the core of the geometry of supersymmetry or {\it supergeometry}.\medskip

Colored supermanifolds, also called $\Z_2^{\times n}$-manifolds or $\mathbb{Z}_2^n$-manifolds, have function sheaves with a $\mathbb{Z}_2^n$-grading and local coordinates of all $\Z_2^n$-degrees that obey the commutation rule induced by the standard scalar product of $\Z_2^n\,$. They have been introduced in a series of papers \cite{tr&ber,sign,local,towards} which investigate their category, their differential calculus and part of their integration theory including the $\Z_2^n$-generalization of the Berezinian. The splitting theorem and the Frobenius theorem for $\Z_2^n$-manifolds are proved in \cite{SplThe} and \cite{FroThe}, respectively, products of $\Z_2^n$-manifolds and related functional analytic questions are studied in \cite{products} and \cite{FunAna}, whereas \cite{SchVor} and \cite{LinAct} clarify the functor of points approach to $\Z_2^n$-manifolds -- which is of fundamental importance in physics -- and use it to study $\Z_2^n$-Lie group actions on $\Z_2^n$-manifolds. Colored supermanifolds and the corresponding {\it higher supergeometry} show significant differences from classical supergeometry, especially in the proofs of standard supergeometric results, which are mostly more subtle in the $\Z_2^n$-case, and in integration theory, which is significantly different from the standard supergeometric situation, the novel aspect being the integration with respect to even non-zero degree parameters.\medskip

The {\it motivation} to introduce and study $\Z_2^n$-geometry is broad. First $\Z_{2}^{n}$-gradings with $n \geq 2$ can be found in the theory of parastatistics \cite{ParSta,Green,ParFie,Yang} and in relation to an alternative approach to supersymmetry \cite{SupPoi}. Higher graded generalizations of the super Schr\"{o}dinger algebra and the super Poincar\'{e} algebra have appeared in \cite{SupSch} and \cite{Z2nSupSym}. Furthermore, such gradings are used in the theory of mixed symmetry tensors as found in string theory and some formulations of supergravity \cite{MixSymTen}. It must also be pointed out that quaternions and more general Clifford algebras can be understood as $\Z_2^n$-graded algebras whose vectors commute according to the above-mentioned $\Z_2^n$-scalar-product rule \cite{Oct, CliAlg,WelPap1,WelPap2}. Finally, \emph{any} `sign rule' can be interpreted in terms of a $\Z_{2}^{n}$-commutation rule \cite{sign}.\medskip

{\it Background} information on various sheaf-theoretical concepts can be found in Hartshorne \cite[Chapter II]{Hartshorne} and  Tennison \cite{Sheaf}. There are several good introductory books on the theory of supermanifolds including Bartocci, Bruzzo \&  Hern\'{a}ndez Ruip\'{e}rez \cite{Bruzzo}, Bernstein, Leites, Molotkov \& Shander \cite{BerLeiMolSha}, Carmeli, Caston \& Fioresi \cite{Rita}, Deligne \& Morgan \cite{DelMor}, Leites \cite{LeiIntSupMan}, Manin \cite{Manin} and Varadarajan \cite{Varadarajan}. For categorical notions we refer to Mac Lane \cite{MacLane}.\medskip

Our text is structured as follows.\medskip 

In the first chapter we show how even and odd supercoordinates occur naturally when we consider a system made of both bosonic and fermionic particles. If we glue such supercoordinate domains together, we get the concept of supermanifold which is reminiscent of a standard smooth `base' manifold surrounded by a `cloud of odd stuff'. Special attention is paid to a careful introduction of a minimum of sheaf-theoretic notions and the definition of supermanifolds as locally ringed spaces of algebras of superfunctions. The question of the invertibility of a superfunction naturally leads to the projection of superfunctions onto base manifold functions and to the kernel $\mathcal J$ of this projection, which plays a prominent role in the theory of supermanifolds $M$. In particular, $\mathcal J$ can be interpreted as a neighborhood of the superfunction $0$ and so it induces a basis of neighborhoods of superfunctions that defines the so-called $\mathcal J$-adic topology on the algebra $\mathcal{O}_M$ of superfunctions. We explain why all supermorphisms $\mathcal{O}_N\to\mathcal{O}_M$ are continuous with respect to this topology and prove the fundamental supermorphism theorem, which makes supergeometry a reasonable theory.

With this short description of the category of supermanifolds in mind, we move to differential calculus on supermanifolds, contextualizing each concept by means of the corresponding concept in differential geometry. After a brief digression on the conditions needed to encode all the information of a sheaf-theoretic geometry (sheaf of vector fields of a manifold) into a geometry that uses mainly global objects (vector fields defined globally on the manifold), we define the sheaf of vector fields or tangent sheaf of a supermanifold, avoiding the problem that supergeometry, unlike differential geometry, lacks a good concept of point. From a local basis of this locally free tangent sheaf of modules over superfunctions or, equivalently this supervector bundle, we derive a basis of the tangent space of a supermanifold at a point $m$ of its base manifold, thus proceeding in reverse order with respect to differential geometry. We are now ready to define the derivative at $m$ of a morphism between supermanifolds in the locally ringed space environment in which we work. Since the superworld is slightly non-commutative (anticommuting coordinates), the Jacobian matrix of a composite of morphisms between supermanifolds turns out to be the product of the Jacobian matrices of the components only if we change the sign of some entries of the Jacobian matrix, which leads to what we call the modified Jacobian matrix. Similar requirements that arise in linear superalgebra are mentioned below. We close this first chapter by a coordinate-dependent but informative approach to the two possible de Rham complexes of a supermanifold, thereby introducing the so-called Deligne and Bernstein-Leites sign conventions for the commutation of super differential forms.\medskip

The second chapter consists of a brief introduction to higher supergeometry, which highlights its relation to other areas of mathematics and physics, and the fact that this non-trivial generalization of standard supergeometry is not only necessary but also sufficient. As said above, $\Z_2^n$-manifolds are, roughly speaking, supermanifolds whose function sheaf carries a $\Z_2^n$-grading and whose local coordinates are $\Z_2^n$-commutative, i.e. commute according to the sign rule given by the standard scalar product of the involved $\Z_2^n$-degrees. Since therefore even coordinates can anticommute, odd coordinates can commute and coordinates with nonzero degree need not be nilpotent, local higher superfunctions are necessarily formal power series in the nonzero degree coordinates with coefficients in the smooth functions with respect to the degree zero coordinates. The fundamental invertibility criterion of standard superfunctions mentioned above is based on nilpotency, but remains valid in colored supergeometry despite the loss of nilpotency, precisely because we use formal power series. Furthermore, the crucial supermorphism theorem goes through in the colored situation, since the colored superfunction sheaf is Hausdorff-complete. We explain in a simple way what this means and how we use it in the proof of this theorem.\medskip

In the last chapter, a discussion of linear $\mathbb{Z}_2^n$-algebra provides a basis for the definition of integrals over $\mathbb{Z}_2^n$-manifolds. 

For instance, linear maps between free modules over a $\Z_2^n$-graded $\Z_2^n$-commutative algebra $\mathcal A$ are represented by $\Z_2^n$-graded block-matrices whose blocks consist of entries belonging to a term of $\mathcal{A}$ whose degree is determined by the position of the block and the $\Z_2^n$-degree of the matrix under consideration. We explain in detail the non-standard definitions of the product of such a matrix by a scalar in $\mathcal{A}$, of the transpose of such a matrix and of its trace. Connected to this colored supertrace is its group analogue - the colored Berezinian determinant, or just $\Z_2^n$-Berezinian. We discover this generalization of the standard Berezinian or $\Z_2$-Berezinian, explain its explicit expression in terms of quasideterminants in the sense of Gelfand and Retakh, and compute through instructive examples. 

The focus of the chapter is on the determination of integrable objects, i.e. objects that are defined over a $\Z_2^n$-manifold $M$ and which we can integrate over $M$ in a coordinate-independent way.

We begin by justifying the definition of oriented smooth manifolds $N$ and by illustrating why we can integrate global smooth differential forms of highest degree coordinate-independently over $N\,.$ We interpret the free module of local top-forms as the determinant module of the free module of local 1-forms, which is the rank 1 free module over functions whose basis element is multiplied by the determinant of the Jacobian matrix when we change the local coordinates. Although there are no top-forms in super- and $\Z_2^n$-geometry, for the free $\Z_2^n$-module of local $\Z_2^n$-1-forms we find a free rank 1 $\Z_2^n$-module over $\Z_2^n$-functions whose basis element is multiplied by the $\Z_2^n$-Berezinian of the modified $\Z_2^n$-Jacobian matrix if we change the local $\Z_2^n$-coordinates. We explicitly construct this determinant or $\Z_2^n$-Berezinian module as the only non-vanishing cohomology module of a cochain complex of $\Z_2^n$-modules. Its elements can be thought of as local replacements for the non-existing $\Z_2^n$-top-forms -- substitutes we call local $\Z_2^n$-Berezinian sections -- and its basis element can be thought of as local $\Z_2^n$-Berezinian volume. The fact that the $\Z_2^n$-Berezinian volume gets multiplied by the $\Z_2^n$-Berezinian of the modified $\Z_2^n$-Jacobian matrix if we change the considered $\Z_2^n$-coordinates, leads to the coherent sheaf condition that we have to encode in the definition that glues global $\Z_2^n$-Berezinian sections from local ones. These global sections are the global substitutes for $\Z_2^n$-top-forms and should be the objects that we can integrate over a $\Z_2^n$-manifold.

In the case $n=1$ the results of the previous paragraph allow us to make the definition of the integral of a compactly supported global Berezinian section over a supermanifold with oriented base appear natural. More specifically, this Berezinian integration consists of a differentiation with respect to the odd or degree 1 formal coordinates and a Lebesgue integration with respect to the even or degree zero ordinary coordinates. We explain why this integral is coordinate-independent.

In the case $n=2$ the Berezinian integration consists in addition to the differentiation with respect to the formal coordinates of the odd degrees $(0,1)$ and $(1,0)$ and the Lebesgue integration with respect to the ordinary coordinates of the even degree $(0,0)$, of an new integration with respect to the formal coordinates of the even degree $(1,1)$. We point out that this new integration has one degree of freedom and show that the natural choice of this parameter leads to a coordinate-free definition of the integral of a global $\Z_2^2$-Berezinian section over a $\Z_2^2$-manifold with oriented base only if the section is in some sense compactly supported with respect to the two even coordinate degrees. We find that the obstruction to coordinate-independence is a universal issue that also appears in standard supergeometry, regardless of which approach to standard supergeometry one chooses. In fact, the problem lies at the heart of Berezinian integration: it is the reason for the shortcoming of this theory, which is that one cannot integrate non-compactly supported sections. As already mentioned, in $\Z_2^2$-geometry a first solution is to integrate only sections that are compactly supported with respect to both even coordinate degrees. A second solution originates in complex analysis, changes the nature of the objects we integrate using their localization and leads to technical problems that we can however solve. 

We conclude the chapter with a short description of the integration theory of $\mathbb{Z}_2^n$-manifolds of arbitrary height $n\,$.
\newpage

\section{Introduction to supergeometry}

\subsection{Supersymmetry}

Symmetry is one of the most fundamental concepts in mathematics and physics. Supersymmetry is a symmetry first proposed in string theory in the 1970s but quickly adopted throughout theoretical physics, particularly to solve several shortcomings of the Standard Model. She assumes that every particle in this model has a so-called supersymmetric partner particle: every fermion, i.e. every particle with a half-integer quantum spin, corresponds to a boson partner, i.e. a particle with an integer spin, and vice versa.\medskip

If this is indeed true, the new symmetry fixes the mass of the Higgs boson -- a particle that gives the particles predicted by the Standard Model their mass, and explains why the mass of the Higgs boson is small and gravity is weak. Also, supersymmetry explains that at high energies, like at the beginning of the universe, all three Standard Model interactions -- the electromagnetic, weak nuclear and strong nuclear interactions -- would have the same intensity, which would be a partial unified theory of forces. Finally, supersymmetry would explain the dark matter, which makes up most of the matter in the universe and holds the galaxies together, but which we cannot see. Furthermore, supersymmetry is needed in string theory, and string theory comes with built-in quantum gravity!\medskip

Despite all these potential successes of supersymmetry, it turns out that the most natural models of supersymmetry cannot exist, implying that if supersymmetry is true nonetheless, it only exists at very high energies, but as the initial universe gets colder, the superpartners are massing and decaying so we can't even observe them at the energies of the Large Hadron Collider before the 2019-2022 revamp work. On the other hand, supersymmetry leads to a lot of beautiful and fascinating mathematics with unifying and simplifying effects. Therefore, regardless of the fate of string theory and supersymmetry in physics, it is definitely worth pursuing supergeometry and related ideas.

\subsection{Supermanifolds}

\subsubsection{Smooth superdomains}

Knowing that we can interpret the quantum state of a particle as a point in a Hilbert space and denoting the Hilbert state space of a fermion (respectively a boson) by $\mathcal{H}_1$ (respectively by $\mathcal{H}_0$) we can model the situation in the following way. Due to the Pauli exclusion principle, which asserts that two or more fermions cannot occupy the same quantum state, a system with $q$ fermions can be represented by the exterior product $\wedge^q\mathcal{H}_1$ and a system of $p$ bosons can be seen as the symmetric product $\vee^p\mathcal{H}_0$. Hence, a system of $p$ bosons and $q$ fermions corresponds to the tensor product
\begin{equation}\label{StaSpaFerBosSys}
    \vee^p\mathcal{H}_0\0\wedge^q\mathcal{H}_1.
\end{equation}
Equivalently, we could take the super vector space $\mathcal{H}_0\oplus\mathcal{H}_1$ and use its supersymmetric tensor algebra
\begin{equation}\label{SymTenAlg}
    \odot(\mathcal{H}_0\oplus\mathcal{H}_1)\cong \odot{\cal H}_0\0\odot{\cal H}_1
\end{equation}
to describe the quantum system. Saying that $\mathcal{H}_0\oplus\mathcal{H}_1$ is a super vector space means that it is $\mathbb{Z}_2$-graded. This entails that each homogeneous element, i.e. an element which is either in $\mathcal{H}_0$ or in $\mathcal{H}_1$, has a parity: the elements in $\mathcal{H}_0$ have parity $0$ and are said to be even while the elements in $\mathcal{H}_1$ have parity $1$ and are referred to as odd. If ${\cal H}_0$ (respectively ${\cal H}_1$) is finite dimensional and has dimension $r$ (respectively dimension $s$), we say that the super vector space $\mathcal{H}_0\oplus\mathcal{H}_1$ is of dimension $r|s$. The supersymmetric algebra structure mentioned above is the supercommutative tensor product
\begin{equation*}
    v\odot w=(-1)^{\tilde{v}\tilde{w}}w\odot v,
\end{equation*}
where $v$, $w$ are homogeneous elements of parity $\tilde v,$ $\tilde w$. Note that the supercommutativity condition implies that odd elements anticommute. Consequently, the square or any higher power of an odd element is equal to $0$. Further, from \eqref{SymTenAlg} we get $$\odot({\cal H}_0\oplus{\cal H}_1)\cong \vee{\cal H}_0\0\wedge{\cal H}_1$$ (see \eqref{StaSpaFerBosSys}).\medskip

We now look at a specific super vector space, namely
\begin{equation*}
    \mathbb{R}^{p|q}=\mathbb{R}^p\oplus\mathbb{R}^q.
\end{equation*}
Let $(e_i^0)_i$ be a basis of even elements for $\mathbb{R}^p$ and $(e_a^1)_a$ a basis of odd elements for $\mathbb{R}^q$. Then, any element in our super vector space can be written uniquely as $$\sum_{i=1}^p c^i_0 e_i^0+\sum_{a=1}^q c^a_1 e_a^1\quad (c^i_0,c^a_1\in\R).$$ The dual space
\begin{equation*}
    (\mathbb{R}^{p|q})^*=\Hom_0(\mathbb{R}^{p|q},\,\mathbb{R})\oplus\Hom_1(\mathbb{R}^{p|q},\,\mathbb{R}),
\end{equation*}
is the super vector space of linear maps of parity 0 and linear maps of parity 1. Since real numbers are always of parity 0 so that $\R\cong \R\oplus \{0\}$, the elements in $\Hom_0(\mathbb{R}^{p|q},\,\mathbb{R})$ send each even basis vector to some real number and each odd basis vector to $0$. The maps in $\Hom_1(\mathbb{R}^{p|q},\,\mathbb{R})$ on the other hand send odd basis vectors to real numbers and even basis vectors to $0$. Therefore it is consistent to define the dual basis $(\varepsilon_l^A)_{l,A}$ (for $l=0,\,1$ and $A=1,...,p$ or $A=1,...,q$ depending on $l$) by $$\varepsilon_l^A(e_B^k)=\delta^A_B\delta_l^k.$$ As usual, we can interpret the basis vectors $\ze^A_l$ of the dual space $(\R^{p|q})^*$ as coordinates in the original space $\mathbb{R}^{p|q}$. When $l=0$ we get even coordinates $x^i:=\varepsilon_0^i$ in $\mathbb{R}^{p|q}$ such that $$x^ix^j = \ze_0^i\odot\ze_0^j= \ze_0^j\odot\ze_0^i=x^jx^i,$$ i.e. we get standard commutative coordinates. When $l=1$ we obtain odd coordinates $\zx^a:=\varepsilon_1^a$ in  $\mathbb{R}^{p|q}$ such that \be\label{ACFP}\zx^a\zx^b=\ze_1^a\odot\ze_1^b=-\ze_1^b\odot\ze_1^a=-\zx^b\zx^a,\ee i.e. we obtain anticommutative coordinates. Of course, even coordinates commute with odd ones: $$x^i\zx^a=\ze_0^i\odot\ze_1^a = \ze_1^a\odot\ze_0^i= \zx^ax^i.$$ When equipped with these supercommutative coordinates $$\zm:=(x,\zx):=(\zm^A):=(x^i,\zx^a):=(x^1,\ldots,x^p,\zx^1,\ldots,\zx^q)$$ the space $\R^{p|q}$ is the prototypical supermanifold or $\Z_2$-manifold (with global coordinates) just as $\R^p$ is the prototypical smooth manifold (with global coordinates). Due to their parity and anticommutativity, the odd coordinates $\zx^a$ can of course not take any real value. Therefore they are often referred to as formal parameters and functions like for instance $\sin(\zx^a)$ do not make sense. Moreover, from \eqref{ACFP} it follows that a monomial like $\zx^1\zx^4\zx^2$ coincides up to a sign with the same monomial $\zx^1\zx^2\zx^4$ in which the parameters are naturally ordered, and that the $\zx^a$ are nilpotent so that a monomial like $\zx^1\zx^2\zx^1$ vanishes just as does every monomial $\zx^{a_1}\ldots\zx^{a_{q+1}}$ with more than $q$ factors. Therefore a superfunction $f$ of the supermanifold $\mathbb{R}^{p|q}$ must be of the form
\begin{align}
    f(x,\,\zx)&=f_0(x)+\sum_af_a(x)\zx^a+\sum_{a_1<a_2}f_{a_1a_2}(x)\zx^{a_1}\zx^{a_2}+\cdot\cdot\cdot+f_{1\cdot\cdot\cdot q}(x)\zx^1\cdot\cdot\cdot\zx^q \label{superfct}\\
    &=\sum_{k=0}^q\sum_{|\alpha|=k}f_\alpha(x)\zx^\alpha\,, \label{superfct2}
\end{align}
where $\za$ is a multi-index and $f_\alpha\in\mathcal{C}^\infty(U)$ for some open subset $U\in{\tt Open}(\mathbb{R}^p)$ of $\R^{p}$. As these superfunctions or $\Z_2$-functions are polynomials in the $\zx^1,\ldots,\zx^q$ with coefficients in $\Ci(U)$, we denote the algebra of these functions by $\mathcal{C}^\infty(U)[\zx^1,...,\zx^q]$. Replacing $U$ by any of its open subsets $V\in{\tt Open}(U)$ we obtain a sheaf
\begin{equation*}
  \Ci_{p|q}:{\tt Open}(U)\ni V\mapsto \mathcal{C}^\infty_{p|q}(V)=\mathcal{C}^\infty(V)[\zx^1,...,\zx^q]
\end{equation*}
of supercommutative associative unital real algebras over $U$, with obvious restrictions and gluings. The pair \be\label{superdom}{\cal U}^{p|q}:=(U,\Ci_{p|q})\ee made of the topological space $U$ and the sheaf of supercommutative rings $\Ci_{p|q}$ is a super ringed space which we will call a superdomain or $\Z_2$-domain.

\subsubsection{Smooth manifolds}\label{LRS}

Usually we define a smooth $n$-dimensional manifold $M$ as a set which comes equipped with an (equivalence class of compatible) atlas(es) whose chart maps are valued in $\R^n$ and whose coordinate transformations are smooth maps. Then the commutative associative unital real algebra $\mathcal{C}^\infty(M)$ of global functions of $M$ allows us to construct a function sheaf $\mathcal{C}^\infty$ that takes open sets $U$ in $M$ and sends them to the corresponding commutative algebra $\mathcal{C}^\infty(U)$. As algebras are in particular rings the pair $(M,\,\mathcal{C}^\infty)$ is a ringed space, i.e. a topological space together with a sheaf of rings on it.\medskip

It is well known that the map $$M\ni x\mapsto \ker(\op{eval}_x):=\{f\in \Ci(M):f(x)=0\}\in\op{Spm}(\Ci(M))$$ that sends every point $x$ of $M$ to the corresponding maximal ideal $\ker(\op{eval}_x)$ in the maximal spectrum $\op{Spm}(\Ci(M))$ of $\Ci(M)$ is a 1:1 correspondence. Hence the points of $M$ `are' the maximal ideals of $\Ci(M)$. Similarly, in Algebraic Geometry the points of an affine variety or affine scheme are the maximal or prime ideals of the global function ring of this variety or scheme. Hence it is crucial to also highlight the maximal ideals of the ringed space $(M,\Ci)$. More precisely, for every point $x$ in $M$ the stalk $\Ci_x$ at $x$ of the sheaf $\mathcal{C}^\infty$ -- the algebra of germs at $x$ of local functions -- is known to have a unique maximal ideal $\mathfrak{m}_x$ given by \be\label{MaxIdeSta}\mathfrak{m}_x=\{[f]_x:f(x)=0\}\subseteq\Ci_x.\ee This means that $(M,\,\mathcal{C}^\infty)$ is a locally ringed space (LRS), i.e. a ringed space where all stalks are local rings. In particular, the trivial smooth $n$-dimensional manifold $\mathbb{R}^n$ with its sheaf of smooth functions $\mathcal{C}_{\mathbb{R}^n}^\infty$ is a LRS. Since $M$ is locally isomorphic to $\mathbb{R}^n$, the LRS $(M,\,\mathcal{C}^\infty)$ and the LRS $(\mathbb{R}^n,\,\mathcal{C}_{\mathbb{R}^n}^\infty)$ are locally isomorphic as well. This motivates the definition of the category of LRS that are locally isomorphic as LRS to the LRS $(\mathbb{R}^n,\,\mathcal{C}_{\mathbb{R}^n}^\infty)$. It can be shown that this category is equivalent to the category of smooth $n$-dimensional manifolds. Thus we have two equivalent ways to define manifolds -- atlases and LRS-s.\medskip

Because the atlas definition of a manifold is strongly based on the concept of point $x\simeq(x^1,\ldots,x^n)$ of a manifold and since supermanifolds do not have a proper notion of point $(x,\zx)$ as the $\zx$-s are not proper coordinates, we will define smooth supermanifolds of dimension $p|q$ as locally super ringed spaces (LSRS) that are locally isomorphic as LSRS to the LSRS $(\R^p,\Ci_{p|q})$. Therefore, we start investigating LSRS-s and their (iso)morphisms.

\subsubsection{Smooth supermanifolds}\label{smoothsupmfd}

Having already mentioned super ringed spaces we now provide a concise definition.

\begin{Definition}
A super ringed space (SRS) is a pair $(M,\,\mathcal{O})$ consisting of a topological space $M$ and a sheaf $\mathcal{O}$ of supercommutative associative unital algebras over $\R$. If additionally, for every $x\in M$ the stalk $\mathcal{O}_x$ of $\mathcal{O}$ at $x$ has a unique homogeneous maximal ideal we say that $(M,\,\mathcal{O})$ is a locally super ringed space (LSRS).
\end{Definition}

Let us recall the concept of a homogenous ideal.

\begin{Definition}
If $R=R_0\oplus R_1$ is a $\mathbb{Z}_2$-graded ring then an ideal $I\subseteq R$ is said to be homogeneous if it is compatible with the grading in the sense that $I=(I\cap R_0)\oplus(I\cap R_1)$.
\end{Definition}

Thus, as said above, every superdomain $\mathcal{U}^{p|q}=(U,\,\mathcal{C}^\infty_{p|q})$ ($U\in{\tt Open}(\R^p)$) is a SRS. Furthermore, it can be shown that for every $x\in U$ the stalk $\mathcal{C}^\infty_{p|q,x}$ of $\mathcal{C}^\infty_{p|q}$ at $x$ has a unique maximal ideal given by
\begin{equation}\label{MaxIdeDom}
    \mathfrak{m}_x=\{[f]_x\,:\,f_0(x)=0\}\subseteq \Ci_{p|q, x}
\end{equation}
(see \cite{lectures}, page 42; see also \eqref{superfct} and \eqref{MaxIdeSta}). As $\mathfrak{m}_x$ is obviously homogeneous, every superdomain $\mathcal{U}^{p|q}$ is a LSRS. This result suggests using $\mathcal{U}^{p|q}$ as prototypical supermanifold that all supermanifolds are modelled onto, analogously to differentiable manifolds that are modelled on the LRS $(\mathbb{R}^n,\,\mathcal{C}_{\mathbb{R}^n}^\infty)$, see paragraph \ref{LRS}. For this, we need to define morphisms between locally super ringed spaces. Since morphisms in all categories preserve the data needed to define the structure of the category's objects, we get the

\begin{Definition}\label{LSRSmor}
A morphism $\Phi=(\phi,\,\phi^*)$ between two (locally) super ringed spaces $(M,\,\mathcal{O}_M)$ and $(N,\,\mathcal{O}_N)$ consists of
\begin{itemize}
    \item a continuous map $\phi:M\to N$ and
    \item a family $\phi^*=\{\phi^*_V\,:\,V\in{\tt Open}(N)\}$ of morphisms $\phi^*_V:\mathcal{O}_N(V)\to\mathcal{O}_M(\phi^{-1}(V))$ of $\Z_2$-graded unital $\R$-algebras such that the following diagram (involving the restriction morphisms $\rho^V_W$ and $r^V_W$ of the sheaves $\mathcal{O}_M$ and $\mathcal{O}_N$ respectively) commutes
    \begin{equation*}
    \begin{tikzcd}
    \phi^*_V:\mathcal{O}_N(V) \arrow{r} \arrow{d}{r^V_W} & \mathcal{O}_M(\phi^{-1}(V)) \arrow{d}{\rho^V_W} \\
    \phi^*_W:\mathcal{O}_N(W) \arrow{r} & \mathcal{O}_M(\phi^{-1}(W))
    \end{tikzcd}
    \end{equation*}
    and, in the case of locally super ringed spaces, such that for every $m\in M$ the induced algebra morphism
    \begin{align*}
        \phi^*_m:\,&\mathcal{O}_{N,\phi(m)}\longrightarrow\mathcal{O}_{M,m}\\
        &[\,g\,]_{\phi(m)}\longmapsto[\,\phi^*_Vg\,]_m
    \end{align*}
    verifies $\phi^*_m(\mathfrak{m}_{N,\phi(m)})\subseteq\mathfrak{m}_{M,m}$.
\end{itemize}
\end{Definition}

Now we are ready to define supermanifolds.

\begin{Definition}\label{smfd}
A smooth supermanifold or $\mathbb{Z}_2$-manifold of dimension $p|q$ is a super ringed space $\mathcal{M}=(M,\,\mathcal{O}_M)$, where $M$ is a second countable Hausdorff topological space, such that for every point $m\in M$ there exist open subsets $m\in U\subseteq M$ and $U^p\subseteq\mathbb{R}^p$ as well as an isomorphism $\Phi=(\phi,\,\phi^*)$ of super ringed spaces between the SRS $(U,\,\mathcal{O}_M\restr{U})$ and the LSRS $(U^p,\,\mathcal{C}^\infty_{p|q})$. The prototypical supermanifolds $(U^p,\,\mathcal{C}^\infty_{p|q})$ are called $\mathbb{Z}_2$-domains.
\end{Definition}

\begin{Remark}
Examining the isomorphism $\Phi:(U,\,\mathcal{O}_M\restr{U})\rightarrow(U^p,\,\mathcal{C}^\infty_{p|q})$ from Definition \ref{smfd} it becomes clear that for every $m\in M$ the induced map $\phi^*_m:\mathcal{C}^\infty_{p|q,\phi(m)}\rightarrow\mathcal{O}_{M,m}$ must be an isomorphism of algebras. Since $\mathcal{C}^\infty_{p|q,\phi(m)}$ contains a unique homogeneous maximal ideal the same must hold for $\mathcal{O}_{M,m}$, which means that any supermanifold $\mathcal{M}=(M,\,\mathcal{O}_M)$ is a LSRS.
\end{Remark}

\begin{Example}\label{vb}
Consider a smooth manifold $M$ of dimension $n$ and its tangent bundle $TM\rightarrow M$. We turn the total space $TM$ into the supermanifold $TM[1]$, where $[1]$ represents a parity shift of the fibre coordinates, i.e. we decide to see them as odd parameters and thereby create a $\mathbb{Z}_2$-grading on $TM[1]$. Letting $U\subseteq M$ be a trivialization domain of $TM$ and denoting the sheaf of functions on $TM[1]$ by $\mathcal{O}_{TM[1]}$ we get
\begin{equation*}
    \mathcal{O}_{TM[1]}(U)=\{\sum^n_{k=0}\sum_{a_1<\cdot\cdot\cdot< a_k}f_{a_1\cdot\cdot\cdot a_k}(x)\,\zx^{a_1}\cdot\cdot\cdot\zx^{a_k}\},
\end{equation*}
where $(\zx^1,\ldots,\zx^n)$ are the odd fibre coordinates, $(x^1,\ldots,x^n)$ are the even base coordinates and $f_{a_1\cdot\cdot\cdot a_k}\in\mathcal{C}^\infty(U)$. On the other hand, the differential forms on $U$ are given by
\begin{equation*}
    \Omega(U)=\Gamma(U,\,\wedge T^*M)=\{\sum^n_{k=0}\sum_{a_1<\cdot\cdot\cdot< a_k}\omega_{a_1\cdot\cdot\cdot a_k}(x)\,dx^{a_1}\wedge\cdot\cdot\cdot\wedge dx^{a_k}\},
\end{equation*}
where $(dx^1,\ldots,dx^n)$ is the local frame of $T^*M$ and $\omega_{a_1\cdot\cdot\cdot a_k}\in\mathcal{C}^\infty(U)$. Since the wedge product between these basis elements behaves similarly as the product between the odd parameters we can identify the two function spaces above and we get that $(M,\,\Omega)\cong TM[1]$ is a supermanifold. More generally, any vector bundle $E\rightarrow M$ over $M$ of rank $k$ can be equipped with a parity shift in the fibre coordinates and can then be seen as a supermanifold of dimension $n|k$. It can even be shown that any supermanifold $\mathcal{M}=(M,\,\mathcal{O}_M)$ is isomorphic to $E[1]=(M,\,\Gamma(\wedge E^*))$ for some vector bundle $E\rightarrow M$. However, this identification is not canonical and the categories of supermanifolds and vector bundles do not coincide, which will become clear during the study of morphisms between supermanifolds.
\end{Example}

Consider now the $\mathbb{Z}_2$-domain $(\mathbb{R}^p,\,\mathcal{C}^\infty_{p|q})$ and for every open subset $U\subseteq\mathbb{R}^p$ define a mapping $\varepsilon_U:\mathcal{C}^\infty_{p|q}(U)\rightarrow\mathcal{C}^\infty(U)$ that sends a superfunction given by
\begin{equation*}
f(x,\,\zx)=f_0(x)+\sum_af_a(x)\zx^a+\sum_{a_1<a_2}f_{a_1a_2}(x)\zx^{a_1}\zx^{a_2}+\cdot\cdot\cdot+f_{1\cdot\cdot\cdot q}(x)\zx^1\cdot\cdot\cdot\zx^q
\end{equation*}
to the function $f_0\in\mathcal{C}^\infty(U)$. Clearly, $\varepsilon_U$ is a surjective unital algebra morphism. Denoting the kernel of $\varepsilon_U$ by $\mathcal{J}(U)$ we get the following short exact sequence of algebras
\begin{equation*}
    0\,\rightarrow\,\mathcal{J}(U)\,\xrightarrow{i}\,\mathcal{C}^\infty_{p|q}(U)\,\xrightarrow{\varepsilon_U}\,\mathcal{C}^\infty(U)\,\rightarrow\,0.
\end{equation*}

\begin{Proposition}\label{inv}
A function $f\in\mathcal{C}^\infty_{p|q}(U)$ is invertible if and only if $\varepsilon_U(f)=f_0\in\mathcal{C}^\infty(U)$ is invertible.
\end{Proposition}
\begin{proof}
If $f\in\mathcal{C}^\infty_{p|q}(U)$ has inverse $f^{-1}$ then the inverse of $f_0=\varepsilon_U(f)$ is given by $$f_0^{-1}=(\varepsilon_U(f))^{-1}=\varepsilon_U(f^{-1})$$ since $\varepsilon_U$ is a unital algebra morphism.

Conversely, assume $f_0\in\mathcal{C}^\infty(U)$ has inverse $f_0^{-1}$. Since $f$ is invertible if and only if $f_0^{-1}f$ is invertible we focus on $f_0^{-1}f=1+t$, where $t$ consists of terms that involve at least one of the odd parameters. Then $t^{q+1}=0$ and therefore the inverse of $1+t$ is given by $1+\sum_{m=1}^qt^m$.
\end{proof}

Let now $U\subseteq\mathbb{R}^p$ be an open subset. Since a function $f\in\mathcal{C}^\infty(U)$ is invertible if and only if $f(x)\neq0$ for all $x\in U$, the value of $f$ at $x$ can be characterized as the unique real number $k$ such that $f-k$ is not invertible in any neighbourhood of $x$. Note that a superfunction $g\in\mathcal{C}^\infty_{p|q}(U)$ cannot be evaluated at a point because the coordinates in $\mathbb{R}^{p|q}$ involve formal parameter. However, in view of Proposition \ref{inv}, for every $x\in U$ there exists a unique real number $l$ such that $g-l$ is not invertible in any neighborhood of $x$. As this is a local property and all supermanifolds are locally isomorphic to a $\mathbb{Z}_2$-domain the same holds for superfunctions on an arbitrary supermanifold. So if $\mathcal{M}=(M,\,\mathcal{O}_M)$ is a supermanifold and $V\subseteq M$ an open subset then for every $s\in\mathcal{O}_M(V)$ and for every $x\in V$ there exists a unique real number $m$ such that $s-m$ is not invertible in any neighborhood of $x$. Now, we can define an algebra morphism $\varepsilon_V$ on $\mathcal{O}_M(V)$ by setting $\varepsilon_V(s)(x):=m$. Denoting its kernel by $\mathcal{J}(V)$ and its image by $\mathcal{F}(V)$ we obtain the following short exact sequence of algebras
\begin{equation*}
    0\,\rightarrow\,\mathcal{J}(V)\,\xrightarrow{i_V}\,\mathcal{O}_M(V)\,\xrightarrow{\varepsilon_V}\,\mathcal{F}(V)\,\rightarrow\,0.
\end{equation*}
In fact the kernel ${\cal J}_M:V\mapsto {\cal J}(V)$ is a subsheaf of ${\cal O}_M$. The presheaf ${\cal F}$ is locally isomorphic to $\Ci_{\R^p}$ and is thus locally a sheaf. Hence ${\cal F}$ generates a sheaf $\mathfrak{F}$ which is locally isomorphic to $\Ci_{\R^p}$ and thus implements a $p$-dimensional smooth manifold structure on $M$ such that $\Ci_M\cong\mathfrak{F}$, see subsection \ref{LRS}. Thus, there exists a short exact sequence
\begin{equation*}
    0\,\rightarrow\,\mathcal{J}_M\,\xrightarrow{i}\,\mathcal{O}_M\,\xrightarrow{\varepsilon}\,\mathcal{C}^\infty_M\,\rightarrow\,0
\end{equation*}
of sheaves of supercommutative associative real algebras over $M$ and the projection $\varepsilon$ of the function sheaf $\mathcal{O}_M$ of the supermanifold $\mathcal{M}$ onto the function sheaf $\mathcal{C}^\infty_M$ of the underlying smooth manifold $M$ can be viewed as an embedding of the base manifold $M$ into the supermanifold $\mathcal{M}$.\medskip

This investigation of the function sheaf of a supermanifold shows, firstly, that a supermanifold structure $(M,\,\mathcal{O}_M)$ always induces a smooth manifold structure on its base topological space $M$ and secondly, that $M$ can be embedded into $\mathcal{M}$, so that supermanifolds can be seen as smooth manifolds with a cloud of odd ``stuff'' around them.\medskip

Let us finally mention that in the next subsection we will further explain the role of the ideals
$$\mathcal{J}(V)=\{s\in\cO_M(V):\ze_V(s)\equiv 0\}\subseteq \cO_M(V)$$
above ($V\in{\tt Open}(M)$) and of the unique homogeneous maximal ideals $\mathfrak{m}_m\subseteq\mathcal{O}_m$ ($m\in M$). In addition, for upcoming applications, we note that, if we choose a supercoordinate chart $(x,\zx)$ centered at $m$ it follows from \eqref{MaxIdeDom} that $\mathfrak{m}_m$ is given by
\begin{equation*}
    \mathfrak{m}_m=\{[s]_m\,:\,\varepsilon(s)(m)=0\}\cong\{[f]_0\,:\,f(x,\,\zx)=0(x)+\sum^q_{k=1}\sum_{a_1<\cdot\cdot\cdot< a_k}f_{a_1\cdot\cdot\cdot a_k}(x)\,\zx^{a_1}\cdot\cdot\cdot\zx^{a_k}\}\subseteq\cO_m,
\end{equation*}
where $0(x)$ are terms of degree at least $1$ in $x$.

\subsection{Morphisms of supermanifolds}
\subsubsection{Continuity}\label{continuity}
A morphism between two supermanifolds $\mathcal{M}=(M,\,\mathcal{O}_M)$ and $\mathcal{N}=(N,\,\mathcal{O}_N)$ (of dimension $p|q$ and $r|s$ respectively) is a morphism $\Phi=(\phi,\,\phi^*)$ of the corresponding locally super ringed spaces.\medskip

We want to investigate continuity properties of such morphisms and start by observing that the projection $\varepsilon$ introduced above commutes with $\phi^*$. We denote the projection of $\mathcal{O}_N$ onto the sheaf $\mathcal{C}^\infty_N$ of smooth functions of $N$ by $\varepsilon_N$ and choose open subsets $V\in{\tt Open}(N)$ and $U=\phi^{-1}(V)\in{\tt Open}(M)$\,. Then, if there exist supercoordinates $(y,\,\eta)$ on $V$ and $(x,\,\zx)$ on $U$\,, we have on the one hand
\begin{equation}\label{einerseits}
    \phi^*_V(\varepsilon_{N,V}(f))=\phi^*_V(f_0)=f_0\circ\phi\restr{U}\in\mathcal{C}_M^\infty(U)
\end{equation}
for every $f\in\mathcal{O}_N(V)$\,. The first equality in \eqref{einerseits} follows from the decomposition of $f$ as in \eqref{superfct} and the second one from the fact that the pullback of a classical function $f_0$ on $V$ by the map $\phi:M\rightarrow N$ is given by $f_0\circ\phi\restr{U}$\,. On the other hand, applying the algebra morphism $\phi^*_V$ to $f$, decomposed as in \eqref{superfct2}, yields
\begin{equation*}
    \phi^*_V(f(y,\,\eta))=\phi^*_V(\sum_{k=0}^s\sum_{|\alpha|=k}f_\alpha(y)\eta^\alpha)=\sum_{k=0}^s\sum_{|\alpha|=k}\phi^*_V(f_\alpha(y))\phi^*_V(\eta^1)^{\alpha_1}\cdot\cdot\cdot\phi^*_V(\eta^s)^{\alpha_s}
\end{equation*}
and since $\phi^*_V$ respects parities $\phi^*_V(\eta^a)$ is odd for all $a\in\{1,...,s\}$ and we get that $\phi^*_V(f(y,\,\eta))$ is equal to the sum of $\phi^*_V(f_0(y))$ and terms that include at least one of the odd parameters $\zx^1,..,\zx^q$. Therefore,
\begin{equation*}
    \varepsilon_{M,U}(\phi^*_V(f))=\phi^*_V(f_0)=f_0\circ\phi\restr{U}\in\mathcal{C}_M^\infty(U),
\end{equation*}
which shows in conjunction with \eqref{einerseits} that the following diagram commutes
\begin{equation*}
\begin{tikzcd}
\mathcal{O}_N(V) \arrow{r}{\phi^*_V} \arrow{d}{\varepsilon_{N,V}} & \mathcal{O}_M(U) \arrow{d}{\varepsilon_{M,U}} \\
\mathcal{C}^\infty_N(V) \arrow{r}{\phi^*_V} & \mathcal{C}^\infty_M(U)\,.
\end{tikzcd}
\end{equation*}

This result can also be proven in a coordinate-free manner (see \cite{lectures}, p. 46) and entails in particular that elements $g\in\mathcal{J}_N(V)$ in the kernel of $\varepsilon_{N,V}$ verify
\begin{equation*}
    \varepsilon_{M,U}(\phi^*_V(g))=\phi^*_V(\varepsilon_{N,V}(g))=0\,.
\end{equation*}
Since $\phi^*_V(g_1\cdot g_2)=\phi^*_V(g_1)\cdot\phi^*_V(g_2)$ this does not only imply $\phi^*_V(\mathcal{J}_N(V))\subseteq\mathcal{J}_M(U)$ but also
\begin{equation}\label{C0}
    \phi^*_V(\mathcal{J}^k_N(V))\subseteq\mathcal{J}_M^k(U)
\end{equation}
for every $k\in\{0,...,s\}$. Passing from superfunctions in $\mathcal{O}_N(V)$ to germs of superfunctions in $\mathcal{O}_{N,\phi(x)}$ for some $x\in M$, \eqref{C0} implies
\begin{equation}\label{C02}
    \phi^*_x(\mathfrak{m}^k_{N,\phi(x)})\subseteq\mathfrak{m}^k_{M,x}\,,
\end{equation}
which means in particular that the requirement concerning the preservation of the unique maximal ideal in Definition \ref{LSRSmor} is redundant when defining morphisms between $\mathbb{Z}_2$-manifolds.

Focusing on the powers of the ideal $\mathcal{J}_N(V)$ we get a decreasing sequence of ideals
\begin{equation}\label{DSI}
    \mathcal{O}_N(V)=\mathcal{J}_N^0(V)\supseteq\mathcal{J}_N^1(V)\supseteq\mathcal{J}_N^2(V)\supseteq\cdot\cdot\cdot\supseteq\mathcal{J}_N^s(V)\supseteq\mathcal{J}_N^{s+1}(V)=\{0\}\,.
\end{equation}
Since the powers of ${\cal J}_N$ are sheaves, a section in ${\cal J}_N^{q+1}(V)$ vanishes if its restrictions to a cover of coordinate domains vanish. Hence assume that on $W\subseteq V$ we have coordinates $(y,\zh)\,.$ While $\mathcal{O}_N(W)$ contains all superfunctions
\begin{equation*}
    f(y,\,\eta)=f_0(y)+\sum_af_a(y)\eta^a+\sum_{a_1<a_2}f_{a_1a_2}(y)\eta^{a_1}\eta^{a_2}+\cdot\cdot\cdot+f_{1\cdot\cdot\cdot s}(y)\eta^1\cdot\cdot\cdot\eta^s\,,
\end{equation*}
the elements of $\mathcal{J}_N(W)$ contain at least one odd parameter in each of their terms. Similarly, the elements of $\mathcal{J}_N^2(W)$ contain at least two odd parameters in each of their terms and the elements of $\mathcal{J}_N^s(W)$ only contain a term in all of the parameters $\zh^1,...,\zh^s$. Since any combination of $s+1$ parameters must contain two copies of the same parameter it follows that $\mathcal{J}_N^{s+1}(W)=\{0\}$ and that ${\cal J}_N^{s+1}(V)=\{0\}\,.$ We interpret the sequence \eqref{DSI} as a sequence of smaller and smaller neighborhoods of $0\in\mathcal{O}_N(V)$. This motivates the definition of the $\mathcal{J}$-adic topology on $\mathcal{O}_N(V)$ by means of the basis
\begin{equation*}
    \{g+\mathcal{J}_N^k(V)\,:\,g\in\mathcal{O}_N(V),\,0\leq k\leq s\}\,.
\end{equation*}
Analogously, $\mathcal{O}_M(U)$ is equipped with the $\mathcal{J}$-adic topology defined by the basis $$\{f+\mathcal{J}_M^k(U)\,:\,f\in\mathcal{O}_M(U),\,0\leq k\leq q\}\,.$$ Hence, $\phi^*_V:\mathcal{O}_N(V)\rightarrow\mathcal{O}_M(U)$ is a map between two topological spaces and we can ask whether it is continuous. We claim that
\begin{equation}\label{open}
    \phi^{*\,-1}_V(f+\mathcal{J}^k_M(U))=\bigcup_{g\in\phi^{*\,-1}_V(f+\mathcal{J}^k_M(U))}(g+\mathcal{J}^k_N(V))\,
\end{equation}
for any element $f+\mathcal{J}^k_M(U)$ in the basis of the $\mathcal{J}$-adic topology of $\mathcal{O}_M(U)$. Since the right-hand side of \eqref{open} is open as union of open sets the claim asserts that $\phi^*_V$ {\it is continuous with respect to the $\mathcal{J}$-adic topology}. It is clear that any element $g\in\phi^{*\,-1}_V(f+\mathcal{J}^k_M(U))$ is included in the union on the right-hand side of \eqref{open} as this union consists of neighborhoods of these very elements. To show the other inclusion we apply $\phi^*_V$ to an arbitrary neighborhood $g+\mathcal{J}^k_N(V)$ of the union and obtain $\phi^*_V(g)+\phi^*_V(\mathcal{J}^k_N(V))$ since $\phi^*_V$ is an algebra morphism. While the first term $\phi^*_V(g)$ is contained in $f+\mathcal{J}^k_M(U)$ by the way $g$ was chosen, Equation \eqref{C0} ensures that the second term verifies $\phi^*_V(\mathcal{J}^k_N(V))\subseteq\mathcal{J}^k_M(U)$. Taking into account that $\mathcal{J}^k_M(U)$ is an ideal we can deduce that $\phi^*_V(g)+\phi^*_V(\mathcal{J}^k_N(V))$ is a subset of $f+\mathcal{J}^k_M(U)$, which concludes the proof of \eqref{open}.

It should be mentioned that in a similar fashion \eqref{C02} can be used to endow $\mathcal{O}_{M,x}$ and $\mathcal{O}_{N,\phi(x)}$ for every $x\in M$ with a topology called $\mathfrak{m}$-adic topology and it can be shown that the map $\phi^*_x$  {\it is continuous with respect to the $\mathfrak{m}$-adic topology}.

Furthermore, the continuous map $\phi$ {\it between the smooth manifolds $M$ and $N$ can be proven to be smooth} by showing that its components $\phi^i=y^i\circ\phi$ defined in a neighborhood of any point $x\in M$ are smooth functions.

\subsubsection{Fundamental theorem of supermorphisms}
Following this discussion of continuity properties of morphisms between supermanifolds we examine the defining elements of such morphisms, which leads us to the fundamental theorem of supermorphisms. For this, let
\begin{equation*}
    \Phi=(\phi,\,\phi^*):\mathcal{M}=(M,\,\mathcal{O}_M)\rightarrow\mathcal{V}^{r|s}=(V,\,\mathcal{C}^\infty_{r,s})
\end{equation*}
be a morphism between a supermanifold $\mathcal{M}$ of dimension $p|q$ and a $\mathbb{Z}_2$-domain $\mathcal{V}^{r|s}$ of dimension $r|s$, the latter being equipped with the global coordinate system $(y,\,\eta)$. Since smooth functions of the even coordinates $y^i$ are even and the $\eta^a$ are odd it is possible to assign a canonical parity to each term of an arbitrary superfunction $f\in\mathcal{C}^\infty_{r|s}(V)$. In particular, $y^i\in\mathcal{C}^\infty_{r|s}(V)_0$ and $\eta^a\in\mathcal{C}^\infty_{r|s}(V)_1$ and since $\phi^*$ respects parities we observe, denoting $\phi_V^*y^i$ by $s^i$ and $\phi_V^*\eta^a$ by $\sigma^a$, that
\begin{align}
    s^i&\in\mathcal{O}_M(M)_0\,,\hspace{1cm} \text{ for }i\in\{1,...,r\}\,,\label{si}\\
    \sigma^a&\in\mathcal{O}_M(M)_1\,,\hspace{0.9cm} \text{ for }a\in\{1,...,s\}\,.\label{sigmaa}
\end{align}
Furthermore, applying the projection map $\varepsilon$ to the $s^i$ yields $$\varepsilon s^i=\varepsilon\phi^*y^i=\phi^*\varepsilon y^i=\phi^* y^i=y^i\circ\phi=\phi^i\in\mathcal{C}^\infty(M)\,,$$ which implies
\begin{equation}\label{third}
(\varepsilon s^1,...,\varepsilon s^r)(M)\subseteq V\,.
\end{equation}
These pullbacks of the coordinates in the superdomain actually completely determine the morphism $\Phi$ as stated by the following theorem.

\begin{Theorem}[Fundamental theorem of supermorphisms]\label{fundathm}
Being given a supermanifold $\mathcal{M}=(M,\,\mathcal{O}_M)$, a superdomain ${\cal V}^{r|s}=(V,\Ci_{r|s})$ with coordinates $(y,\,\eta)$ and elements $$s^1,...,s^r,\,{\sigma}^1,...,{\sigma}^s\in\mathcal{O}_M(M)$$ that verify \eqref{si}, \eqref{sigmaa} and \eqref{third} then there exists a unique morphism of supermanifolds $$\Phi=(\phi,\,\phi^*):\mathcal{M}\rightarrow\mathcal{V}^{r|s}\,,$$ such that
\begin{equation*}
     s^i=\phi_V^*y^i\hspace{1cm}\text{ and }\hspace{1cm}\sigma^a=\phi_V^*\eta^a\,.
\end{equation*}
\end{Theorem}

While we do not provide a rigorous proof for Theorem \ref{fundathm} (see \cite{lectures}, page 51), we explain the idea behind the construction of the morphism $\Phi$ after making some useful observations.

Based on the relation
\begin{equation}\label{pullback}
\psi^*y^i=y^i\circ\psi=y^i(\psi(x))=y^i(x)
\end{equation}
for a morphism $\psi$ between classical smooth manifolds with local coordinates $x=(x^1,...,x^m)$ respectively $y=(y^1,...,y^n)$ and adopting the notation $y^i=y^i(x)$, common in Physics, we decide to sometimes omit the pullback in expressions like \eqref{pullback} and in similar ones for morphisms between smooth supermanifolds. So, for instance, if $\Phi=(\phi,\,\phi^*):\mathbb{R}^{p|q}\rightarrow\mathbb{R}^{r|s}$ is a morphism between superdomains endowed with coordinates $(x,\,\zx)$ respectively $(y,\,\eta)$ then we can write
\begin{align}
    y^i=\phi^*y^i&=y^i_0(x)+\sum_{\alpha_1<\alpha_2}y^i_{\alpha_1\alpha_2}(x)\zx^{\alpha_1}\zx^{\alpha_2}+\cdot\cdot\cdot\label{locsupmor}\\
    \eta^a=\phi^*\eta^a&=\sum_\za\eta^a_\alpha(x)\zx^\alpha+\sum_{\alpha_1<\alpha_2<\alpha_3}\eta^a_{\alpha_1\alpha_2\alpha_3}(x)\zx^{\alpha_1}\zx^{\alpha_2}\zx^{\alpha_3}+\cdot\cdot\cdot\nonumber
\end{align}

\begin{Remark}
In example \ref{vb} we discovered that any $\mathbb{Z}_2$-manifold can be identified with some vector bundle and vice versa. However, we also mentioned that the categories of supermanifolds and vector bundles do not coincide, which we can justify by the fact that the former one has much more morphisms. Indeed, any smooth supermanifold is locally isomorphic to an appropriate $\mathbb{Z}_2$-domain and thus any supermorphism locally reads as in \eqref{locsupmor}, whereas a morphism between two vector bundles equipped with local coordinates $(x,\,\zx)$ and $(y,\,\eta)$ is locally given by
\begin{align*}
    y^i&=y^i(x)\\
    \eta^a&=\sum_b\eta^a_b(x)\zx^b\,.
\end{align*}
\end{Remark}

\begin{Example}\label{exmorph}
Consider a morphism $\Phi=(\phi,\,\phi^*)$ between supermanifolds that locally reads as
\begin{align}
    y&=x+\zx^1\zx^2\label{SCT}\\
    \eta^1&=\zx^1\nonumber\\
    \eta^2&=\zx^2\nonumber\,.
\end{align}
Using this morphism we want to pull back a superfunction $f$ in the variables $(y,\,\eta)$ to a superfunction in the variables $(x,\,\zx)$. If $f$ is given by $f(y,\,\eta)=y\eta^1$ then
\begin{equation*}
    \phi^*f=(\phi^*y)(\phi^*\eta^1)=(x+\zx^1\zx^2)\zx^1=x\zx^1
\end{equation*}
clearly is a superfunction in $(x,\,\zx)$. However, if $f(y,\,\eta)=\sin y$ then the expression
\begin{equation*}\label{sin}
    \phi^*f=\phi^*(\sin y)=\sin(x+\zx^1\zx^2)
\end{equation*}
is not a superfunction since for this we need it to be a smooth function in $x$ multiplied by a polynomial in $\zx^1$ and $\zx^2$. Recalling that the Taylor series of $\sin$ is given by
\begin{equation*}
    \sin(z+h)=\sum_{k=0}^\infty\frac{1}{k!}\sin^{(k)}(z)h^k
\end{equation*}
for any $z,\,h\in\mathbb{R}$ and taking into account that in a superfunction any term in which appear two or more copies of the same odd parameter vanishes it seems reasonable to define
\begin{equation*}
    \sin(x+\zx^1\zx^2)=\sin x+(\cos x)\zx^1\zx^2\,.
\end{equation*}
This process is called formal Taylor expansion and allows us thanks to nilpotency of odd parameters to transform classical functions into superfunctions.
\end{Example}

\begin{Remark}
In paragraph \ref{smoothsupmfd} we established for an arbitrary $\mathbb{Z}_2$-manifold $\mathcal{M}=(M,\,\mathcal{O}_M)$ the projection $\varepsilon:\mathcal{O}_M\rightarrow\mathcal{C}^\infty_M$ and thus an embedding $M\xhookrightarrow{}\mathcal{M}$\,. However, there does not exist a canonical projection ${\cal M}\rightarrow M$, i.e. a canonical embedding $\mathcal{C}^\infty_M(U)\xhookrightarrow{}\mathcal{O}_M(U)$ for any $U\in{\tt Open}(M)\,.$ Even if $U$ is a coordinate domain and $\mathcal{O}_M(U)\cong\Ci_{p|q}(U)\,,$ the embedding is not coordinate-independent. Indeed, the supercoordinate transformation \eqref{SCT} induces in the base the standard coordinate transformation $y=x$ and the classical function $\sin x = \sin y$ could be associated with the superfunctions $\sin x$ or $\sin y=\sin x+(\cos x)\zx^1\zx^2$\,. However, there is a non-canonical embedding of the sheaf $\Ci_M$ into the sheaf $\cO_M\,,$ as stated by the Batchelor-Gaw\c{e}dzki theorem. 
\end{Remark}

Now we construct a morphism $\Phi=(\phi,\,\phi^*):\mathcal{M}=(M,\,\mathcal{O}_M)\rightarrow\mathcal{V}^{r|s}=(V,\,\mathcal{C}^\infty_{r|s})$ on the basis of some elements $s^1,...,s^r,\,\eta^1,...,\eta^s\in\mathcal{O}_M(M)$ which satisfy the conditions of Theorem \ref{fundathm} thus capturing the main idea of the theorem's proof. On the one hand, the map $\phi:M\rightarrow V$ is defined by $\phi:=(\varepsilon s^1,...,\varepsilon s^r)\in\mathcal{C}^\infty(M,\,V)$\,. On the other hand, $\phi^*$ should be a morphism of $\mathbb{Z}_2$-graded unital $\mathbb{R}$-algebras, so applying it to an arbitrary superfunction must yield
\begin{equation*}
    \phi^*\left(\sum_\alpha f_\alpha(y)\eta^\alpha\right):=\sum_\alpha\phi^*(f_\alpha(y))(\phi^*\eta^1)^{\alpha_1}\cdot\cdot\cdot(\phi^*\eta^s)^{\alpha_s}\,.
\end{equation*}
Furthermore, we have to set $\phi^*\eta^a:=\sigma^a$ for all $a\in\{1,...,s\}$ to fulfill the assertion of the theorem and thus focus on the factors $\phi^*(f_\alpha(y))$, which we define to mean
\begin{equation*}
    \phi^*(f_\alpha(y)):=f_\alpha(\phi^*y)=f_\alpha(\phi^*y^1,...,\phi^*y^r)=f_\alpha(s^1,...,s^r)\,,
\end{equation*}
setting $\phi^*y^i:=s^i$ for $i\in\{1,...,r\}$ for the same reason as above. Each $s^i$ is assumed to be even so if for the sake of simplicity we take $\mathcal{M}=\mathbb{R}^{p|q}$ with coordinates $(x,\,\zx)$ we can write $s^i=s^i_0(x)+n^i$ for some smooth functions $s^i_0$ and some nilpotent elements $n^i$ featuring an even number of the odd parameters $\zx^1,...,\zx^q$ in each of their terms. Applying formal Taylor expansion, which has been introduced in Example \ref{exmorph} and can also be used in the case of several variables based on the Taylor series for functions of several variables, we finally set
\begin{equation*}
    f_\alpha(s_0(x)+n):=\sum_\beta\frac{1}{\beta!}(\partial^\beta_yf_\alpha)(s_0(x))n^\beta\,,
\end{equation*}
where the sum is finite due to nilpotency. Therefore, we finally obtain
\begin{equation*}
    \phi^*\left(\sum_\alpha f_\alpha(y)\eta^\alpha\right)=\sum_\alpha\sum_\beta\frac{1}{\beta!}(\partial^\beta_yf_\alpha)(s_0(x))n^\beta\sigma^\alpha
\end{equation*}
and $\phi^*$ defined in this way is an algebra morphism that respects parities as can easily be checked. Furthermore it can be shown that it commutes with the restriction maps and that any two morphisms satisfying the conditions of Theorem \ref{fundathm} must coincide and thus our definition of $\phi$ and $\phi^*$ provides the unique supermorphism whose existence is stated in the fundamental theorem of supermorphisms.

\subsection{Differential calculus on supermanifolds}

\subsubsection{Sheaves versus global sections}

Even though differential geometry is sheaf-theoretic often it is not necessary to use sheaf theory in order to deal with problems in this domain because global sections and morphisms between them encode all necessary information and are typically easier to work with than sheaves and sheaf morphisms. For instance, let $M$ be a smooth manifold and denote by $\Omega(M)$ the globally defined differential forms on $M$, i.e. the global sections of the exterior bundle of $M$. Adding the usual restriction and gluing we can reconstruct the sheaf $(M,\,\Omega)$ of differential forms. Moreover, in this case the reconstruction of the sheaf morphisms from the morphisms between global sections works as follows. Any local operator $\tau:\Omega(M)\rightarrow\Omega(M)$ can be restricted to an open subset $U\in{\tt Open}(M)$ thanks to the existence of bump functions. More precisely, for every point $p\in U$ we are able to choose a bump function $\gamma$ that is equal to $1$ in a neighbourhood of $p$ and vanishes in a neighbourhood of the complement of $U$ in order to define the restriction of $\tau$ to $U$ by setting for all $\omega_U\in\Omega(U)$
\begin{equation*}
    \tau\restr{U}(\omega_U)(p):=\tau(\gamma\omega_U)(p)\,.
\end{equation*}
Then the restriction of $\tau$ verifies for all $\omega\in\Omega(M)$
\begin{equation*}
    \tau\restr{U}(\omega\restr{U})=\tau(\omega)\restr{U}
\end{equation*}
and defining $\tau\restr{V}$ analogously for some open set $V\subseteq U$ we obtain the following commutative diagram, which means that from $\tau$ we constructed the associated sheaf morphism.
\begin{equation*}
    \begin{tikzcd}
    \tau\restr{U}:\Omega(U) \arrow{r} \arrow{d}{\rho} & \Omega(U) \arrow{d}{\rho} \\
    \tau\restr{V}:\Omega(V) \arrow{r} & \Omega(V)
    \end{tikzcd}
\end{equation*}

When working with real-analytic or holomorphic functions we cannot resort to partitions of unity as they do not exist and consequently sheaf theory is indispensable in these cases.

The definition of partitions of unity can be adapted to $\mathbb{Z}_2$-manifolds and their existence can be proven. Therefore, in supergeometry it is sometimes possible to work with global sections rather than using sheaves similarly as in standard differential geometry. Even though sheaves are in many cases indispensable we can observe that the existence of partitions of unity enables in certain cases the reconstruction of a sheaf morphism from the corresponding morphism between global sections. A result that illustrates this observation is Theorem $9$ in \cite{products} which in particular asserts that for every pair of supermanifolds $\mathcal{M}=(M,\,\mathcal{O}_M)$ and $\mathcal{N}=(N,\,\mathcal{O}_N)$ there exists a bijection
\begin{equation*}
    \beta:\Hom_{\mathbb{Z}_2\text{-Man}}(\mathcal{M},\,\mathcal{N})\ni\Phi=(\phi,\,\phi^*)\mapsto\phi^*_N\in\Hom_{\mathbb{Z}_2\text{-Alg}}(\mathcal{O}_N(N),\,\mathcal{O}_M(M))\,.
\end{equation*}

\subsubsection{Super tangent bundle}
In differential geometry a vector field $X\in\Gamma(TM)$ on a smooth manifold $M$ assigns to every point $m\in M$ a tangent vector $X_m\in T_mM\subseteq TM$\,. Since the coordinates on a supermanifold involve formal parameters there is no good concept of a point in supergeometry, which implies that the aforementioned definition of vector fields on standard manifolds cannot simply be transferred to supermanifolds. However, it is well known that the space of vector fields on $M$ is isomorphic to the space of derivations of smooth funtions on $M$. Thus, for any $U\in{\tt Open}(M)$ we can set
\begin{equation}\label{TMU}
    TM(U):=\Gamma(U,\,TM)\cong \op{Der}\,\mathcal{C}^\infty(U)
\end{equation}
and note that $TM(U)$ is a real vector space, a $\mathcal{C}^\infty(U)$-module as well as a Lie algebra over $\mathbb{R}$. This identification of vector fields with derivations enables us to define $\mathbb{Z}_2$-vector fields in accordance with the definition from standard differential geometry, adapting it slightly in terms of parity.\medskip

From now on let $\mathcal{M}=(M,\,\mathcal{O})$ be a supermanifold of dimension $p|q$ and $U\in{\tt Open}(M)$ an open set in the underlying base manifold. Analogously to \eqref{TMU} we set
\begin{equation*}
    T\mathcal{M}(U):=\mathbb{Z}_2\op{Der}\,\mathcal{O}(U)=\mathbb{Z}_2\op{Der}_0\,\mathcal{O}(U)\oplus\mathbb{Z}_2\op{Der}_1\,\mathcal{O}(U)\,,
\end{equation*}
whose meaning is clarified in the

\begin{Definition}
A homogeneous superderivation  $X\in\mathbb{Z}_2\op{Der}_{\tilde{X}}\,\mathcal{O}(U)$ of parity $\tilde{X}\in\{0,\,1\}$ is an $\mathbb{R}$-linear map $X:\mathcal{O}(U)_i\rightarrow\mathcal{O}(U)_{i+\tilde{X}}$\,, $i\in\{0,\,1\}$, that verifies the graded Leibniz rule
\begin{equation*}
    X(st)=(Xs)t+(-1)^{\tilde{X}\tilde{s}}s(Xt)
\end{equation*}
for all $s,\,t\in\mathcal{O}(U)$ and where $\tilde{s}$ denotes the parity of $s$.
\end{Definition}

Clearly both $\mathbb{Z}_2\op{Der}_0\,\mathcal{O}(U)$ and $\mathbb{Z}_2\op{Der}_1\,\mathcal{O}(U)$ are real vector spaces, which means that $T\mathcal{M}(U)=\mathbb{Z}_2\op{Der}\,\mathcal{O}(U)$ is a real super vector space. Moreover, $T\mathcal{M}(U)$ can be endowed with a super $\mathcal{O}(U)$-module structure and with a super Lie algebra structure, for more details see \cite{lectures}, page 54.

Thanks to the existence of super bump functions in supergeometry, which are defined analogously as bump functions in differential geometry, it can be proven that any superderivation $X\in T\mathcal{M}(U)$ is a local operator and can be restricted to $\mathcal{O}(V)$ for any $V\in{\tt Open}(U)$ such that the restriction $X\restr{V}$ verifies
\begin{equation*}
    X\restr{V}(s\restr{V})=(Xs)\restr{V}
\end{equation*}
for all $s\in\mathcal{O}(U)$. 
Then the assignment
\begin{equation*}
    T\mathcal{M}:{\tt Open}(M)\ni U\mapsto\mathbb{Z}_2\op{Der}\,\mathcal{O}(U)\in\mathbb{Z}_2\op{Mod}(\mathcal{O}(U))
\end{equation*}
together with the restriction maps $\rho^U_V:\mathbb{Z}_2\op{Der}\,\mathcal{O}(U)\ni X\mapsto X\restr{V}\in\mathbb{Z}_2\op{Der}\,\mathcal{O}(V)$ defines a presheaf and even a sheaf of $\mathbb{Z}_2$-modules over $\mathcal{O}$ and $\mathbb{Z}_2$-Lie algebras over $\mathbb{R}$.

\begin{Definition}
The sheaf $T\mathcal{M}$ is referred to as tangent sheaf of the supermanifold $\mathcal{M}$ and the elements in the $\mathcal{O}(M)$-module $T\mathcal{M}(M)$ are called vector fields of $\mathcal{M}$.
\end{Definition}

In order to establish the local form of super vector fields we first recall what is meant by a supercoordinate chart with coordinates $(x,\,\zx)$ around some point $x\in M$. We thereby indicate the existence of an open subset $U\in{\tt Open}(M)$ containing $x$ such that $(U,\,\mathcal{O}\restr{U})$, the restriction of $\mathcal{M}$ to $U$, is isomorphic as super ringed space to the super domain $(U,\,\mathcal{C}^\infty_{p|q}\restr{U})$ where the open subset of  $\mathbb{R}^p$ diffeomorphic to $U\in{\tt Open}(M)$ is also denoted by $U$. This entails the following isomorphism between $\mathbb{Z}_2$-algebras
\begin{equation*}
    \mathcal{O}(V)\cong\mathcal{C}^\infty_{p|q}(V)=\mathcal{C}^\infty(V)[\zx^1,...,\zx^q]
\end{equation*}
for every $V\in{\tt Open}(U)$, which implies in particular that elements in $\mathcal{O}(V)$ can be viewed as superfunctions of the form $f(x,\,\zx)=\sum_\alpha f_\alpha(x)\zx^a$ for some $f_\alpha\in\mathcal{C}^\infty(V)$.\\

Now let $(U,\,(x,\,\zx))$ be a super coordinate chart. We define $\partial_{x^i}\in\mathbb{Z}_2\op{Der}_0\,\mathcal{O}(U)$ for $i\in\{1,..,p\}$ and $\partial_{\zx^a}\in\mathbb{Z}_2\op{Der}_1\,\mathcal{O}(U)$ for $a\in\{1,...,q\}$ by setting
\begin{align*}
    \partial_{x^i}\left(\sum_\alpha f_\alpha(x)\zx^\alpha\right)&:=\sum_\alpha(\partial_{x^i} f_\alpha(x))\zx^\alpha \\
    \partial_{\zx^a}\left(\sum_\alpha f_\alpha(x)\zx^\alpha\right)&:=\sum_\alpha f_\alpha(x)\partial_{\zx^a}\zx^\alpha
\end{align*}
for all $\sum_\alpha f_\alpha(x)\zx^a\in\mathcal{O}(U)$. Morevover, in order to complete the above definition we set $\partial_{\zx^a}\zx^b:=\delta^b_a$ and illustrate what this means for $\partial_{\zx^a}\zx^\alpha$ on the example
\begin{equation*}
    \partial_{\zx^a}(\zx^b\zx^a)=(\partial_{\zx^a}\zx^b)\zx^a-\zx^b(\partial_{\zx^a}\zx^a)=-\zx^b\,.
\end{equation*}
It can be shown (see \cite{lectures}, page 57) that $\partial_{x^1},...,\partial_{x^p},\,\partial_{\zx^1},...,\partial_{\zx^q}$ form a basis of the $\mathcal{O}(U)$-module $T\mathcal{M}(U)$. Firstly, this result implies the existence of a unique decomposition of any $X\in T\mathcal{M}(U)$ into
\begin{equation*}
    X=\sum_{i=1}^p\mathcal{X}^i\partial_{x^i}+\sum_{a=1}^q\mathscr{X}^a\partial_{\zx^a}
\end{equation*}
for some $\mathcal{X}^i,\,\mathscr{X}^a\in\mathcal{O}(U)$. Secondly, we obtain that $T\mathcal{M}$ is a locally free sheaf of super $\mathcal{O}$-modules over $M$, which in conjunction with the fact that there exists a $1$-to-$1$ correspondence between locally free sheaves of $\mathcal{C}^\infty$-modules over a standard manifold $M$ and vector bundles over $M$ motivates the

\begin{Definition}\label{supervb}\cite{SupVecBun}
A super vector bundle over a supermanifold $\cM=(M,\,\cO)$ is a locally free sheaf of $\cO$-modules over $M$.
\end{Definition}
In particular, the tangent sheaf $T\cM$ of $\cM$ is a super vector bundle over $\cM$ that we call super tangent bundle of $\cM\,.$

\subsubsection{Super tangent spaces}
Starting again with the well-known corresponding concept in differential geometry we recall that there exists an isomorphism between the tangent space $T_mM$ to a standard manifold $M$ at one of its points $m\in M$ and the derivations at $m$ of the stalk $\mathcal{C}^\infty_m$ given by
\begin{equation*}
    L:T_mM\ni X_m\mapsto L_{X_m}\in \op{Der}_m\,\mathcal{C}^\infty_m\,,\hspace{2cm}
    L_{X_m}:\mathcal{C}^\infty_m\ni[f]\mapsto(d_mf)(X_m)\in\mathbb{R}\,.
\end{equation*}
The choice of the stalk $\mathcal{C}^\infty_m$ as source space of $L_{X_m}$ is based on the fact that $d_m$ is a local operator, so that $d_mf$ only depends on $f$ in an arbitrarily small neighbourhood of $m$.\medskip

Similarly, for a supermanifold $\cM=(M,\,\cO)$ we have the

\begin{Definition}
The super tangent space $T_m\cM$ of $\cM$ at $m\in M$ is given by the real super vector space $\mathbb{Z}_2\op{Der}_m\,\cO_m$ of superderivations at $m$ of the $\mathbb{Z}_2$-algebra $\cO_m$, which is defined in terms of the vector spaces of homogeneous superderivations of parity $0$ and $1$:
\begin{equation*}
    \mathbb{Z}_2\op{Der}_m\,\cO_m=\mathbb{Z}_2\op{Der}_{m,0}\,\cO_m\oplus\mathbb{Z}_2\op{Der}_{m,1}\,\cO_m\,.
\end{equation*}
A homogeneous super tangent vector $X_m$ at $m$ to $\cM$ of parity $\tilde{X}_m\in\{0,\,1\}$ is a homogeneous superderivation of parity $\tilde{X}_m$ at $m$ of $\cO_m$, i.e. $X_m$ is an $\R$-linear map $X_m:\cO_m\rightarrow\R$ verifying
\begin{equation*}
    X_m([s]\cdot[t])=X_m[s](\varepsilon[t])(m)+(-1)^{\tilde{X}_m\tilde{s}}(\varepsilon[s])(m)\cdot X_m[t]
\end{equation*}
for all $[s],[t]\in\mathcal{O}_m$ and where $\tilde{s}$ denotes the parity of $s$, the map $\varepsilon:\cO_m\rightarrow\mathcal{C}_m^\infty$ is induced by the projection $\varepsilon:\mathcal{O}\rightarrow\mathcal{C}^\infty$ and the germ of $s$ at $m$ is denoted by $[s]$.
\end{Definition}

Considering a point $m\in M$ and a neighborhood $U$ of $m$ we observe that any vector field $X\in T\cM(U)$ induces a tangent vector $X_m\in T_mX$, which is of the same parity if $X$ is homogeneous. Indeed, this tangent vector is given by
\begin{equation*}
    X_m=\text{ev}_m\circ\varepsilon\circ X\,
\end{equation*}
where $\text{ev}_m:\mathcal{C}_m^\infty\rightarrow\R$ is the evaluation morphism at $m$ and $\varepsilon:\cO_m\rightarrow\mathcal{C}_m^\infty$ is as above.

Therefore, the basis $(\partial_{x^i},\,\partial_{\zx^a})$ induces a basis $(\partial_{x^i,m},\,\partial_{\zx^a,m})$ of the super tangent space at $m$. This implies in particular that $T_m\mathcal{M}$ has the same dimension as $\mathcal{M}$ and that each super tangent vector $X_m\in T_m\mathcal{M}$ can be written uniquely as
\begin{equation*}
    X_m=\sum_{i=1}^p\mathcal{X}^i_m\partial_{x^i,m}+\sum_{a=1}^q\mathscr{X}^a_m\partial_{\zx^a,m}
\end{equation*}
for some $\mathcal{X}^i_m,\,\mathscr{X}^a_m\in\mathbb{R}$.\medskip

In standard differential geometry the tangent map $T_mf$ of a map $f\in\mathcal{C}^\infty(M,\,N)$ between two smooth manifolds at a point $m\in M$ is a linear map between the tangent spaces $T_m M$ and $T_{f(m)}N$, which are isomorphic to $\op{Der}_m\mathcal{C}^\infty_{M,m}$ and $\op{Der}_{f(m)}\mathcal{C}^\infty_{N,f(m)}$ respectively. It is given by
\begin{equation*}
    T_mf(X_m)=X_m\circ f^*_m
\end{equation*}
for any tangent vector $X_m:\mathcal{C}^\infty_{M,m}\rightarrow\R$ and where $f^*_m:\mathcal{C}^\infty_{N,f(m)}\rightarrow\mathcal{C}^\infty_{M,m}$ denotes the pullback by $f$.

Transferring this concept to super geometry we define super tangent maps as follows.

\begin{Definition}
The tangent map $T_m\Phi$ of a morphism $\Phi=(\phi,\,\phi^*):\mathcal{M}\rightarrow\mathcal{N}$ between supermanifolds at a point $m\in M$ is the super vector space morphism given by
\begin{align*}
    T_m\Phi:T_m\mathcal{M}&\rightarrow T_{\phi(m)}\mathcal{N}\\
    X_m&\mapsto X_m\circ \phi^*\,,
\end{align*}
where $\phi^*$ is the induced pullback morphism between stalks.
\end{Definition}

The tangent map of a supermorphism behaves similarly as the tangent map of a morphism between smooth manifolds when it comes to composition of morphisms. Indeed, let $\Phi=(\phi,\,\phi^*):\mathcal{M}\rightarrow\mathcal{N}$ and $\Psi=(\psi,\,\psi^*):\mathcal{N}\rightarrow\mathcal{P}$ be morphisms between supermanifolds and consider a point $m\in M$. The tangent map $T_m\Phi$ acts on a tangent vector in $T_m\mathcal{M}$ by composing it with the pullback between stalks $\phi^*$ and similarly for $T_{\phi(m)}\Psi$. Since the tangent map of their composite $T_m(\Psi\circ\Phi)$ acts on a tangent vector in $T_m\mathcal{M}$ by composing it with $\phi^*\circ\psi^*$ and since composition is associative we obtain
\begin{equation*}
    T_m(\Psi\circ\Phi)=T_{\phi(m)}\Psi\circ T_m\Phi\,.
\end{equation*}
If in differential geometry we have a map $z=z(y)$, where $y=y(x)$, then $z$ also depends on $x$ and for the partial derivative with respect to $x^i$ we obtain
\begin{equation*}
    \partial_{x^i}z=\sum_j\partial_{y^j}z\,\partial_{x^i}y^j=\sum_j\partial_{x^i}y^j\,\partial_{y^j}z\,.
\end{equation*}
Now consider a morphism of supermanifolds $\Phi=(\phi,\,\phi^*):(M,\,\mathcal{O})\rightarrow(N,\,\mathcal{R})$ and assume that $V\in{\tt Open}(N)$ is a supercoordinate domain with coordinates $\nu=(y,\,\eta)$ such that $U\subset\phi^{-1}(V)\in{\tt Open}(M)$ is a supercoordinate domain with coordinates $\mu=(x,\,\zx)$. Picking an element $t\in\mathcal{R}(V)$ and calculating the partial derivative of its pullback $\phi^*t\in\mathcal{O}(U)$ with respect to $\mu^A$ it can be verified that
\begin{equation}
    \partial_{\mu^A}(\phi^*t)=\sum_B\partial_{\mu^A}(\phi^*\nu^B)\phi^*(\partial_{\nu^B}t)\,,
\label{DerComp}\end{equation}
which coincides with the corresponding result in differential geometry in view of the convention to omit pullbacks.\medskip

Next, we would like to investigate how to represent the tangent map $T_m\Phi:T_m\mathcal{M\rightarrow}T_{\phi(m)}\mathcal{N}$ by means of a matrix. Here, $\Phi$ is a morphism between the $\mathbb{Z}_2$-manifolds $\mathcal{M}$ and $\mathcal{N}$ of dimension $p|q$ and $r|s$ respectively and we consider supercoordinate charts around $m\in M$ and around $\phi(m)\in\mathcal{N}$ with coordinates $\mu=(x,\,\zx)$ and $\nu=(y,\,\eta)$ respectively. These supercoordinates induce the bases $$\partial_{\mu^A,m}=(\partial_{x^i,m},\,\partial_{\zx^a,m})\quad\text{and}\quad \partial_{\nu^B,\,\zvf(m)}=(\partial_{y^j,\,\zvf(m)},\,\partial_{\eta^b,\,\zvf(m)})$$ of $T_m\mathcal{M}$ and $T_{\phi(m)}\mathcal{N}$ and $\Phi$ is locally given by $y=y(x,\,\zx)$ and $\eta=\eta(x,\,\zx)$. It is easy to check that the matrix of $T_m\Phi$ in the bases $\partial_{\zm^A,m}$ and $\partial_{\zn^B,\,\zvf(m)}$ is as expected the $(r+s)\times(p+q)$ matrix
\begin{equation}\label{jac}
    \partial_\mu\nu\restr{m}=
    \begin{pmatrix}
    \partial_xy\restr{m}&\partial_\zx y\restr{m}\\
    \partial_x\eta\restr{m}&\partial_\zx\eta\restr{m}
    \end{pmatrix}
    =
    \begin{pmatrix}
    \varepsilon(\partial_xy)(m)&0\\
    0&\varepsilon(\partial_\zx\eta)(m)
    \end{pmatrix}\;,
\end{equation}
where $$\ze(\partial_\zx y)(m)=\ze(\partial_x \zh)(m)=0\,,$$ as $\ze$ preserves the parity.\medskip 

We consider now a second morphism $\Psi:\mathcal{N}\rightarrow\mathcal{P}$ and a coordinate chart around $\psi(\phi(m))$ with coordinates $\omega=(z,\,\theta)$. Since $$T_m(\Psi\circ\Phi)=T_{\phi(m)}\Psi\circ T_m\Phi$$ and since the composite of super vector space morphisms is represented by the product of their representative matrices, we have
\begin{equation*}
\partial_\mu\omega\restr{m}=\partial_\nu\omega\restr{\phi(m)}\cdot\partial_\mu\nu\restr{m}\,.
\end{equation*}
It is natural to ask whether the same result holds for the Jacobian matrices, i.e. {\it whether}
\begin{equation*}
\partial_\mu\omega=\partial_\nu\omega\cdot\partial_\mu\nu\,.
\end{equation*}
From \eqref{DerComp} it follows that
\begin{align*}
    (\partial_\mu\omega)^C_A&=\partial_{\mu^A}\omega^C\\
    &=\sum_B\partial_{\mu^A}\nu^B\partial_{\nu^B}\omega^C\\
    &=\sum_B\pm\partial_{\nu^B}\omega^C\partial_{\mu^A}\nu^B\\
    &=\sum_B\pm(\partial_\nu\omega)^C_B(\partial_\mu\nu)^B_A\,,
\end{align*}
so that \be\label{Wrong}\partial_\mu\omega\neq\partial_\nu\omega\cdot\partial_\mu\nu\;\;.\ee However, the hindering signs can be included in the Jacobian matrix:

\begin{Definition}
The modified super Jacobian matrix of a supermorphism $\Phi$ between $\mathbb{Z}_2$-domains $\mathcal{U}^{p|q}$ and $\mathcal{V}^{r|s}$ given by $y=y(x,\,\zx)$ and $\eta=\eta(x,\,\zx)$ is defined as the $(r+s)\times(p+q)$ matrix
\begin{equation*}
    \mathbb{Z}_2\op{Jac}\,\Phi=
    \begin{pmatrix}
    \partial_xy&-\partial_\zx y\\
    \partial_x\eta&\partial_\zx\eta
    \end{pmatrix}\,.
\end{equation*}
\end{Definition}

With this definition the result \eqref{Wrong} becomes true, i.e. the modified Jacobian matrix of the composite of two supermorphisms {\it is} the product of the two modified Jacobian matrices: \be\label{True}
    \mathbb{Z}_2\op{Jac}\,(\Psi\circ\Phi)=\mathbb{Z}_2\op{Jac}\,\Psi\cdot\mathbb{Z}_2\op{Jac}\,\Phi\;.
\ee
Note that the representative matrix of the tangent map in the induced bases of the tangent spaces is given by \be\label{RepTan} T_m\Phi\cong \partial_{\zm}\zn|_m = \Z_2\op{Jac}\Phi|_m\;,\ee as the difference between the two matrices disappears in the projection onto the base.

\subsubsection{Super differential forms}

The $\mathcal{C}^\infty(M)$-module of differential $1$-forms on a smooth manifold $M$ is given by $$\Omega^1(M)=\Gamma(T^*M)=\Hom_{\mathcal{C}^\infty(M)}(\Gamma(TM),\,\mathcal{C}^\infty(M))\;.$$ We also set $\Omega^0(M)=\mathcal{C}^\infty(M)$ and define the linear map
\begin{align*}
    d:\Omega^0(M)&\rightarrow\Omega^1(M)\\
    f&\mapsto df\,,
\end{align*}
where $df$ associates each $X\in\Gamma(TM)$ with the Lie derivative of $f$ in the direction of $X$. The map $d$ can be uniquely extended to a degree $1$ linear map on the differential $k$-forms on $M$ ($k\ge 1$) that verifies the graded derivation property with respect to the wedge product of differential forms and the equation $d^2=0$\,.\medskip

This suggests defining the super differential $1$-forms on a supermanifold $\mathcal{M}=(M,\,\mathcal{O})$ as $$\Omega^1\mathcal{M}:=\mathscr{H}om_{\mathcal{O}}(T\mathcal{M},\,\mathcal{O})\;.$$ It should be noted that even though $T\mathcal{M}$ and $\mathcal{O}$ are sheaves $\mathscr{H}om_{\mathcal{O}}(T\mathcal{M},\,\mathcal{O})$ is not made of sheaf morphisms but is itself a sheaf that associates to every $U\in{\tt Open}(M)$ the super $\mathcal{O}(U)$-module $\Omega^1\mathcal{M}(U)$\, that consists of sheaf morphisms as detailed in the following definition.

\begin{Definition}\label{diffform}
    A $\mathbb{Z}_2$-differential $1$-form $\omega\in\Omega^1\mathcal{M}(U)$ over $U$ is an $\mathcal{O}(U)$-linear map 
    \begin{equation*}
        \omega:T\mathcal{M}(U)\rightarrow\mathcal{O}(U)
    \end{equation*}
    along with its $\mathcal{O}(V)$-linear restrictions $\omega\restr{V}:T\mathcal{M}(V)\rightarrow\mathcal{O}(V)$ for every $V\in{\tt Open}(U)$ that verify $\omega(X)\restr{V}=\omega\restr{V}(X\restr{V})$ for all $X\in T\mathcal{M}(U)$\,.
\end{Definition}

Furthermore, we set $\Omega^0\cM:=\mathcal{O}$ and define the morphism of sheaves of super $\mathcal{O}$-modules $d:\Omega^0\mathcal{M}\rightarrow\Omega^1\mathcal{M}$ as the family of maps
\begin{align*}
    d_U:\Omega^0\mathcal{M}(U)&\rightarrow\Omega^1\mathcal{M}(U)\\
    s&\mapsto d_Us\,
\end{align*}
where $U\in{\tt Open}(M)$ and the differential of a section $s\in\mathcal{O}(U)$ of parity $\tilde{s}$ is given by
\begin{equation*}
(d_Us)(X):=(-1)^{\tilde{X}\tilde{s}}Xs
\end{equation*}
for all $X\in T\mathcal{M}(U)$ of parity $\tilde{X}$\,. It is easily checked that the maps $d_U$ are $\mathcal{O}(U)$-linear, commute with restrictions and preserve the parities, so that they define a morphism of sheaves of $\cO$-modules of parity zero.\medskip

In search of the coordinate expression of a $\mathbb{Z}_2$-differential $1$-form $\omega\in\Omega^1\mathcal{M}(U)$ for some super coordinate chart $U\in{\tt Open}(M)$ with coordinates $\mu=(x,\,\zx)$ we consider the differential $1$-forms $d\mu^A$, or more explicitly $dx^i$ and $d\zx^a$, induced by the local supercoordinate functions (for the sake of simplicity we write $d$ instead of $d_U$). They can be shown to form a basis for $\Omega^1\mathcal{M}(U)$ (see \cite{lectures}, page 66), which means that every $\omega\in\Omega^1\mathcal{M}(U)$ can uniquely be written as
\begin{equation}\label{omega}
    \omega=\sum_idx^i\text{w}_i(x,\,\zx)+\sum_ad\zx^aw_a(x,\,\zx)
\end{equation}
for some $\text{w}_i,\,w_a\in\mathcal{O}(U)$\,. Moreover, the existence of such a basis implies that $\Omega^1\mathcal{M}$ is a locally free sheaf of super $\mathcal{O}$-modules, which means in view of Definition \ref{supervb} that $\Omega^1\mathcal{M}$ is a $\mathbb{Z}_2$-vector bundle over $\mathcal{M}$ of rank $p|q$ and taking into account its relation with $T\mathcal{M}$ we often denote this vector bundle by $T^*\mathcal{M}$\,.

\begin{Example}
Applying $\omega$, decomposed as in \eqref{omega}, to $\partial_{\zx^b}$ yields
\begin{align*}
    \omega(\partial_{\zx^b})&=\sum_idx^i\text{w}_i(x,\,\zx)(\partial_{\zx^b})+\sum_ad\zx^aw_a(x,\,\zx)(\partial_{\zx^b})\\
    &=\sum_i(-1)^{\tilde{\omega}\cdot1}dx^i(\partial_{\zx^b})\text{w}_i(x,\,\zx)+\sum_a(-1)^{(\tilde{\omega}+1)\cdot1}d\zx^a(\partial_{\zx^b})w_a(x,\,\zx)\\
    &=(-1)^{\tilde{\omega}}w_b(x,\,\zx)\,,
\end{align*}
where the reason for the appearance of the signs $(-1)^{\tilde{\omega}\cdot1}$ and $(-1)^{(\tilde{\omega}+1)\cdot1}$ is supercommutativity and the fact that all $\text{w}_i$ must be of parity $\tilde{\omega}$, while all $w_a$ must be of parity $\tilde{\omega}+1$\,.
\end{Example}

A similar calculation leads to $\omega(\partial_{x^i})=\text{w}_i(x,\,\zx)$\,, hence we can conclude that the sections $\text{w}_i,\,w_a\in\mathcal{O}(U)$ that appear in \eqref{omega} are given by
\begin{align*}
    \text{w}_i(x\,\zx)&=\omega(\partial_{x^i})\\
    w_a(x,\,\zx)&=(-1)^{\tilde{\omega}}\omega(\partial_{\zx^a})\,.
\end{align*}
It follows that $d_U$ can be decomposed as $$d_U=\sum_idx^i\partial_{x^i}+\sum_ad\zx^a\partial_{\zx^a}=\sum_Ad\mu^A\partial_{\mu^A}\,.$$ Indeed if $f=f(x,\,\zx)$ is a superfunction, we obtain 
\begin{align*}
    d_Uf&=\sum_idx^i(d_Uf)(\partial_{x^i})+\sum_a(-1)^{\tilde{f}}d\zx^a(d_Uf)(\partial_{\zx^a})\\
    &=\sum_idx^i\partial_{x^i}f+\sum_a(-1)^{\tilde{f}}d\zx^a(-1)^{\tilde{f}}\partial_{\zx^a}f\\
    &=\left(\sum_idx^i\partial_{x^i}+\sum_ad\zx^a\partial_{\zx^a}\right)f\,.
\end{align*}\medskip

Moving on to the definition of super differential $2$-forms, or more generally super differential $k$-forms for some $k\geq0$\,, we begin by formally extending the operator $d:\Omega^0\mathcal{M}\rightarrow\Omega^1\mathcal{M}$ to act on a $\mathbb{Z}_2$-differential $1$-form of the form $df$ for some $f\in\mathcal{O}(U)$ and making sure this yields $0$ as should be expected in view of the definition of the de Rham differential in standard differential geometry. In the following equation the parity of an element is denoted by the same symbol as the element itself and Deligne sign convention is used. More details on this convention and an alternative will be discussed below. We compute
\begin{align*}
    d(df)&=\sum_Ad\mu^A\otimes\partial_{\mu^A}\left(\sum_Bd\mu^B\otimes\partial_{\mu^B}f\right)\\
    &=\sum_{AB}(-1)^{\mu^A\cdot\mu^B}d\mu^Ad\mu^B\otimes\partial_{\mu^A}\partial_{\mu^B}f\\
    &=\sum_{AB}(-1)^{\mu^A\cdot\mu^B}(-(-1)^{\mu^A\mu^B}d\mu^Bd\mu^A)\otimes((-1)^{\mu^A\mu^B}\partial_{\mu^B}\partial_{\mu^A})f\\
    &=-\sum_{AB}(-1)^{\mu^A\cdot\mu^B}d\mu^Bd\mu^A\otimes\partial_{\mu^B}\partial_{\mu^A}f\\
    &=-\sum_{AB}(-1)^{\mu^A\cdot\mu^B}d\mu^Ad\mu^B\otimes\partial_{\mu^A}\partial_{\mu^B}f\\
    &=0\,.
\end{align*}
The tensor product symbol $\otimes$ is used to stress that $d\mu^A$ is a map whose argument is a vector field and $\partial_{\mu^A}$ a map whose argument is a function. The Koszul sign $(-1)^{\mu^A\cdot\mu^B}$ appears on the second line of the equation due to the commutation of $\partial_{\mu^A}$ and $d\mu^B$ and the fact that $\partial_{\mu^A}$ is of parity $\mu^A$ and $d\mu^B$ is of parity $\mu^B$ by definition. The commutation of $\partial_{\mu^A}$ and $\partial_{\mu^B}$ causes the sign $(-1)^{\mu^A\cdot\mu^B}$ to appear since the basis elements $\partial_{\mu^A}$ are super commutative as can easily be checked. The basis elements $d\mu^A$ however are chosen to be super anticommutative, which is part of the Deligne sign convention mentioned above and leads to the apparition of the sign $-(-1)^{\mu^A\cdot\mu^B}$\,. Simplifying the resulting expression and interchanging the roles of $A$ and $B$ it becomes clear that the super differential $2$-form $d(df)$ is equal to its opposite and hence must be zero as required.

Above we made use of the Deligne sign convention by letting $d$ be even and letting the $\mathbb{Z}_2$-differential $1$-forms $d\mu^A$ be $\mathbb{Z}_2$-anticommutative. It can be shown that this convention is one of two possible settings in which the differential squares to $0$\,. The alternative is called Bernstein-Leites sign convention and involves defining $d$ to be odd and the $1$-forms $d\mu^A$ to be $\mathbb{Z}_2$-commutative.\medskip

To conclude this introduction to supergeometry we specify the local form of a general super differential $2$-form $\omega\in\Omega^2\mathcal{M}(U)$ for some super coordinate chart $U\in{\tt Open}(M)$ with coordinates $\mu=(x,\,\zx)$, namely
\begin{align*}
    \omega&=\sum_{AB}d\mu^Ad\mu^B\omega_{AB}(\mu)\\
    &=\sum_{i<j}dx^idx^jf_{ij}(x,\,\zx)+\sum_{i,a}dx^id\zx^ag_{ia}(x,\,\zx)+\sum_{a\leq b}d\zx^ad\zx^bh_{ab}(x,\,\zx)\,,
\end{align*}
for some $\omega_{AB},\,f_{ij},\,g_{ia},\,h_{ab}\in\mathcal{O}(U)\,,$ and the local form of a general super differential $k$-form $\omega\in\Omega^k\mathcal{M}(U)$, i.e.
\begin{equation*}
    \omega=\sum_{|\alpha|+|\beta|=k}(dx)^\alpha(d\zx)^\beta\omega_{\alpha\beta}(x,\,\zx)\,,
\end{equation*}
for some $\omega_{\alpha\beta}\in\mathcal{O}(U)$ and where $\alpha_1,...,\alpha_{p}\in\{0,\,1\}$ and $\beta_1,...,\beta_{q}\in\mathbb{N}$\,. The fact that the same differential of a formal parameter $d\zx^a$ can appear multiple times in the same term while the square of any basis element $dx^i$ vanishes follows from the $\mathbb{Z}_2$-anticommutativity of the elements $d\mu^A$\,.\medskip

It will prove important that the super anticommutivity of the differentials $d\mu^A$ reads \begin{equation*}
    d\mu^Ad\mu^B=-(-1)^{\mu^A\mu^B}d\mu^Bd\mu^A=(-1)^{1\cdot1+\mu^A\mu^B}d\mu^Bd\mu^A\,,
\end{equation*}
where the exponent in the last term is the sum of the products of the cohomological degrees of $d\mu^A$ and $d\mu^B$ and the parities of $d\mu^A$ and $d\mu^B$ respectively. More generally, the product $\odot$ (so far we have omitted the symbol $\odot$) of a super differential $k$-form $\omega_1\in\Omega^k\mathcal{M}(U)$ and a super differential $l$-form $\omega_2\in\Omega^l\mathcal{M}(U)$ satisfies
\begin{equation*}
    \omega_1\odot\omega_2=(-1)^{k\cdot l+\tilde{\omega}_1\tilde{\omega}_2}\omega_2\odot\omega_1\,,
\end{equation*}
where the exponent can be interpreted as the scalar product $\la(k,\,\tilde{\omega}_1),\,(l,\,\tilde{\omega}_2)\rangle\,,$ so that -- when taking the integers $k,l$ modulo 2 -- we have an example of a $\mathbb{Z}_2^2$-commutative algebra, which will be discussed in more detail in the next chapter. Using the Bernstein-Leites sign convention we obtain
\begin{equation*}
    \omega_1\odot\omega_2=(-1)^{(k+\tilde{\omega}_1)(l+\tilde{\omega}_2)}\omega_2\odot\omega_1\,.
\end{equation*}

\section{Introduction to higher supergeometry}

Having given an overview of the most important concepts in supergeometry we now move on to a more general setting, where the $\mathbb{Z}_2$-grading is replaced with a $\mathbb{Z}_2^n$-grading for an arbitrary $1\le n\in\mathbb{N}$\,. Here $\Z_2^n$ means $\Z_2^{\times n}=\Z_2\times\ldots\times\Z_2$ ($n$ factors). More precisely, coordinates in $\mathbb{Z}_2^2$-geometry may have the degree
\begin{equation*}
    (0,\,0),\,(0,\,1),\,(1,\,0)\text{ or }(1,\,1)\,,
\end{equation*}
the degrees of the coordinates in $\mathbb{Z}_2^3$-geometry are given by
\begin{equation*}
    (0,\,0,\,0),\,(0,\,0,\,1),\,(0,\,1,\,0),\,(1,\,0,\,0),\,(0,\,1,\,1),\,(1,\,0,\,1),\,(1,\,1,\,0)\text{ and }(1,\,1,\,1)
\end{equation*}
and hence in $\mathbb{Z}_2^n$-geometry coordinates can have $2^n$ different degrees, each with $n$ components in $\Z_2$. If the sum of the components of a $\Z_2^n$-degree equals $0$ modulo $2$ then the corresponding coordinate is even and otherwise it is odd. The commutation rule for coordinates in $\mathbb{Z}_2^n$-geometry generalizes the one in $\mathbb{Z}_2$-geometry since the product of the parities is replaced by the scalar product of the $\Z_2^n$-degrees. For instance this means that if $y$ and $\eta$ are of degree $(1,\,0,\,1)$ and $(0,\,0,\,1)$ respectively then we get
\begin{equation*}
y\cdot\eta=(-1)^{\langle(1,\,0,\,1),\,(0,\,0,\,1)\rangle}\eta\cdot y=-\eta\cdot y\,.
\end{equation*}
This new scalar product commutation rule does not have the same properties as the sign rule in classical supergeometry. Indeed, even coordinates may anticommute, odd coordinates may commute and non-zero degree even parameters are not nilpotent, all of which can easily be verified by means of the degrees in $\mathbb{Z}_2^3$-geometry.

\subsection{Motivation}

It is sufficient to study $\mathbb{Z}_2^n$-gradings with the above commutation rule since any sign rule for any finite number $m$ of coordinates has the form of a $\Z_2^n$-scalar-product commutation rule for some $n\leq2m$ (see \cite{sign}, page $4$). And it is necessary to study $\mathbb{Z}_2^n$-gradings since they appear among others in Physics, Algebra and Geometry as illustrated by the following examples.

\subsubsection{Physics}

String theory does not only make use of classical supergeometry but also benefits from results in $\mathbb{Z}_2^n$-geometry for $n>1$\,. Furthermore, $\mathbb{Z}_2^n$-gradings can be found in parastatistical supersymmetry. More precisely, in classical mechanics the distribution of particles over energy states is described by the Maxwell-Boltzmann statistics. If quantum effects must be taken into account, one uses the Bose-Einstein statistics and the Fermi-Dirac statistics when dealing with bosons and fermions respectively. Parastatistics is one of several alternative statistics and leads to paraparticles -- parabosons and parafermions -- and parastatistical supersymmetry.

\subsubsection{Algebra}

A $\mathbb{Z}_2^n$-commutative algebra for $n=2$ can be found when considering super differential forms on a smooth supermanifold $\mathcal{M}=(M,\cO_M)$. Indeed, using the Deligne sign convention the commutation of $\omega_1\in\Omega^k\mathcal{M}(M)$ and $\omega_2\in\Omega^l\mathcal{M}(M)$ is given by
\begin{equation*}
    \omega_1\odot\omega_2=(-1)^{k\cdot l+\tilde{\omega}_1\cdot\tilde{\omega}_2}\omega_2\odot\omega_1=(-1)^{\langle(k',\,\tilde{\omega}_1),\,(l',\,\tilde{\omega}_2)\rangle}\omega_2\odot\omega_1\,,
\end{equation*}
where $k'=k\mod2,\,l'=l\mod2$ and thus $(k',\,\tilde{\omega}_1),\,(l',\,\tilde{\omega}_2)\in\mathbb{Z}_2^2$\,.\medskip

Another example is the algebra $\mathbb{H}=\mathbb{R}\oplus i\mathbb{R}\oplus j\mathbb{R}\oplus k\mathbb{R}$ of quaternions. The products of the basis elements are defined by the relations
\begin{equation*}
    i^2=j^2=-1\,,\quad-ji=ij=k
\end{equation*}
together with the fact that $1$ is the multiplicative identity. Associativity can then be used to obtain the remaining product rules
\begin{equation*}
    ijk=k^2=-1\,,\quad-kj=jk=i\,,\quad-ik=ki=j\,.
\end{equation*}
The basis elements $\{1,\,i,\,j,\,k\}$ verify the scalar product commutation rule introduced above when assigning them the following even $\mathbb{Z}_2^3$-degrees:
\begin{equation*}
    \deg 1:=(0,\,0,\,0)\,,\quad\deg i:=(0,\,1,\,1)\,,\quad\deg j:=(1,\,0,\,1)\,,\quad\deg k:=(1,\,1,\,0)\,.
\end{equation*}
Therefore, if we denote by $(\Z_2^3)_{\op{ev}}$ the purely even part of the group $\Z_2^3$, the algebra $\mathbb{H}$ is $(\mathbb{Z}_2^3)_{\op{ev}}$-graded and $(\mathbb{Z}_2^3)_{\op{ev}}$-commutative in the sense of the scalar product commutation rule.\medskip

More generally, we can define the Clifford algebra $\mathscr{C}l_{p,q}(\mathbb{R})$ of signature $(p,\,q)$ over $\mathbb{R}$ (for some natural numbers $p$ and $q$ whose sum is denoted by $n$) as the associative unital $\mathbb{R}$-algebra generated by $(e_1,...,e_n)\in(\mathbb{R}^n)^n$ modulo the relations
\begin{align*}
    e_ie_j&=-e_je_i\quad\text{for all }i\neq j\\
    e_i^2&=1\quad\text{for }i\leq p\\
    e_j^2&=-1\quad\text{for }j>p\,.
\end{align*}
Then
\begin{equation*}
\mathscr{C}l_{p,q}(\mathbb{R})=\left\{\sum_{k=0}^n\sum_{i_1<\cdot\cdot\cdot <i_k}\mathbb{R}e_{i_1}\cdot\cdot\cdot e_{i_k}\right\}\,,
\end{equation*}
which is isomorphic as vector space to the exterior algebra $\wedge\mathbb{R}^n$ but not as algebra since for instance $e_i^2=\pm1$ for all $i\in\{1,...,n\}$ while $e_i\wedge e_i=0$ for all $i\in\{1,...,n\}$\,. Defining the degree of $e_i$ for every $i\in\{1,...,n\}$ as
\begin{equation*}
    \deg e_i:=(0,...,0,1,0,...,0,1)\,,
\end{equation*}
where the ones are in positions $i$ and $n+1$ of the vector, we can see that $\mathscr{C}l_{p,q}(\mathbb{R})$ becomes a $(\mathbb{Z}_2^{n+1})_{\op{ev}}$-commutative associative unital $\mathbb{R}$-algebra. This generalizes the previous example since the algebra $\mathbb{H}$ of quaternions is nothing more than the algebra $\mathscr{C}l_{0,3}(\mathbb{R})$\,.

\subsubsection{Geometry}

In geometry $\mathbb{Z}_2^n$-manifolds arise naturally as illustrated by the following example.
We start with a smooth supermanifold $\mathcal{M}$ of dimension $p|q$ with supercoordinates $(x,\,\zx)$\,, i.e. coordinates $x$ of parity $0$ and formal parameters $\zx$ of parity $1$\,. Since a basis of the dual gives coordinates on the original space, we denote the supercoordinates of the tangent bundle $T\mathcal{M}$ of $\cM$ by $(x,\,\zx,\,dx,\,d\zx)\,$. If we adopt the Bernstein-Leites sign convention, we consider $d$ odd and use the $\Z_2$-commutation rule. This leads to coordinates $(x,\,\zx,\,dx,\,d\zx)\,$ of $\Z_2$-degrees $(0,1,1,0)$ and to a $\mathbb{Z}_2$-manifold structure on $T\cM$ whose function sheaf is over the coordinate domain $U$ given by
\begin{equation*}
    \mathcal{C}^\infty_{p+q|p+q}(U)=\mathcal{C}^\infty(x,\,d\zx)[\zx,\,dx]\,.
\end{equation*}
On the other hand, if we use the Deligne sign convention, we consider $d$ even and use the $\Z_2^2$-commutation rule for the bidegree made of the cohomological degree modulo 2 and the parity. This leads to coordinates $(x,\,\zx,\,dx,\,d\zx)$ of $\Z_2^2$-degrees $$((0,0),(0,1),(1,0),(1,1))$$ and to a $\Z_2^2$-manifold structure on $T\cM$ whose function sheaf is over $U$ given by
\begin{equation*}
    \mathcal{C}^\infty_{p|(q,q,p)}(U)=\mathcal{C}^\infty(x)\llbracket d\zx,\,\zx,\,dx\rrbracket\,,
\end{equation*}
where $\llbracket d\zx,\,\zx,\,dx\rrbracket$ represents formal power series in $d\zx,\,\zx$ and $dx$\,. Reasons for the use of formal power series will be given below. Notice that the $\Z_2^2$-degrees carry richer information than the corresponding $\Z_2$-degrees and that in the $\Z_2^2$-manifold we do not need consider the differential $d\zx$ of a parameter as a standard base variable as in the corresponding $\Z_2$-manifold.

\subsection{Smooth $\mathbb{Z}_2^n$-manifolds}\label{z2nman}

We start by explaining why in the local representations of superfunctions in higher supergeometry there appear formal series in the parameters $y:=d\zx\,$, $\zx$ and $\eta:=dx\,$. As mentioned before non-zero degree even coordinates are not nilpotent in $\mathbb{Z}_2^n$-geometry. In the case of $\mathbb{Z}_2^2$-coordinates as described above for instance we have
\begin{equation*}
    y^2=(-1)^{\langle(1,\,1),\,(1,\,1)\rangle}y^2=y^2\,,
\end{equation*}
which means that $y$ is not nilpotent. Consider now the coordinate transformation given by
\begin{align*}
    &x'=x+y^2&\zx'=\zx\\
    &y'=y&\eta'=\eta
\end{align*}
and apply the formal Taylor expansion to express a function $F$ in $x'$ as a function in the original coordinates\,:
\begin{equation*}
    F(x')=F(x+y^2)=\sum_\alpha\frac{1}{\alpha!}(\partial_y^\alpha F)(x)y^{2\alpha}\,,
\end{equation*}
where the pullback has been omitted. Since $y$ is not nilpotent the sum on the right-hand side is not necessarily finite and is therefore a power series in $y$\,. Combining this with the fact that the pullback of a superfunction on the target space must be a superfunction on the source space it becomes clear that superfunctions in higher geometry must be represented by power series. It should be noted that these power series are formal and thus there is no need to question whether they converge.\medskip

The most general form of a $\mathbb{Z}_2^2$-morphism can be found observing that $\zx$ and $\eta$ are nilpotent and checking which degree corresponds to different powers of $y$ and to different combinations of the parameters. It is given by
\begin{align*}
    &x'=\sum_rf_r^{x'}(x)y^{2r}+\sum_rg_r^{x'}(x)y^{2r+1}\zx\eta&\zx'=\sum_rf_r^{\zx'}(x)y^{2r}\zx+\sum_rg_r^{\zx'}(x)y^{2r+1}\eta\\
    &y'=\sum_rf_r^{y'}(x)y^{2r+1}+\sum_rg_r^{y'}(x)y^{2r}\zx\eta&\eta'=\sum_rf_r^{\eta'}(x)y^{2r}\eta+\sum_rg_r^{\eta'}(x)y^{2r+1}\zx\;.
\end{align*}

Concerning notation we observe that the abelian group $\mathbb{Z}_2^n$ has $2^n$ elements, $2^{n-1}$ of them are even and the remaining $2^{n-1}$ elements are odd. We order these $2^n$ elements by first ordering the $2^{n-1}$ even elements lexicographically and then ordering the $2^{n-1}$ odd elements lexicographically. For instance in the case of $\Z_2^2$ this {\it standard ordering} leads to $$((0,0),(1,1),(0,1),(1,0))\;.$$ Further we denote the $i$-th element of $\Z_2 ^n$ by $\gamma_i$ for $i\in\{0,\,1,...,2^n-1\}$. As explained above a $\Z_2^n$-manifold can have supercoordinates of all $\Z_2^n$-degrees $\zg_i\,.$ The standard base coordinates $x=(x^1,...,x^p)\in\mathbb{R}^p$ are all of degree $\gamma_0=(0,...,0)$ while the formal parameters are summarized as $\zx=(\zx^1,...,\zx^q)$ and if we denote by $q_i$ the number of parameters that have degree $\gamma_i$ then $\underline{q}=(q_1,...,q_{2^n-1})$ is a tuple of $2^n-1$ natural numbers whose sum is $q\,.$ Thus the sheaf of superfunctions on a $\mathbb{Z}_2^n$-domain $\mathbb{R}^{p|\underline{q}}$ of dimension $p|\underline{q}$ is defined as
\begin{equation*}
    \mathcal{C}^\infty_{p|\underline{q}}(U):=\mathcal{C}^\infty(U)\llbracket\zx^1,...,\zx^q\rrbracket
\end{equation*}
for every $U\in{\tt Open}(\mathbb{R}^p)$\,.\medskip

Similarly to super ringed spaces and supermanifolds we now define locally $\mathbb{Z}_2^n$-ringed spaces and $\mathbb{Z}_2^n$-manifolds.

\begin{Definition}
    A $\mathbb{Z}_2^n$-ringed space is a pair $(M,\,\mathcal{O}_M)$ consisting of a topological space $M$ and a sheaf $\mathcal{O}_M$ of $\mathbb{Z}_2^n$-graded $\mathbb{Z}_2^n$-commutative (in the sense of the scalar product commutation rule) associative unital $\mathbb{R}$-algebras over $M$. If additionally, for every $x\in M$\,, the stalk $\mathcal{O}_x$ has a unique homogeneous maximal ideal we say that $(M,\,\mathcal{O}_M)$ is a locally $\mathbb{Z}_2^n$-ringed space.
\end{Definition}

\begin{Definition}
    A smooth $\mathbb{Z}_2^n$-manifold of dimension $p|\underline{q}$ is a locally $\mathbb{Z}_2^n$-ringed space $\mathcal{M}=(M,\,\mathcal{O}_M)$\,, where $M$ is a second countable Hausdorff topological space, that is locally isomorphic to the smooth $\mathbb{Z}_2^n$-domain $\mathbb{R}^{p|\underline{q}}=(\mathbb{R}^p,\,\mathcal{C}^\infty_{p|\underline{q}})$\,.
\end{Definition}

\subsection{Fundamental results in higher supergeometry}

Even though most results from supergeometry are also valid in higher supergeometry they often require different or more subtle proofs, which will be illustrated in this section by means of two important theorems. Furthermore it should be remarked that while the theory of supergeometry originates from a model in Physics and thus contains some developments that are not entirely precise (or even wrong), higher supergeometry has been designed carefully from scratch using mathematical tools. The main difference between $\mathbb{Z}_2$-geometry and $\mathbb{Z}_2^n$-geometry can be found in integration theory, which will be introduced in Chapter \ref{inttheo}.

\subsubsection{Invertibility of $\Z_2^n$-functions}

In Proposition \ref{inv} we proved that a superfunction $f\in\mathcal{C}^\infty_{p|q}(U)$ is invertible if and only if its parameter-independent term $\varepsilon_U(f)=f_0\in\mathcal{C}^\infty(U)$ is invertible. The corresponding fundamental result of $\mathbb{Z}_2^n$-geometry reads as follows.

\begin{Theorem}\label{inv2}
A $\Z_2^n$-function $$f\in\mathcal{C}^\infty_{p|\underline{q}}(U)=\mathcal{C}^\infty(U)\llbracket\zx^1,...,\zx^q\rrbracket$$ is invertible if and only if $f_0\in\mathcal{C}^\infty(U)$, the term of $f$ that does not contain any of the parameters $\zx^a$\,, is invertible.
\end{Theorem}
\begin{proof}
Similarly to the proof of Proposition \ref{inv} it suffices to show that $1-t$ is invertible for any element $t\in\mathcal{C}^\infty_{p|\underline{q}}(U)$ that only consists of terms that contain at least one of the parameters $\zx^a$\,. Since the proof of Proposition \ref{inv} relies on the fact that the parameters $\zx^a$ are nilpotent and in $\mathbb{Z}_2^n$-geometry there exist parameters that are not nilpotent it has to be adapted in order to hold in the $\Z_2^n$-context.\medskip

We claim that the inverse of $1-t$ is given by $\sum_{l=0}^\infty t^l\in\mathcal{C}^\infty_{p|\underline{q}}(U)$ and start by showing that $\sum_{l=0}^\infty t^l$ is indeed an element of $\mathcal{C}^\infty_{p|\underline{q}}(U)$\,. If $t$ is given by
\begin{equation*}
    t=\sum_{k=1}^\infty\sum_{|\alpha|=k}f_\alpha(x)\zx^\alpha=\sum_{|\alpha|\geq1}f_\alpha(x)\zx^\alpha\;,
\end{equation*}
we have
\begin{align*}
    \sum_{l=0}^\infty t^l&=\sum_{l=0}^\infty\left(\sum_{|\alpha_1|\geq1}f_{\alpha_1}(x)\zx^{\alpha_1}\cdot...\cdot\sum_{|\alpha_l|\geq1}f_{\alpha_l}(x)\zx^{\alpha_l}\right)\\
    &=\sum_{l=0}^\infty\sum_{|\alpha_i|\geq1,\forall i}f_{\alpha_1}(x)\cdot...\cdot f_{\alpha_l}(x)\zx^{\alpha_1}\cdot...\cdot\zx^{\alpha_l}\\
    &=\sum_{l=0}^\infty\sum_{|\beta|=l}^\infty\left(\sum_{\substack{\alpha_1+...+\alpha_l=\beta\\|\alpha_i|\geq1,\forall i}}\pm f_{\alpha_1}(x)\cdot...\cdot f_{\alpha_l}(x)\right)\zx^\beta\\
    &=\sum_{|\beta|=0}^\infty\left(\sum_{l=0}^{|\beta|}F_\beta^l(x)\right)\zx^\beta\\
    &=\sum_\beta F_\beta(x)\zx^\beta\in\mathcal{C}^\infty_{p|\underline{q}}(U)\,,
\end{align*}
where $F_\beta^l\in\mathcal{C}^\infty(U)$ since the sum over all $\alpha_1,...,\alpha_l$ such that $\alpha_1+...+\alpha_l=\beta$ and $|\alpha_i|\geq1,\forall i$ is finite and $f_{\alpha_i}\in\mathcal{C}^\infty(U)$ for every $\alpha_i$, which in turn implies that $F_\beta\in\mathcal{C}^\infty(U)$ since the sum $\sum_{l=0}^{|\beta|}$ is finite. Moreover $\zx^\beta$ means that the powers $\zx^{a^{\alpha_{i,a}}}$ of parameters have been regrouped taking into account first the index $a$ and then the index $\alpha_i$\,, which might change the sign of some of the terms due to $\mathbb{Z}_2^n$-commutativity. To conclude the proof that $\sum_{l=0}^\infty t^l$ is the inverse of $1-t$ we observe that
\begin{equation*}
    (1-t)\sum_{l=0}^\infty t^l=\sum_{l=0}^\infty t^l-\sum_{l=1}^\infty t^l=t^0=1
\end{equation*}
and analogously $\sum_{l=0}^\infty t^l(1-t)=1$. Hence, while in the super case nilpotency allowed us to conclude, it is here the fact that we replaced polynomials with formal power series.
\end{proof}

\subsubsection{Higher morphism theorem}

In order to extend Theorem \ref{fundathm} to higher supergeometry we need to use the fact that $\mathcal{O}_M$\,, the structure sheaf of the source space $\mathcal{M}=(M,\,\mathcal{O}_M)$ of the considered supermorphism, is Hausdorff-complete. What this means and how it can be used to prove the fundamental theorem of supermorphisms in $\mathbb{Z}_2^n$-geometry is discussed in the following.\medskip

To show that the field of rational numbers $\mathbb{Q}$ is not complete we can resort to the sequence $(x_n)$ of rational numbers defined by
\begin{equation*}
x_1=1,\quad x_{n+1}=\frac{x_n}{2}+\frac{1}{x_n}\,.
\end{equation*}
It can easily be verified that $(x_n)$ is a Cauchy sequence with respect to the standard norm on $\mathbb{Q}$ given by the absolute value of the difference and that the limit $x$ of $(x_n)$\,, if it exists, must satisfy $x^2=2$\,, which leads to $x=\pm\sqrt{2}\not\in\mathbb{Q}$\,. Therefore there exist Cauchy sequences of rational numbers that do not converge in $\mathbb{Q}$\,.\medskip

To show that the ring $\mathbb{R}[x]$ of polynomials in $x$ with coefficients in $\mathbb{R}$ evaluated at $x\in[0,\,1]$ is not complete consider the sequence of polynomials $(p_n)$ given by
\begin{equation*}
    p_n(x)=\sum_{k=0}^n\left(\frac{x}{2}\right)^k\,.
\end{equation*}
Then $(p_n)$ is clearly a Cauchy sequence with respect to the norm $||-||_\infty$ defined by
\begin{equation*}
    ||p(x)||_\infty=\sup_{x\in[0,1]}|p(x)|\,.
\end{equation*}
Since $(p_n)$ is a geometric series and $|\frac{x}{2}|<1$ the limit of $(p_n)$ is $(1-\frac{x}{2})^{-1}\not\in\mathbb{R}[x]$\,, proving the existence of Cauchy sequences in $\mathbb{R}[x]$ that do not converge in $\mathbb{R}[x]$\,.\medskip

Since rational functions are real analytic, the algebra $\R\llbracket x\rrbracket$ of formal power series should be complete. Likewise, for every $U\in{\tt Open}(M)\,,$ the model $\mathbb{Z}_2^n$-function algebra $\mathcal{C}^\infty_{p|\underline{q}}(U)$ should be complete. However, we first need to equip it with a norm, or equivalently with a topology, and define Cauchy sequences and convergence of sequences with respect to this norm in order to allow for a notion of completeness on $\mathcal{C}^\infty_{p|\underline{q}}(U)$ and thereby on the $\mathbb{Z}_2^n$-function algebra $\mathcal{O}_M(U)$\,. Denoting $\mathcal{C}^\infty_{p|\underline{q}}(U)=\mathcal{C}^\infty(U)\llbracket\zx\rrbracket$ by $\mathcal{A}$ and the kernel $\mathcal{J}(U)$ of the projection $\varepsilon_U$ by $\mathcal{I}\,,$ we consider the $\mathcal{I}$-adic topology introduced in Section \ref{continuity} by means of the basis
\begin{equation*}
    \{\rho+\mathcal{I}^k\,:\,\rho\in\mathcal{A},\,k\in\mathbb{N}\}\,.
\end{equation*}

\begin{Definition}
    A sequence $(a_n)_{n\in\mathbb{N}}\subseteq\mathcal{A}$ is a Cauchy sequence if for every $k\in\mathbb{N}$ there exists $l\in\mathbb{N}$ such that $a_r-a_s\in\mathcal{I}^k$ for all $r,\,s\geq l\,.$
\end{Definition}

\begin{Definition}
    A sequence $(a_n)_{n\in\mathbb{N}}\subseteq\mathcal{A}$ converges to $a\in\mathcal{A}$ if for every $k\in\mathbb{N}$ there exists $l\in\N$ such that $a_n-a\in\mathcal{I}^k$ for all $n\geq l$\,.
\end{Definition}

Now consider the decreasing sequence of ideals
\begin{equation*}
    \mathcal{A}\supseteq\mathcal{I}\supseteq\mathcal{I}^2\supseteq\mathcal{I}^3\supseteq\cdot\cdot\cdot
\end{equation*}
and take quotients of $\mathcal{A}$ to obtain
\begin{equation}\label{invsys}
    \mathcal{A}/\mathcal{A}\leftarrow\mathcal{A}/\mathcal{I}\leftarrow\mathcal{A}/\mathcal{I}^2\leftarrow\mathcal{A}/\mathcal{I}^3\leftarrow\cdot\cdot\cdot\,,
\end{equation}
where $\mathcal{A}/\mathcal{I}$ represents the superfunctions that do not contain any formal parameters, $\mathcal{A}/\mathcal{I}^2$ represents the superfunctions consisting of terms with at most one formal parameter and the arrows denote the natural projections. Then \eqref{invsys} is an inverse system and it can be shown that its inverse limit is given by
\begin{equation*}
    \lim_{\overleftarrow{\hspace{2mm}k\hspace{2mm}}}\mathcal{A}/\mathcal{I}^k\cong\mathcal{A}\,,
\end{equation*}
which constitutes the definition of Hausdorff-completeness: the algebra $\mathcal{A}$ is Hausdorff-complete with respect to the $\mathcal{I}$-adic topology. For more details see \cite{sign}, page 13.  We use without proof the result that Hausdorff-completeness implies standard completeness, which allows us to make use of the fact that every Cauchy sequence in $\mathcal{A}$ converges to a limit in $\mathcal{A}$ in the following proof of the fundamental theorem of $\Z_2^n$-morphisms.

\begin{Theorem}
We consider a $\mathbb{Z}_2^n$-manifold $\mathcal{M}=(M,\,\mathcal{O}_M)$\,, a $\mathbb{Z}_2^n$-domain $\mathcal{V}^{r|\underline{s}}=(V,\,\mathcal{C}^\infty_{r|\underline{s}})$ with coordinates $(y,\,\eta)$ and $\Z_2^n$-functions
\begin{equation*}
    s^1,...,\,s^r,\,\sigma^1,...,\sigma^s\in\mathcal{O}_M(M)
\end{equation*}
that verify
\begin{align*}
    \deg(s^i)&=\deg(y^i),\quad\text{for }i\in\{1,...,r\}\,,\\
    \deg(\sigma^a)&=\deg(\eta^a),\quad\text{for }a\in\{1,...,s\}\,
\end{align*}
and
\begin{equation*}
    (\varepsilon s^1,...,\varepsilon s^r)(M)\subseteq V\;.
\end{equation*}
Then there exists a unique morphism of $\mathbb{Z}_2^n$-manifolds
\begin{equation*}
    \Phi=(\phi,\,\phi^*):\mathcal{M}\rightarrow\mathcal{V}^{r|\underline{s}}\,,
\end{equation*}
such that
\begin{equation*}
    s^i=\phi_V^*y^i\hspace{2cm}\text{and}\hspace{2cm}\sigma^a=\phi^*_V\eta^a\,.
\end{equation*}
\end{Theorem}
\begin{proof}
To begin with we show how uniqueness of the algebra morphism
\begin{equation*}
\phi^*_W:\mathcal{C}^\infty_{r|s}(W)\rightarrow\mathcal{O}_M(\phi^{-1}(W))
\end{equation*}
for all $W\in{\tt Open}(V)$ can be proved in the case of $\mathbb{Z}_2$-manifolds in order to highlight the similarities and differences between both cases. If the required algebra morphism $\phi^*_W$ exists then its value on a superfunction $\sum_\alpha f_\alpha(y)\eta^\alpha\in\mathcal{C}^\infty_{r|s}(W)$ must necessarily be given by
\begin{equation*}
    \phi^*_W\left(\sum_\alpha f_\alpha(y)\eta^\alpha\right)
    =\phi^*_W\left(\sum_{k=0}^n\sum_{|\alpha|=k}f_\alpha(y)\eta^\alpha\right)
    =\sum_{k=0}^n\sum_{|\alpha|=k}\phi^*_W(f_\alpha(y))(\phi^*_W\eta)^\alpha\,.
\end{equation*}
The pullback $\phi^*_W\eta$ is $\sigma$ by the requirements of the theorem and if $f_\alpha(y)$ is a polynomial $$\sum_\beta r^\za_\beta y^\beta=\sum_{l=0}^{N_\za}\sum_{|\beta|=l}r^\za_\beta y^\beta\;,$$ then we necessarily have
\begin{equation*}
    \phi^*_W(f_\alpha(y))
    =\phi^*_W\left(\sum_{l=0}^{N_\za}\sum_{|\beta|=l}r^\za_\beta y^\beta\right)
    =\sum_{l=0}^{N_\za}\sum_{|\beta|=l}r^\za_\beta(\phi^*_Wy)^\beta
\end{equation*}
with $\phi^*_Wy=s$. Hence $\phi^*_W$\,, if it exists, is uniquely determined on polynomials in $\eta$ with coefficients in polynomials in $y$ and in view of polynomial approximation (see \cite{lectures}, page 51) $\phi^*_W$ is unique on all superfunctions in $\mathcal{C}^\infty_{r|s}(W)$\,.\medskip

Switching to $\mathbb{Z}_2^n$-geometry, we assume again that the required algebra morphism
\begin{equation*}
    \phi^*_W:\mathcal{C}^\infty_{r|\underline{s}}(W)\rightarrow\mathcal{O}_M(U)\,,
\end{equation*}
where $U=\phi^{-1}(W)$\,, exists for all $W\in{\tt Open}(V)$ and show that it is uniquely determined on an arbitrary $\mathbb{Z}_2^n$-function $\sum_\alpha f_\alpha(y)\eta^\alpha\in\mathcal{C}^\infty_{r|\underline{s}}(W)$\,. In this case the fact that $\phi^*_W$ is an algebra morphism cannot be used to bring it inside the sum since we are dealing with series. Therefore, we adopt the following notation for the time being: 
\begin{equation*}
    \phi^*_W\left(\sum_\alpha f_\alpha(y)\eta^\alpha\right)
    =\phi^*_W\left(\sum_{k=0}^\infty\sum_{|\alpha|=k}f_\alpha(y)\eta^\alpha\right)=:a\,.
\end{equation*}
However, for any $n\in\mathbb{N}$ we can apply $\phi^*_W$ to the above $\mathbb{Z}_2^n$-function truncated at its $(n+1)$-th term to obtain
\begin{equation}\label{a_n}
    \phi^*_W\left(\sum_{k=0}^n\sum_{|\alpha|=k}f_\alpha(y)\eta^\alpha\right)
    =\sum_{k=0}^n\phi^*_W\left(\sum_{|\alpha|=k}f_\alpha(y)\eta^\alpha\right)
    =\sum_{k=0}^n\sum_{|\alpha|=k}\phi^*_W(f_\alpha(y))(\phi^*_W\eta)^\alpha\,,
\end{equation}
where the right-hand side is a section in $\mathcal{O}_M(U)$ and will be denoted by $a_n$. The sequence $(a_n)_{n\in\mathbb{N}}\subseteq\mathcal{O}_M(U)$ is Cauchy, which can be seen by considering for $r>s$ the difference
\begin{align*}
    a_r-a_s&=\sum_{k=0}^r\sum_{|\alpha|=k}\phi^*_W(f_\alpha(y))(\phi^*_W\eta)^\alpha-\sum_{k=0}^s\sum_{|\alpha|=k}\phi^*_W(f_\alpha(y))(\phi^*_W\eta)^\alpha\\
    &=\sum_{k=s+1}^r\sum_{|\alpha|=k}\phi^*_W(f_\alpha(y))(\phi^*_W\eta)^\alpha\,.
\end{align*}
Looking back on Equation \eqref{a_n} we note that $\sum_{|\alpha|=k}f_\alpha(y)\eta^\alpha\in\mathcal{J}^k(W)$\,, which implies due to continuity of $\phi^*_W$ that
\begin{equation*}
    \phi^*_W\left(\sum_{|\alpha|=k}f_\alpha(y)\eta^\alpha\right)
    =\sum_{|\alpha|=k}\phi^*_W(f_\alpha(y))(\phi^*_W\eta)^\alpha\in\mathcal{J}^k(U)\,.
\end{equation*}
Since $\mathcal{J}^k(U)\subseteq\mathcal{J}^{s+1}(U)$ for all $k\in\{s+1,...,r\}$ we have $a_r-a_s\in\mathcal{J}^{s+1}(U)$\,, which can be reformulated by saying that $a_r-a_s\in\mathcal{J}^N(U)$ whenever $r>s\geq N-1$\,. As $\mathcal{O}_M(U)$ is complete the Cauchy sequence $(a_n)$ has a unique limit in $\mathcal{O}_M(U)$\,, which we denote by
\begin{equation*}
    \lim_na_n=:\sum_{k=0}^\infty\sum_{|\alpha|=k}\phi^*_W(f_\alpha(y))(\phi^*_W\eta)^\alpha\,.
\end{equation*}
But arguing similarly as above we have
\begin{align*}
    a-a_n&=\phi^*_W\left(\sum_{k=0}^\infty\sum_{|\alpha|=k}f_\alpha(y)\eta^\alpha\right)-\sum_{k=0}^n\sum_{|\alpha|=k}\phi^*_W(f_\alpha(y))(\phi^*_W\eta)^\alpha\\
    &=\phi^*_W\left(\sum_{k=n+1}^\infty\sum_{|\alpha|=k}f_\alpha(y)\eta^\alpha\right)\in\mathcal{J}^{n+1}(U)\,,
\end{align*}
so that $a-a_n\in\mathcal{J}^N(U)$ whenever $n\geq N-1$ and by uniqueness of the limit we obtain
\begin{equation*}
    a=\phi^*_W\left(\sum_{k=0}^\infty\sum_{|\alpha|=k}f_\alpha(y)\eta^\alpha\right)
    =\sum_{k=0}^\infty\sum_{|\alpha|=k}\phi^*_W(f_\alpha(y))(\phi^*_W\eta)^\alpha\,.
\end{equation*}
Arguing similarly as in the $\mathbb{Z}_2$-case and applying the $\mathbb{Z}_2^n$-version of polynomial approximation (see \cite{sign}, page 14) we can thus state that $\phi^*_W$ is uniquely determined on all $\mathbb{Z}_2^n$-functions in $\mathcal{C}^\infty_{r|\underline{s}}(W)$\,.
The remaining part of the theorem can be proved as in the $\mathbb{Z}_2$-case (see \cite{sign}, page 14).
\end{proof}

\section{Integration theory}\label{inttheo}

\subsection{Linear $\mathbb{Z}_2$-algebra}

\subsubsection{$\Z_2$-modules and linear maps}

Let $\mathcal{A}$ be a $\mathbb{Z}_2$-algebra over $\mathbb{R}$\,, i.e. a $\mathbb{Z}_2$-vector space over $\mathbb{R}$ equipped with a $\mathbb{Z}_2$-commutative associative unital $\mathbb{R}$-bilinear multiplication $\cdot$ that is compatible with the $\mathbb{Z}_2$-grading in the sense that $\mathcal{A}_i\cdot\mathcal{A}_j\subseteq\mathcal{A}_{i+j}$\,. Let $M$ be a $\mathbb{Z}_2$-module over $\mathcal{A}$\,, i.e. a $\mathbb{Z}_2$-abelian group together with an $\mathcal{A}$-action $\triangleleft$ that is compatible with the $\mathbb{Z}_2$-grading in the sense that $\mathcal{A}_i\triangleleft M_j\subseteq M_{i+j}$\,.
\begin{Remark}
Recall that a left action $\triangleleft$ verifies for all $\alpha,\,\beta\in\mathcal{A}$ and all $m,\,m'\in M\,,$
\begin{enumerate}
    \item $\alpha\triangleleft(\beta\triangleleft m)=(\alpha\cdot\beta)\triangleleft m\,,$
    \item $1_{\mathcal{A}}\triangleleft m=m\,,$
    \item $(\alpha+\beta)\triangleleft m=\alpha\triangleleft m+\beta\triangleleft m\,,$
    \item $\alpha\triangleleft(m+m')=\alpha\triangleleft m+\alpha\triangleleft m'$
\end{enumerate}
and that due to supercommutivity there is a one-to-one correspondence between left and right actions, for instance each left action $\triangleleft$ induces a right action $\triangleright$ by setting
$$m\triangleright\alpha:=(-1)^{\tilde{\alpha}\tilde{m}}\alpha\triangleleft m$$ for all $\alpha\in\mathcal{A}$ and all $m\in M$\,.
\end{Remark}

\begin{Definition}
    The set of linear maps between two $\mathbb{Z}_2$-modules $M$ and $N$ over $\mathcal{A}$ is defined as
    \begin{equation*}
        \Hom_\mathcal{A}(M,\,N):=\Hom_{\mathcal{A},0}(M,\,N)\oplus\Hom_{\mathcal{A},1}(M,\,N)\,,
    \end{equation*}
    where a linear map $\lambda\in\Hom_{\mathcal{A},\tilde{\lambda}}(M,\,N)$ of degree $\tilde{\lambda}\in\{0,\,1\}$ is an additive map $\lambda:M_i\rightarrow N_{i+\tilde{\lambda}}$ that satisfies
    \begin{equation*}
        \lambda(\alpha\triangleleft m)=(-1)^{\tilde{\zl}\tilde{\alpha}}\alpha\triangleleft\lambda(m)
    \end{equation*}
    or, equivalently, in terms of the corresponding right action $$\lambda(m\triangleright\alpha)=\lambda(m)\triangleright\alpha\;.$$
\end{Definition}

Then $\Hom_\mathcal{A}(M,\,N)$ is a $\mathbb{Z}_2$-abelian group as direct sum of abelian groups. The action $\za\triangleleft\zl$ of $\za\in\mathcal{A}$ on $\zl\in\Hom_{\mathcal{A}}(M,N)$ defined by
\begin{equation*}
    (\alpha\triangleleft\lambda)(m):=\alpha\triangleleft\lambda(m)
\end{equation*}
for all $m\in M,$ is a new linear map $\za\triangleleft \zl\in\Hom_{\mathcal{A}}(M,N)$ in view of the $\Z_2$-commutativity of the multiplication $\cdot$ in $\mathcal{A}$\,. Hence the group $\Hom_\mathcal{A}(M,\,N)$ of linear maps between $\mathbb{Z}_2$-modules over $\mathcal{A}$ is itself a $\mathbb{Z}_2$-module over $\mathcal{A}$\,.

\begin{Remark}
In the following the symbols $\cdot\,,\,\triangleleft$ and $\triangleright$ will be omitted.
\end{Remark}

In standard non-graded linear algebra an element $m$ in a free module $M$ over some commutative
algebra $\mathcal{A}$ of rank $p$ can be identified with a vector
\begin{equation*}
    m\cong
    \begin{pmatrix}
    m^1\\\vdots\\m^p
    \end{pmatrix}
    \in\mathcal{A}^p\,.
\end{equation*}
A linear map $l\in\Hom_\mathcal{A}(M,\,N)$ between free modules of rank $p$ and $r$ can then be identified with a matrix $L\in\text{gl}(r\times p,\,\mathcal{A})$\,, where $\text{gl}(r\times p,\,\mathcal{A})$ denotes the space of $r\times p$ matrices with entries in $\mathcal{A}$\,, so that multiplying $L$ with the representative vector of $m$ we obtain the representative vector of $l(m)$\,.\medskip

We have similar vector and matrix representations in linear $\mathbb{Z}_2$-algebra. Let $M$ be a free $\mathbb{Z}_2$-module of rank $p|q$ over a $\mathbb{Z}_2$-commutative associative unital $\mathbb{R}$-algebra $\mathcal{A}$\,. If $M$ has the basis $(e_1,...,e_p,\,e_{p+1},...,e_{p+q})$\,, where the first $p$ elements are even and the remaining elements are odd, then every $m\in M$ reads uniquely as
\begin{equation*}
    m=\sum_{i=1}^pe_im^i+\sum_{a=1}^qe_{p+a}m^{p+a}=\sum_Ae_Am^A
\end{equation*}
for some $m^1,...,m^{p+q}\in\mathcal{A}$\,. Therefore, $m$ can be represented by the vector
\begin{equation*}\renewcommand*{\arraystretch}{1.4}
m\cong\left(
\begin{array}{c}
m^1\\\vdots\\m^p\\\hline m^{p+1}\\\vdots\\m^{p+q}
\end{array}\right)
\in\mathcal{A}^{p|q}\,,
\end{equation*}
where
\begin{equation*}
    m^1,...,m^p\in\mathcal{A}_0\text{ and }m^{p+1},...,m^{p+q}\in\mathcal{A}_1
\end{equation*}
when $m$ is even, whereas
\begin{equation*}
    m^1,...,m^p\in\mathcal{A}_1\text{ and }m^{p+1},...,m^{p+q}\in\mathcal{A}_0
\end{equation*}
when $m$ is odd. As indicated above the space containing such vectors is denoted by $\mathcal{A}^{p|q}$\,.\medskip

Moreover, a linear map $\lambda\in\Hom_\mathcal{A}(M,\,N)$ between free $\Z_2$-modules of rank $p|q$ and $r|s$ has a representative $\mathbb{Z}_2$-matrix
\begin{equation*}\renewcommand*{\arraystretch}{1.4}
    \Lambda=\left(
    \begin{array}{c|c}
    A&B\\\hline
    C&D
    \end{array}\right)
    \in\mathbb{Z}_2\,\text{gl}(r|s\times p|q,\,\mathcal{A})
\end{equation*}
with
\begin{equation*}
    A\in\text{gl}(r\times p,\,\mathcal{A}_0)\,, B\in\text{gl}(r\times q,\,\mathcal{A}_1)\,, C\in\text{gl}(s\times p,\,\mathcal{A}_1)\text{ and } D\in\text{gl}(s\times q,\,\mathcal{A}_0)
\end{equation*}
when $\lambda$ is even and with
\begin{equation*}
    A\in\text{gl}(r\times p,\,\mathcal{A}_1)\,, B\in\text{gl}(r\times q,\,\mathcal{A}_0)\,, C\in\text{gl}(s\times p,\,\mathcal{A}_0)\text{ and } D\in\text{gl}(s\times q,\,\mathcal{A}_1)
\end{equation*}
when $\lambda$ is odd. Depending on the parity of $\lambda$ we refer to $\Lambda$ as an even respectively as an odd $\mathbb{Z}_2$-matrix. As indicated above the space of $\mathbb{Z}_2$-matrices of size $r|s\times p|q$ with entries in $\mathcal{A}$ is denoted by $\mathbb{Z}_2\,\text{gl}(r|s\times p|q,\,\mathcal{A})$\,. Furthermore, the representation of linear maps by $\mathbb{Z}_2$-matrices preserves addition, multiplication by scalars and composition:
\begin{align*}
    \lambda+\lambda'&\cong\Lambda+\Lambda'\;,\\
    \alpha\lambda&\cong\alpha\Lambda\;,\\
    \lambda''\circ\lambda&\cong\Lambda''\Lambda\;,
\end{align*}
where $\Lambda,\,\Lambda'\in\mathbb{Z}_2\,\text{gl}(r|s\times p|q,\,\mathcal{A})$ are the representative $\mathbb{Z}_2$-matrices of $\lambda,\,\lambda'\in\Hom_\mathcal{A}(M,\,N)\,,$ $\alpha\in\mathcal{A}$ and $\zL''\in\mathbb{Z}_2\,\text{gl}(u|v\times r|s,\,\mathcal{A})$ is the $\Z_2$-matrix of $\lambda''\in\Hom_\mathcal{A}(N,\,P)\,.$ The sum and product of two supermatrices are defined as for standard matrices but the definition of $\alpha\Lambda$ deviates from the standard definition. More precisely, to ensure that the representation of linear maps by matrices preserves multiplication by scalars in the context of supercommutativity, we have to set 
\begin{equation}\renewcommand*{\arraystretch}{1.4}\label{scalarmult}
\alpha\left(
    \begin{array}{c|c}
    A&B\\\hline
    C&D
    \end{array}\right)
    :=\left(
    \begin{array}{c|c}
    \alpha A&\alpha B\\\hline
    (-1)^{\tilde{\alpha}}\alpha C&(-1)^{\tilde{\alpha}}\alpha D
    \end{array}\right)\,.
\end{equation}\medskip

Analogously, the adjoint operator $\lambda^*\in\Hom_\mathcal{A}(N^*,\,M^*)$ of some linear map $\lambda\in\Hom_\mathcal{A}(M,\,N)$ is a linear map between the dual of $N$ and the dual of $M$\,. Taking into account the $\mathbb{Z}_2$-grading we define it by setting
\begin{equation*}
    \lambda^*(n^*)(m):=(-1)^{\tilde{\lambda}\tilde{n}^*}n^*(\lambda(m))\in\mathcal{A}
\end{equation*}
for any $n^*\in N^*=\Hom_\mathcal{A}(N,\,\mathcal{A})$ and any $m\in M$\,. If
\begin{equation*}\renewcommand*{\arraystretch}{1.4}
    \Lambda=\left(
    \begin{array}{c|c}
    A&B\\\hline
    C&D
    \end{array}\right)
\end{equation*}
is the representative $\mathbb{Z}_2$-matrix of $\lambda$ then the representative $\mathbb{Z}_2$-matrix of $\lambda^*$ is given by
\begin{equation*}
    \prescript{\mathbb{Z}_2t}{}{\Lambda}:=
    \begin{cases}\left(
     \begin{array}{c|c}
    \prescript{t}{}{A}&\prescript{t}{}{C}\\\hline
    -\prescript{t}{}{B}&\prescript{t}{}{D}
    \end{array}\right)
    &\text{if }\lambda\text{ is even},\\
    &\\\left(
     \begin{array}{c|c}
    \prescript{t}{}{A}&-\prescript{t}{}{C}\\\hline
    \prescript{t}{}{B}&\prescript{t}{}{D}
    \end{array}\right)
    &\text{if }\lambda\text{ is odd}.
    \end{cases}
\end{equation*}
We refer to $\prescript{\mathbb{Z}_2t}{}{\Lambda}$ as the supertranspose of $\zL\,.$ Similarly, the $\mathbb{Z}_2$-trace of $\Lambda$ must be defined as
\begin{equation*}
    \mathbb{Z}_2\,\text{tr}\,(\Lambda):=\text{tr}\,A-(-1)^{\tilde{\Lambda}}\text{tr}\,D\,.
\end{equation*}

\subsubsection{$\mathbb{Z}_2$-Berezinian}\label{bersection}

One of the main properties of the classical determinant for standard matrices is multiplicativity, i.e. if $A$ and $B$ are matrices over a commutative ring then $$\det(A\cdot B)=\det A\cdot\det B\;.$$ However, if $$A=\begin{pmatrix}a&b\\c&d\end{pmatrix}\quad\text{and}\quad B=\begin{pmatrix}\alpha&\beta\\\gamma&\delta\end{pmatrix}$$ are $2\times 2$ matrices with entries in a non-commutative ring then
\begin{align*}
    \det\left(
    \begin{pmatrix}
        a&b\\c&d
    \end{pmatrix}
    \begin{pmatrix}
        \alpha&\beta\\\gamma&\delta
    \end{pmatrix}
    \right)&=\det
    \begin{pmatrix}
    a\alpha+b\gamma&a\beta+b\delta\\
    c\alpha+d\gamma&c\beta+d\delta
    \end{pmatrix}\\
    &=a\alpha c\beta+a\alpha d\delta+b\gamma c\beta+b\gamma d\delta-a\beta c\alpha-a\beta d\gamma-b\delta c\alpha-b\delta d\gamma
\end{align*}
and
\begin{align*}
    \det
    \begin{pmatrix}
        a&b\\c&d
    \end{pmatrix}
    \det
    \begin{pmatrix}
        \alpha&\beta\\\gamma&\delta
    \end{pmatrix}
    &=(ad-bc)(\alpha\delta-\beta\gamma)\\
    &=ad\alpha\delta-ad\beta\gamma-bc\alpha\delta+bc\beta\gamma\,,
\end{align*}
which shows that the classical determinant is not multiplicative in a non-commutative context. Since in linear $\mathbb{Z}_2$-algebra we are working with matrices over a $\mathbb{Z}_2$-commutative algebra -- so a (slightly) non-commutative algebra -- the above example highlights the necessity of introducing a new map that replaces the determinant in the case of matrices over $\mathbb{Z}_2$-commutative algebras. This {\it new determinant}, which {\it shares some important properties} with the standard determinant and {\it will play a fundamental role in $\Z_2$-integration theory}, will be called $\mathbb{Z}_2$-Berezinian.\medskip

According to I. Gelfand and V. Retakh {\it every good notion of a determinant} is made of quasideterminants (see for example \cite{quasidet}, page 58). Therefore, we briefly introduce quasideterminants. Let $A$ and $D$ be square matrices of size $p$ and $q$ respectively and assume $D$ to be invertible. Then the block matrix 
\begin{equation*}\renewcommand*{\arraystretch}{1.4}
    \Lambda=\left(\begin{array}{c|c}
        A & B \\\hline
        C & D
    \end{array}\right)
\end{equation*}
can be decomposed into
\begin{equation}\renewcommand*{\arraystretch}{1.4}\label{UDL}
    \Lambda=\left(
    \begin{array}{c|c}
       A&B\\\hline C&D 
    \end{array}\right)
    =\left(
    \begin{array}{c|c}
       \mathbbm{1}&BD^{-1}\\\hline
       0&\mathbbm{1}
    \end{array}\right)\left(
    \begin{array}{c|c}
       A-BD^{-1}C&0\\\hline
       0&D
    \end{array}\right)\left(
    \begin{array}{c|c}
       \mathbbm{1}&0\\\hline
       D^{-1}C&\mathbbm{1}
    \end{array}\right)
\end{equation}
and this decompostion is referred to as UDL decomposition since on the right-hand side we have an upper unitriangular, a diagonal and a lower unitriangular block matrix. If $\Lambda$ has entries in a commutative ring then it makes sense to apply the standard determinant and we obtain
\begin{equation*}
    \det\Lambda=\det(A-BD^{-1}C)\cdot\det D\,.
\end{equation*}
Building on this observation we make the following definition.

\begin{Definition}\label{defquasi}
    Let
    \begin{equation*}\renewcommand*{\arraystretch}{1.4}
    \Lambda=\left(\begin{array}{c|c}
        A & B \\\hline
        C & D
    \end{array}\right)
    \end{equation*}
    be a square block matrix with entries in a unital not necessarily commutative ring $R\,$. The quasideterminant of $\Lambda$ with respect to the block entry $11$\,, i.e. with respect to the block $A$\,, is given by
    \begin{equation*}\renewcommand*{\arraystretch}{1.4}\left|
        \begin{array}{c|c}
            A&B\\\hline
            C&D
        \end{array}\right|_{11}
        :=A-BD^{-1}C\,,
    \end{equation*}
provided $D$ is invertible over $R\,.$
\end{Definition}

\begin{Example}\label{qdet}
Dividing the matrix
\begin{equation*}
    \begin{pmatrix}
       x&a&b\\
       c&y&d\\
       e&f&z
    \end{pmatrix}
\end{equation*}
over $R$ into blocks in two different ways and calculating the quasideterminant with respect to the respective upper left-hand block entry yields
\begin{align*}\left|
    \begin{array}{cc|c}
        x&a&b\\
        c&y&d\\\hline
        e&f&z
    \end{array}\right|_{11}
    &=\begin{pmatrix}x&a\\c&y\end{pmatrix}-\begin{pmatrix}b\\d\end{pmatrix}z^{-1}\begin{pmatrix}e&f\end{pmatrix}
\end{align*}
and
\begin{align*}\left|
    \begin{array}{c|cc}
        x&a&b\\\hline
        c&y&d\\
        e&f&z
    \end{array}\right|_{11}
    &=x-\begin{pmatrix}a&b\end{pmatrix}\begin{pmatrix}y&d\\f&z\end{pmatrix}^{-1}\begin{pmatrix}c\\e\end{pmatrix}\,,
\end{align*}
where the inverse of the $2\times2$ matrix in the second line can be shown to equal
\begin{equation}\label{invformula}
    \begin{pmatrix}
       (y-dz^{-1}f)^{-1}&-(y-dz^{-1}f)^{-1}dz^{-1}\\
       -z^{-1}f(y-dz^{-1}f)^{-1}&z^{-1}+z^{-1}f(y-dz^{-1}f)^{-1}dz^{-1}
    \end{pmatrix}\;,
\end{equation}
if all the inverses exist.
\end{Example}

\begin{Remark}
As can be seen in Example \ref{qdet} quasideterminant consist of rational functions, not necessarily polynomials. It follows that, as already mentioned above, certain inverses must exist in order to allow for a certain quasideterminant to be defined.
\end{Remark}

Collecting some {\it important properties} of the classical determinant, which we would also like the $\mathbb{Z}_2$-Berezinian to verify, we obtain for all matrices $X,\,Y\in\text{gl}(n,\,\mathbb{R})$\,, $A\in\text{gl}(p,\,\mathbb{R})$\,, $B\in\text{gl}(p\times q,\,\mathbb{R})$\,, $C\in\text{gl}(q\times p,\,\mathbb{R})$ and $D\in\text{gl}(q,\,\mathbb{R})$\,:
\begin{enumerate}
    \item $\det(X\cdot Y)=\det X\cdot\det Y\;,$
    \item $\det\left(\begin{array}{c|c}A&0\\\hline 0&D\end{array}\right)=\det A\cdot\det D\;,$
    \item $\det\left(\begin{array}{c|c}\mathbbm{1}&B\\\hline 0&\mathbbm{1}\end{array}\right)=1=\det\left(\begin{array}{c|c}\mathbbm{1}&0\\\hline C&\mathbbm{1}\end{array}\right)\;,$
    \item $\det e^X=e^{\text{tr}X}$\;.\label{4}
\end{enumerate}
For a matrix $X$ in the Lie algebra $\text{gl}(n,\,\mathbb{R})$ over $\mathbb{R}$ we have that $$e^X=\sum_{k=0}^\infty\frac{X^k}{k!}$$ is an element of the Lie group $$\text{GL}(n,\,\mathbb{R})=\{X\in\text{gl}(n,\,\mathbb{R})\,|\,\det X\neq0\}\;,$$ so that Property iv can be summarized by saying that the determinant is the group analogue of the trace.\medskip

Concerning the {\it usefulness of determinants in integration theory}, we recall that if $y=y(x)$ is a standard coordinate transformation between open subsets $U$ and $V$ of $\R^p$ and $\partial_xy$ is the corresponding Jacobian matrix, a function $f(y)$ is integrable over $V$ (with respect to the Lebesgue measure) if and only if the function $f(y(x))|\det\partial_xy|$ is integrable over $U$ and in this case 
\begin{equation*}
    \int_Vdy\,f(y)=\int_Udx\,f(y(x))\,|\det\partial_xy|\,.
\end{equation*}

Now that we have specified our conclusions from the first paragraph of this subsection, let us recall that we are currently working towards the definition of a $\mathbb{Z}_2$-Berezinian determinant that has properties similar to Properties i - iv and is defined for certain matrices $\Lambda\in\text{gl}(p|q,\,\mathcal{A})$ with entries in a $\mathbb{Z}_2$-algebra $\mathcal{A}$ over $\mathbb{R}$\,. Since a $\mathbb{Z}_2$-coordinate transformation
\begin{equation*}
    y=y(x,\,\zx)\hspace{3cm}\eta=\eta(x,\,\zx)
\end{equation*}
in a superdomain $\mathcal{U}^{p|q}=(U,\Ci_{p|q})$ preserves the parities and is invertible, its Jacobian matrix is the even invertible matrix
\begin{equation*}\renewcommand*{\arraystretch}{1.4}\left(
    \begin{array}{c|c}
        \partial_xy&\partial_\zx y\\\hline
        \partial_x\eta&\partial_\zx\eta
    \end{array}\right)
    \in\mathbb{Z}_2\,\text{GL}_0(p|q,\,\mathcal{C}^\infty_{p|q}(U))\,.
\end{equation*}
This suggests that for our application in integration theory it is sufficient to define the $\mathbb{Z}_2$-Berezinian on the group $\mathbb{Z}_2\,\text{GL}_0(p|q,\,\mathcal{A})$ of even invertible $\mathbb{Z}_2$-matrices of size $p|q\times p|q$ with entries in a super $\mathbb{R}$-algebra $\mathcal{A}$. It should be valued in the group $\mathcal{A}^\times_0$ of even invertible elements of $\mathcal{A}$ and hence we are looking for a group morphism
\begin{equation*}
    \mathbb{Z}_2\,\text{Ber}:\mathbb{Z}_2\,\text{GL}_0(p|q,\,\mathcal{A})\rightarrow\mathcal{A}^\times_0
\end{equation*}
that also verifies properties similar to ii - iv.\medskip

First note that similarly to the result proved in Proposition \ref{inv}, which states that a $\mathbb{Z}_2$-function is invertible if and only if its parameter-free even part is invertible, it can be shown that an even matrix
\begin{equation*}\renewcommand*{\arraystretch}{1.4}
    \Lambda=\left(\begin{array}{c|c}
        A & B \\\hline
        C & D
    \end{array}\right)
    \in\mathbb{Z}_2\,\text{gl}_0(p|q,\,\mathcal{A})
\end{equation*}
is invertible if and only if $A\in\text{gl}(p,\,\mathcal{A}_0)$ and $D\in\text{gl}(q,\,\mathcal{A}_0)$ are invertible. We refer to \cite{tr&ber}, page 24, where a more general result is proved. Considering that we want to define the $\mathbb{Z}_2$-Berezinian on the group $\mathbb{Z}_2\,\text{GL}_0(p|q,\,\mathcal{A})$ of even invertible matrices we can therefore always assume that the blocks $A$ and $D$ are invertible. Since the classical determinant works well for blocks consisting exclusively of even elements this is equivalent to assuming that
\begin{equation}\label{detAdetD}
    \det A,\,\det D\in\mathcal{A}^\times_0\,.
\end{equation}

Moreover, we observe that if Property iv, adapted to the $\mathbb{Z}_2$-graded context, holds for $\mathbb{Z}_2\,\text{Ber}$ and $\mathbb{Z}_2\,\text{tr}$ then we have
\begin{align}\renewcommand*{\arraystretch}{1.4}\notag
    \mathbb{Z}_2\,\text{Ber}\left(\begin{array}{c|c}e^A&0\\\hline 0&e^D\end{array}\right)
    &=\mathbb{Z}_2\,\text{Ber}\,e^{\left(\begin{array}{c|c}A&0\\\hline 0&D\end{array}\right)}
    =e^{\mathbb{Z}_2\,\text{tr}\left(\begin{array}{c|c}A&0\\\hline 0&D\end{array}\right)}\\
    &=e^{\text{tr}\,A-\text{tr}\,D}
    =e^{\text{tr}\,A}\cdot(e^{\text{tr}\,D})^{-1}
    =\det e^A\cdot{\det}^{-1} e^D\,,\label{berdiag}
\end{align}
where the second equality follows from the $\mathbb{Z}_2$-version of Property iv and the last equality follows from the original version of this property.\medskip

Hence if we assume that the $\mathbb{Z}_2$-analogues of the properties i, iii and iv hold, then the UDL decomposition \eqref{UDL}, the fact \eqref{detAdetD} that $A$ and $D$ are invertible and the $\Z_2$-analogue \eqref{berdiag} of Property ii imply that the $\mathbb{Z}_2$-Berezinian of a matrix
\begin{equation*}\renewcommand*{\arraystretch}{1.4}
    \Lambda=\left(\begin{array}{c|c}
        A & B \\\hline
        C & D
    \end{array}\right)\in\mathbb{Z}_2\,\text{GL}_0(p|q,\,\mathcal{A})
\end{equation*}
must necessarily be given by
\begin{align*}\renewcommand*{\arraystretch}{1.4}
    \mathbb{Z}_2\,\text{Ber}\left(\begin{array}{c|c}A&B\\\hline C&D\end{array}\right)
    &\renewcommand*{\arraystretch}{1.4}
    =1\cdot\mathbb{Z}_2\,\text{Ber}\left(\begin{array}{c|c}A-BD^{-1}C&0\\\hline 0&D\end{array}\right)\cdot1\\
    &=\det(A-BD^{-1}C)\,{\det}^{-1} D\,.
\end{align*}
So defined the $\mathbb{Z}_2$-Berezinian of $\Lambda$ is invertible since
\begin{equation*}
    \mathbb{Z}_2\,\text{Ber}\,\Lambda\cdot\mathbb{Z}_2\,\text{Ber}\,\Lambda^{-1}=\mathbb{Z}_2\,\text{Ber}\,(\Lambda\cdot\Lambda^{-1})=\mathbb{Z}_2\,\text{Ber}\,\mathbbm{1}=1\;,
\end{equation*}
so that $\mathbb{Z}_2\,\text{Ber}\,\Lambda^{-1}$ is the inverse of $\mathbb{Z}_2\,\text{Ber}\,\Lambda$\,.

\begin{Theorem}
For every $\mathbb{Z}_2$-commutative associative unital $\,\mathbb{R}$-algebra $\mathcal{A}$ there exists a unique group morphism
$$\mathbb{Z}_2\op{Ber}:\mathbb{Z}_2\op{GL}_0(p|q,\,\mathcal{A})\rightarrow\mathcal{A}^\times_0$$ such that
\begin{enumerate}[label=(\roman*)]
    \item $\mathbb{Z}_2\op{Ber}\left(\begin{array}{c|c}A&0\\\hline 0&D\end{array}\right)=\det A\cdot{\det}^{-1}D$ and
    \item $\mathbb{Z}_2\op{Ber}\left(\begin{array}{c|c}\mathbbm{1}&B\\\hline 0&\mathbbm{1}\end{array}\right)
    =1=\mathbb{Z}_2\op{Ber}\left(\begin{array}{c|c}\mathbbm{1}&0\\\hline C&\mathbbm{1}\end{array}\right)$\,.
\end{enumerate}
It is given by
\begin{equation}\label{defber}
    \mathbb{Z}_2\op{Ber}\left(\begin{array}{c|c}A&B\\\hline C&D\end{array}\right)=\det(A-BD^{-1}C)\,{\det}^{-1}D\,.
\end{equation}
\end{Theorem}
\begin{proof}
It can easily be verified that the $\mathbb{Z}_2$-Berezinian when defined as in \eqref{defber} has the properties $(i)$ and $(ii)$. The proof of multiplicativity is more involved and will not be given here (see \cite{tr&ber}, page 24 for the proof of a more general result). The above approach shows that a map that has all the required properties must necessarily be given by \eqref{defber} and thus solves the problem of uniqueness. 
\end{proof}

\subsection{Linear $\mathbb{Z}_2^n$-algebra}

\subsubsection{$\mathbb{Z}_2^n$-modules and linear maps}

We consider $1\leq n\in\mathbb{N}$ and as explained in Section \ref{z2nman} we assume the $\mathbb{Z}_2^n$-degrees $\gamma_0,...,\gamma_{2^n-1}$ to be given in standard order. Let $\mathcal{A}$ be a real $\mathbb{Z}_2^n$-algebra and define linear maps $$\lambda\in\Hom_{\mathcal{A},\tilde{\lambda}}(M,\,N)$$ of degree $\tilde{\lambda}\in\{\gamma_0,...,\gamma_{2^n-1}\}$ between $\mathbb{Z}_2^n$-modules over $\mathcal{A}$ analogously to the $\mathbb{Z}_2$-case. Then set
\begin{equation*}
    \Hom_\mathcal{A}(M,\,N):=\bigoplus_{i=0}^{2^n-1}\Hom_{\mathcal{A},\gamma_i}(M,\,N)=\bigoplus_{i=1}^{2^n}\Hom_{\mathcal{A},\Gamma_i}(M,\,N)\,,
\end{equation*}
where we introduce the alternative notation $\Gamma_i=\gamma_{i-1}$ for the $\mathbb{Z}_2^n$-degrees in order to simplify some of the results below.\medskip

If $M$ and $N$ are free $\mathbb{Z}_2^n$-modules over $\mathcal{A}$ of rank $p|\underline{q}$ and $r|\underline{s}$ respectively, where $\underline{q}=(q_1,...,q_{2^n-1})$ and $\underline{s}=(s_1,...,s_{2^n-1})$\,, then their elements can be represented by column vectors and linear maps between them by matrices. For instance, for any $m\in M$ of degree $\gamma_0$ we have the identification
\begin{equation*}\renewcommand*{\arraystretch}{1.4}
    m\cong\left(\begin{array}{c}m^1\\\vdots\\m^p\\\hline m^{p+1}\\\vdots\\ m^{p+q_1}\\\hline\vdots\\\hline m^{p+\cdots+q_{2^n-2}+1}\\\vdots\\m^{p+\cdots+q_{2^n-1}}\end{array}\right)\in\mathcal{A}^{p|\underline{q}}
\end{equation*}
for some $m^1,...,m^p\in\mathcal{A}_{\gamma_0}$\,, $m^{p+1},...,m^{p+q_1}\in\mathcal{A}_{\gamma_1}$ and $m^{p+\cdots+q_{i-1}+1},...,m^{p+\cdots+q_i}\in\mathcal{A}_{\gamma_i}$ for $i\in\{2,...,2^n-1\}$\,.\medskip

Now fix $n=2$ and consider a linear map $\lambda\in\Hom_{\mathcal{A},\Gamma_1}(M,\,N)$\,. Taking into account that $\lambda$ must in particular preserve the parity of degree $(0,\,0)$ elements
\begin{equation*}\renewcommand*{\arraystretch}{1.4}
    m\cong\left(\begin{array}{c}(0,\,0)\\\hline(1,\,1)\\\hline(0,\,1)\\\hline(1,\,0)\end{array}\right)
    \in\mathcal{A}^{p|\underline{q}}_{\gamma_0}
\end{equation*}
we obtain the identification
\begin{equation}\renewcommand*{\arraystretch}{1.4}\label{z22lambda}
    \lambda\cong\Lambda=\left(\begin{array}{c|c||c|c}(0,\,0)&(1,\,1)&(0,\,1)&(1,\,0)\\\hline
    (1,\,1)&(0,\,0)&(1,\,0)&(0,\,1)\\\hline\hline
    (0,\,1)&(1,\,0)&(0,\,0)&(1,\,1)\\\hline
    (1,\,0)&(0,\,1)&(1,\,1)&(0,\,0)\end{array}\right)
    \in\mathbb{Z}_2^2\op{gl}_{\Gamma_1}(r|\underline{s}\times p|\underline{q},\,\mathcal{A})\,,
\end{equation}
where each block contains elements of $\mathcal{A}$ that have the $\mathbb{Z}_2^2$-degree specified in the corresponding part of the vector or matrix. For instance, setting $\underline{q}=(q_1,\,q_2,\,q_3)$ and $\underline{s}=(s_1,\,s_2,\,s_3)$\,, the elements in the $r\times q_3$ block in the top right-hand corner of $\Lambda$ are of degree $(1,\,0)$\,. Note that dividing $\Lambda$ into four blocks by means of the double lines in \eqref{z22lambda} the blocks in the top left-hand and the bottom right-hand corner only contain elements of even degree whereas the two other blocks consist of odd elements.\medskip

Proceeding similarly for $n=3$ we obtain that a linear map $\lambda\in\Hom_{\mathcal{A},\Gamma_1}(M,\,N)$ can be identified with a matrix $\Lambda\in\mathbb{Z}_2^3\op{gl}_{\Gamma_1}(r|\underline{s}\times p|\underline{q},\,\mathcal{A})$\,, where
\begin{equation*}\renewcommand*{\arraystretch}{1.4}
    \Lambda=\left(\begin{array}{c|c|c|c||c|c|c|c}(0,\,0,\,0)&(0,\,1,\,1)&(1,\,0,\,1)&(1,\,1,\,0)&(0,\,0,\,1)&(0,\,1,\,0)&(1,\,0,\,0)&(1,\,1,\,1)\\\hline
    (0,\,1,\,1)&(0,\,0,\,0)&(1,\,1,\,0)&(1,\,0,\,1)&(0,\,1,\,0)&(0,\,0,\,1)&(1,\,1,\,1)&(1,\,0,\,0)\\\hline
    (1,\,0,\,1)&(1,\,1,\,0)&(0,\,0,\,0)&(0,\,1,\,1)&(1,\,0,\,0)&(1,\,1,\,1)&(0,\,0,\,1)&(0,\,1,\,0)\\\hline
    (1,\,1,\,0)&(1,\,0,\,1)&(0,\,1,\,1)&(0,\,0,\,0)&(1,\,1,\,1)&(1,\,0,\,0)&(0,\,1,\,0)&(0,\,0,\,1)\\\hline\hline
    (0,\,0,\,1)&(0,\,1,\,0)&(1,\,0,\,0)&(1,\,1,\,1)&(0,\,0,\,0)&(0,\,1,\,1)&(1,\,0,\,1)&(1,\,1,\,0)\\\hline
    (0,\,1,\,0)&(0,\,0,\,1)&(1,\,1,\,1)&(1,\,0,\,0)&(0,\,1,\,1)&(0,\,0,\,0)&(1,\,1,\,0)&(1,\,0,\,1)\\\hline
    (1,\,0,\,0)&(1,\,1,\,1)&(0,\,0,\,1)&(0,\,1,\,0)&(1,\,0,\,1)&(1,\,1,\,0)&(0,\,0,\,0)&(0,\,1,\,1)\\\hline
    (1,\,1,\,1)&(1,\,0,\,0)&(0,\,1,\,0)&(0,\,0,\,1)&(1,\,1,\,0)&(1,\,0,\,1)&(0,\,1,\,1)&(0,\,0,\,0)\end{array}\right)
\end{equation*}
and we can observe again that the double lines divide $\Lambda$ into two even and two odd blocks.\medskip

\begin{Remark}\label{blockdeg}
These observations can be generalized, i.e. if $\Lambda\in\mathbb{Z}_2^n\op{gl}_{\Gamma_i}(r|\underline{s}\times p|\underline{q},\,\mathcal{A})$ then its block $\Lambda_{kl}$ exclusively contains elements of $\mathbb{Z}_2^n$-degree $\Gamma_k+\Gamma_l+\Gamma_i$\,.\medskip
\end{Remark}

As in the $\mathbb{Z}_2$-case the identification
\begin{equation*}
    \Hom_\mathcal{A}(\mathcal{A}^{p|\underline{q}},\,\mathcal{A}^{r|\underline{s}})\cong\mathbb{Z}_2^n\op{gl}(r|\underline{s}\times p|\underline{q},\,\mathcal{A})
\end{equation*}
between linear maps and matrices preserves the $\mathbb{Z}_2^n$-degree, addition, multiplication and external multiplication by scalars $\alpha\in\mathcal{A}$ provided we set
\begin{equation*}\renewcommand*{\arraystretch}{1.4}
    \alpha\Lambda:=\left(\begin{array}{c|c||c|c}(-1)^{\langle\tilde{\alpha},\,\Gamma_1\rangle}\alpha\Lambda_{11}&\cdots&\cdots&(-1)^{\langle\tilde{\alpha},\,\Gamma_1\rangle}\alpha\Lambda_{12^n}\\\hline
    &\vdots&\vdots&\\\hline\hline
    &\vdots&\vdots&\\\hline
    (-1)^{\langle\tilde{\alpha},\,\Gamma_{2^n}\rangle}\alpha\Lambda_{2^n1}&\cdots&\cdots&(-1)^{\langle\tilde{\alpha},\,\Gamma_{2^n}\rangle}\alpha\Lambda_{2^n2^n}\end{array}\right)
\end{equation*}
for any $\Lambda\in\mathbb{Z}_2^n\op{gl}(r|\underline{s}\times p|\underline{q},\,\mathcal{A})$\,. Note that this definition is consistent with the $\mathbb{Z}_2$-case as it reduces to \eqref{scalarmult} if $n=1$\,.\medskip

Furthermore, the $\mathbb{Z}_2$-trace can be generalized to the $\mathbb{Z}_2^n$-context as stated in the following theorem.

\begin{Theorem}\label{z2ntrace}
There exists an $\mathcal{A}$-linear graded Lie algebra morphism of degree $\gamma_0$
\begin{equation*}
    \mathbb{Z}_2^n\op{tr}:\mathbb{Z}_2^n\op{gl}(p|\underline{q},\,\mathcal{A})\rightarrow\mathcal{A}\;.\end{equation*}
It is unique up to multiplication by $\alpha\in\mathcal{A}_0$ and it is given for $\zL$ of degree $\zG_i$ by
\begin{equation*}
    \mathbb{Z}_2^n\op{tr}\left(\begin{array}{c|c||c|c}\Lambda_{11}&\cdots&\cdots&\Lambda_{12^n}\\\hline
    &\vdots&\vdots&\\\hline\hline
    &\vdots&\vdots&\\\hline
    \Lambda_{2^n1}&\cdots&\cdots&\Lambda_{2^n2^n}\end{array}\right)
    =\sum_{k=1}^{2^n}(-1)^{\langle\Gamma_k+\Gamma_i,\Gamma_k\rangle}\op{tr}\Lambda_{kk}\,,
\end{equation*}
where $\op{tr}$ denotes the usual trace.
\end{Theorem}

Note that the usual trace is a Lie algebra morphism as it satisfies, for any two matrices $A$ and $B$ with entries in a field, $\op{tr}(B\cdot A)=\op{tr}(A\cdot B)$\,, which implies
\begin{equation*}
    \op{tr}[A,\,B]_c=0=[\op{tr}A,\,\op{tr}B]_c\,,
\end{equation*}
where $[-,-]_c$ denotes the commutator bracket. Moreover, it can easily be verified that the $\mathbb{Z}_2$-trace coincides with the $\mathbb{Z}_2^n$-trace for $n=1$\,. For a proof of Theorem \ref{z2ntrace} see \cite{tr&ber}, page 9.

\subsubsection{$\mathbb{Z}_2^n$-Berezinian}

With the objective of generalizing the $\mathbb{Z}_2$-Berezinian to a $\mathbb{Z}_2^n$-Berezinian we formulate the

\begin{Theorem}\label{z2nber}
For every $\mathbb{Z}_2^n$-commutative associative unital $\mathbb{R}$-algebra $\mathcal{A}$ there exists a unique group morphism
\begin{equation*}
    \mathbb{Z}_2^n\op{Ber}:\mathbb{Z}_2^n\op{GL}_{\gamma_0}(p|\underline{q},\,\mathcal{A})\rightarrow\mathcal{A}^\times_{\gamma_0}
\end{equation*}
such that
\begin{enumerate}[label=(\roman*)]
    \item $\mathbb{Z}_2^n\op{Ber}\left(\begin{array}{c||c}A&0\\\hline\hline0&D\end{array}\right)=\mathbb{Z}_2^n\det A\cdot\mathbb{Z}_2^n\det^{-1}D$ and
    \item $\mathbb{Z}_2^n\op{Ber}\left(\begin{array}{c||c}\mathbbm{1}&B\\\hline\hline0&\mathbbm{1}\end{array}\right)=1=\mathbb{Z}_2^n\op{Ber}\left(\begin{array}{c||c}\mathbbm{1}&0\\\hline\hline C&\mathbbm{1}\end{array}\right)$\,.
\end{enumerate}
It is given by
\begin{equation*}
    \mathbb{Z}_2^n\op{Ber}\left(\begin{array}{c||c}A&B\\\hline\hline C&D\end{array}\right)=\mathbb{Z}_2^n\det(A-BD^{-1}C)\cdot\mathbb{Z}_2^n{\det}^{-1}D\,.
\end{equation*}
\end{Theorem}

As indicated by the use of double lines and by Remark \ref{blockdeg}, the blocks $A$ and $D$ in the above theorem are made of even elements, i.e. $$A,\,D\in(\mathbb{Z}_2^n)_{\op{ev}}\op{gl}_{\gamma_0}(p|\underline{q}_{\op{ev}},\,\mathcal{A})\,.$$ However, it does not make sense to apply the classical determinant to them as their entries do not necessarily commute. So before we can prove or even formulate the above theorem, we have to look for a suitable replacement for the classical determinant. We keep the axioms of the previous theorem motivated in Section \ref{bersection}.

\begin{Theorem}\label{z2ndet}
There exists a unique map
\begin{equation*}
    \mathbb{Z}_2^n\det:(\mathbb{Z}_2^n)_{\op{ev}}\op{gl}_{\gamma_0}(p|\underline{q}_{\op{ev}},\,\mathcal{A})\rightarrow\mathcal{A}_{\gamma_0}
\end{equation*}
such that
\begin{enumerate}[label=(\roman*)]
    \item $\mathbb{Z}_2^n\det$ is multiplicative,
    \item $\mathbb{Z}_2^n\det\left(\begin{array}{c|c|c|c}\Lambda_{11}&0&\cdots&0\\\hline
    0&\Lambda_{22}&\cdots&0\\\hline
    \vdots&\vdots&\ddots&\vdots\\\hline
    0&0&\cdots&\Lambda_{2^{n-1}2^{n-1}}\end{array}\right)={\displaystyle \prod_{k=1}^{2^{n-1}} \det\Lambda_{kk}}\in\mathcal{A}_{\gamma_0}$ and
    \item applying $\mathbb{Z}_2^n\det$ to an upper unitriangular or lower unitriangular matrix yields $1$.
\end{enumerate}
\end{Theorem}
Note that all blocks $\zL_{kk}$ are of $\Z_2^n$-degree $\zg_0$ and therefore have commutative entries, so their classical determinant makes sense.
\begin{proof}
The proof makes use of the fact that every matrix $\Lambda\in(\mathbb{Z}_2^n)_{\op{ev}}\op{gl}_{\gamma_0}(p|\underline{q}_{\op{ev}},\,\mathcal{A})$ has a UDL decomposition, which can be shown to equal
\begin{equation*}\renewcommand*{\arraystretch}{1.4}
    \Lambda=UDL=U\left(\begin{array}{c|c|c|c|c}|\Lambda|_{11}&0&0&\cdots&0\\\hline
    0&|\Lambda^{1:1}|_{22}&0&\cdots&0\\\hline
    0&0&|\Lambda^{12:12}|_{33}&\cdots&0\\\hline
    \vdots&\vdots&\vdots&\ddots&0\\\hline
    0&0&0&0&\Lambda_{2^{n-1}2^{n-1}}\end{array}\right)L\,,
\end{equation*}
for some upper respectively lower unitriangular matrices $U$ and $L$ and where $|\Lambda^{1:1}|_{22}$ denotes the quasi-determinant with respect to the block entry $\Lambda_{22}$ of the matrix obtained from $\Lambda$ by omitting block row $1$ and block column $1$\,.
Based on this decomposition we can then argue that if the $\Z_2^n$-graded determinant exists it must be given by
\begin{equation}\label{formuladet}
    \mathbb{Z}_2^n\det\Lambda=\det|\Lambda|_{11}\cdot\det|\Lambda^{1:1}|_{22}\cdot...\cdot\det\Lambda_{2^{n-1}2^{n-1}}\in\mathcal{A}_{\gamma_0}\;.
\end{equation}
In view of the fact that quasi-determinants are made of rational functions a crucial and challenging part of the proof is to show that $\mathbb{Z}_2^n\det\Lambda$ is a polynomial after simplification and that $\mathbb{Z}_2^n\det$ is multiplicative.
\end{proof}

For a complete proof of Theorem \ref{z2ndet} we refer to \cite{tr&ber}, page $10$. We limit ourselves here to a couple of examples that illustrate what has just been said.

\begin{Example}
Let
\begin{equation*}\renewcommand*{\arraystretch}{1.4}
    \Lambda=\left(\begin{array}{c|c|c|c}x&a&b&c\\\hline
    d&y&e&f\\\hline
    g&h&z&l\\\hline
    m&n&p&w\end{array}\right)\in(\mathbb{Z}_2^3)_{\op{ev}}\op{gl}_{(0,0,0)}(1|(1,\,1,\,1),\,\mathcal{A})
\end{equation*}
be a matrix over a real $\mathbb{Z}_2^3$-algebra $\mathcal{A}$\,. According to \eqref{formuladet} and taking into account that each block of $\Lambda$ consists of a single element the graded determinant of $\Lambda$ is given by
\begin{equation*}
    \mathbb{Z}_2^3\det\Lambda=|\Lambda|_{11}\cdot|\Lambda^{1:1}|_{22}\cdot|\Lambda^{12:12}|_{33}\cdot\Lambda_{44}\,.
\end{equation*}
Clearly, we have $$\Lambda_{44}=w$$ and applying Definition \ref{defquasi} we get $$|\Lambda^{12:12}|_{33}=z-lw^{-1}p\,.$$ Hence it remains to calculate two quasi-determinants. Setting $\alpha:=w^{-1}$ and $\beta:=(z-lw^{-1}p)^{-1}$ we have
\begingroup
\addtolength{\jot}{1em}
\begin{align*}
    |\Lambda^{1:1}|_{22}&=\left|\begin{array}{c|c|c}y&e&f\\\hline h&z&l\\\hline n&p&w\end{array}\right|_{11}\\
    &=y-\left(\begin{array}{cc}e&f\end{array}\right)\left(\begin{array}{cc}z&l\\p&w\end{array}\right)^{-1}\left(\begin{array}{c}h\\n\end{array}\right)\\
    &=y-\left(\begin{array}{cc}e&f\end{array}\right)\left(\begin{array}{cc}(z-lw^{-1}p)^{-1}&-(z-lw^{-1}p)^{-1}lw^{-1}\\-w^{-1}p(z-lw^{-1}p)^{-1}&w^{-1}+w^{-1}p(z-lw^{-1}p)^{-1}lw^{-1}\end{array}\right)\left(\begin{array}{c}h\\n\end{array}\right)\\
    &=y-e\beta h+e\beta l\alpha n+f\alpha p\beta h-f\alpha n-f\alpha p\beta lw^{-1}n\\
    &=\alpha\beta[y(zw-lp)+fph+eln-ehw-fnz]\,,
\end{align*}
\endgroup
where formula $\eqref{invformula}$ is used to compute the inverse matrix and the $\mathbb{Z}_2^3$-commutation rule is applied in order to simplify the resulting expression. Observing that $\Lambda_{44}=\alpha^{-1}$ and $|\Lambda^{12:12}|_{33}=\beta^{-1}$ we obtain that multiplying the three last factors of $\mathbb{Z}_2^3\det\Lambda$ yields
\begin{equation*}
    v:=y(zw-lp)+fph+eln-ehw-fnz\,.
\end{equation*}
Concerning the first factor of $\mathbb{Z}_2^3\det\Lambda$ we compute
\begin{align*}
    |\Lambda|_{11}&=\left|\begin{array}{c|c|c|c}x&a&b&c\\\hline d&y&e&f\\\hline g&h&z&l\\\hline m&n&p&w\end{array}\right|_{11}\\[1em]
    &=x-\left(\begin{array}{ccc}a&b&c\end{array}\right)\left(\begin{array}{ccc}y&e&f\\h&z&l\\n&p&w\end{array}\right)^{-1}\left(\begin{array}{c}d\\g\\m\end{array}\right)\\[1em]
    &=x-\left(\begin{array}{ccc}a&b&c\end{array}\right)\left(\begin{array}{ccc}v^{-1}(zw-lp)&v^{-1}(fp-ew)&v^{-1}(el-fz)\\v^{-1}(ln-hw)&v^{-1}(yw-fn)&v^{-1}(hf-ly)\\v^{-1}(ph-zn)&v^{-1}(ne-py)&v^{-1}(yz-eh)\end{array}\right)\left(\begin{array}{c}d\\g\\m\end{array}\right)\\[1em]
    &=v^{-1}[xv-(a(zw-lp)+b(ln-hw)+c(ph-zn))d\\
    &\hspace{5mm}-(a(fp-ew)+b(yw-fn)+c(ne-py))g\\
    &\hspace{5mm}-(a(el-fz)+b(hf-ly)+c(yz-eh))m]\,,
\end{align*}
where the calculation of the inverse of the involved $3\times 3$ matrix, that can among others be done using its UDL decomposition, is omitted. Finally, multiplying by $v$ and expanding we obtain
\begin{align*}
    \mathbb{Z}_2^3\det\Lambda\ =&\hspace{5mm}xyzw\ -\ xylp\ -\ xehw\ -\ xfhp\ +\ xeln\ -\ xfzn\\
    &-adzw\ +\ adlp\ +\ aegw\ +\ afgp\ -\ aelm\ +\ afzm\\
    &-bdhw\ +\ bdln\ -\ bygw\ +\ bfgn\ +\ bylm\ +\ bfhm\\
    &-cdhp\ -\ cdzn\ -\ cygp\ +\ cegn\ -\ cyzm\ +\ cehm\,.
\end{align*}
\end{Example}

\begin{Example}
Consider the matrix
\begin{equation*}\renewcommand*{\arraystretch}{1.4}
    \Lambda=\left(\begin{array}{cc|c|c}x&a&b&c\\d&y&e&f\\\hline g&h&z&l\\\hline m&n&p&w\end{array}\right)
    \in(\mathbb{Z}_2^3)_{\op{ev}}\op{gl}_{(0,0,0)}(0|(2,\,1,\,1),\,\mathcal{A})\,,
\end{equation*}
where $\mathcal{A}$ is a real $\mathbb{Z}_2^3$-algebra. Its graded determinant is given by
\begin{equation*}
    \mathbb{Z}_2^3\det\Lambda=\det|\Lambda|_{11}\cdot|\Lambda^{1:1}|_{22}\cdot\Lambda_{33}
\end{equation*}
and we immediately obtain
\begin{equation*}
    \Lambda_{33}=w\quad\text{and}\quad|\Lambda^{1:1}|_{22}=z-lw^{-1}p\;.
\end{equation*}
Denoting once again $w^{-1}$ by $\alpha$ and $(z-lw^{-1}p)^{-1}$ by $\beta$ the remaining factor of $\mathbb{Z}_2^3\det\Lambda$ can be computed as follows:
\begin{align*}
    &|\Lambda|_{11}=\left|\begin{array}{cc|c|c}x&a&b&c\\d&y&e&f\\\hline g&h&z&l\\\hline m&n&p&w\end{array}\right|_{11}\\[1em]
    &=\left(\begin{array}{cc}x&a\\d&y\end{array}\right)-\left(\begin{array}{cc}b&c\\e&f\end{array}\right)\left(\begin{array}{cc}z&l\\p&w\end{array}\right)^{-1}\left(\begin{array}{cc}g&h\\m&n\end{array}\right)\\[1em]
    &=\left(\begin{array}{cc}x&a\\d&y\end{array}\right)-\left(\begin{array}{cc}b&c\\e&f\end{array}\right)\left(\begin{array}{cc}(z-lw^{-1}p)^{-1}&-(z-lw^{-1}p)^{-1}lw^{-1}\\-w^{-1}p(z-lw^{-1}p)^{-1}&w^{-1}+w^{-1}p(z-lw^{-1}p)^{-1}lw^{-1}\end{array}\right)^{-1}\left(\begin{array}{cc}g&h\\m&n\end{array}\right)\\[1em]
    &=\left(\begin{array}{cc}x-b\beta g+b\beta l\alpha m+c\alpha p\beta g-c\alpha m-c\alpha p\beta l\alpha m
    &a-b\beta h+b\beta l\alpha n+c\alpha p\beta h-c\alpha n-c\alpha p\beta l\alpha n\\
    d-e\beta g+e\beta l\alpha m+f\alpha p\beta g-f\alpha m-f\alpha p\beta l\alpha m
    &y-e\beta h+e\beta l\alpha n+f\alpha p\beta h-f\alpha n-f\alpha p\beta l\alpha n\end{array}\right)\,,
\end{align*}
so that
\begin{align*}
    \det|\Lambda|_{11}=&(x-b\beta g+b\beta l\alpha m+c\alpha p\beta g-c\alpha m-c\alpha p\beta l\alpha m)\\
    \cdot&(y-e\beta h+e\beta l\alpha n+f\alpha p\beta h-f\alpha n-f\alpha p\beta l\alpha n)\\
    -&(d-e\beta g+e\beta l\alpha m+f\alpha p\beta g-f\alpha m-f\alpha p\beta l\alpha m)\\
    \cdot&(a-b\beta h+b\beta l\alpha n+c\alpha p\beta h-c\alpha n-c\alpha p\beta l\alpha n)\,.
\end{align*}
After multiplication with $w=\alpha^{-1}$ and $z-lw^{-1}p=\beta^{-1}$ the resulting expression can be simplified taking into account the $\mathbb{Z}_2^3$-degrees of the involved components and we obtain
\begin{align*}
    \mathbb{Z}_2^3\det\Lambda\ =&\hspace{5mm}xyzw\ -\ xylp\ -\ xehw\ -\ xfhp\ +\ xeln\ -\ xfzn\\
    &-adzw\ +\ adlp\ +\ aegw\ +\ afgp\ -\ aelm\ +\ afzm\\
    &+bdhw\ -\ bdln\ -\ bygw\ -\ bfgn\ +\ bylm\ +\ bfhm\\
    &+cdhp\ +\ cdzn\ -\ cygp\ -\ cegn\ -\ cyzm\ +\ cehm\,.
\end{align*}
\end{Example}

\begin{Remark}
It should be noted that $\mathbb{Z}_2^n\op{Ber}$ coincides with the $\mathbb{Z}_2$-Berezinian if $n=1$ and thus constitutes a generalization of the standard Berezinian. Furthermore $\mathbb{Z}_2^3\op{Ber}$ coincides -- except for its sign -- with the Dieudonn\'e determinant if we set $\mathcal{A}=\mathbb{H}$ (where $\mathbb{H}$ denotes the algebra of quaternions) and it can be verified that $\mathbb{Z}_2^n\op{Ber}$ is the group analogue of $\mathbb{Z}_2^n\op{tr}$. All these properties confirm that the $\mathbb{Z}_2^n$-Berezinian is a suitable replacement for the classical determinant in $\mathbb{Z}_2^n$-algebra. For a proof of Theorem \ref{z2nber} we refer to \cite{tr&ber}, page 24.
\end{Remark}

\subsection{Integration on smooth manifolds}\label{intsmooth}

On our way towards integration on $\Z_2^n$-manifolds we first deal with integration on smooth manifolds as integration on colored supermanifolds generalizes this theory.\medskip

Let $N$ be a smooth manifold of dimension $p$ and $(U,\,\varphi=(x^1,...,x^p))$ a coordinate chart from an atlas $\mathscr{A}_N$ of $N$\,. Any differential (smooth) top-form $\omega\in\Omega^p(N)$ is locally given by
\begin{equation*}
    \omega\restr{U}=f\ dx^1\wedge\cdots\wedge dx^p
\end{equation*}
for some $f\in\mathcal{C}^\infty(U)$\,, whose support we assume to be compact and contained in $U$ for the time being. Due to this assumption we can set
\begin{equation}\label{intN}
    \int_N\omega=\int_U\omega\restr{U}=\int_Uf\ dx^1\wedge\cdots\wedge dx^p:=\int_{\varphi(U)}f(x)\ dx^1\cdots dx^p\,,
\end{equation}
where the right-hand side denotes the Lebesgue integral of $f\circ\varphi$ over $\varphi(U)\subseteq\mathbb{R}^p$\,.\medskip

Requiring the integral of $\omega$ over $N$ to be well-defined means that $\int_N\omega$ defined as in \eqref{intN} should be independent of the choice of coordinates in $U$\,. In order to prove coordinate-independence we need another assumption, namely that $N$ is orientable. What it means for a smooth manifold to be orientable becomes clear when considering the non-orientable M\"obius strip $M$.

\begin{figure}[htbp]
\centering
\includegraphics[width=10cm, keepaspectratio=true]{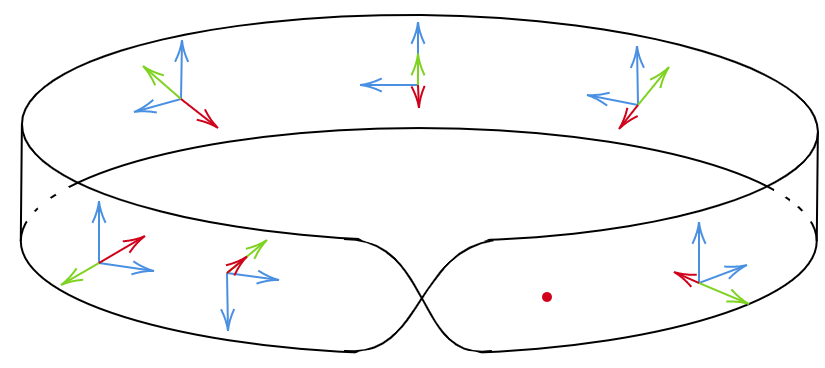}
\caption{non-orientable manifold}
\end{figure}

The blue arrows represent bases of the corresponding tangent spaces. Since the two leftmost bases $(\partial_{x^1},\,\partial_{x^2})$ and $(\partial_{y^1},\,\partial_{y^2})$, where the first (resp. second) vectors are horizontal (resp. vertical), are direct bases their transition matrix, which equals the Jacobian matrix of the coordinate transformation from $x$- to $y$-coordinates, satisfies $\det\partial_xy>0$\,. However, as indicated in the above figure we cannot equip the whole manifold with bases that verify this condition. This means that there does not exist any atlas $\mathscr{A}_M=(U_\alpha,\,\varphi_\alpha)_\alpha$ satifying $$\det(\varphi_\beta\circ\varphi_\alpha^{-1})'(x)>0$$ for all $x\in\varphi_\alpha(U_\alpha\cap U_\beta)$ and for all indices $\alpha$ and $\beta$\,, which is a defining criterion for orientability. Moreover, it can be observed that on non-orientable manifolds such as $M$ there does not exist any nowhere vanishing top-form, which constitutes an equivalent criterion for orientability. Indeed, the top-form represented by the green arrows is not smooth and the one indicated by the red arrows vanishes. We conclude that orientable smooth manifolds admit nowhere vanishing (smooth) top-forms and atlases whose Jacobian matrices have strictly positive determinants.\medskip

Hence we formulate our additional hypothesis as follows. We assume $N$ to be orientable and let $\Omega$ be a nowhere vanishing top-form on $N$\,, which we call volume form. Then we fix an orientation, either $\Omega$ or $-\Omega$\,, and choose a compatible atlas $\mathscr{A}_N$\,, i.e. an atlas that is compatible with the chosen orientation and where the determinant of each Jacobian matrix is strictly positive. For example, the Cartesian space $\mathbb{R}^p$ is orientable with $\Omega=dx^1\wedge\cdots\wedge dx^p$ as volume form.\medskip

Picking two coordinate charts $(U,\,\varphi=(x^1,...,x^p))$ and $(U,\,\psi=(y^1,...,y^p))$\,, where for simplicity we assume the coordinate domains to coincide, the integral $\int_N\omega$ can be expressed as
\begin{equation}\label{2ints}
    \int_N\omega=\int_U\omega\restr{U}=\begin{cases}\int_Uf\ dx^1\wedge\cdots\wedge dx^p=\int_{\varphi(U)}f(x)\ dx^1\cdots dx^p\\[1em]
    \int_Ug\ dy^1\wedge\cdots\wedge dy^p=\int_{\psi(U)}g(y)\ dy^1\cdots dy^p\end{cases}\,.
\end{equation}
We need to show that the Lebesgue integrals on the right-hand side of \eqref{2ints} coincide. First, we observe that the coordinate transformation between $x$- and $y$-coordinates allows us to express $\omega$ locally as
\begin{align}
    g(y)\ dy^1\wedge\cdots\wedge dy^p& =\omega\restr{\psi(U)}\notag\\[8pt]&=f(x(y))\ dx^1\wedge\cdots\wedge dx^p\label{dxi}\\[8pt]
    &=f(x(y))\sum_{(\sigma_1\cdots\sigma_p)=\sigma\in\mathbb{S}_p}\partial_{y^{\sigma_1}}x^1\cdots\partial_{y^{\sigma_p}}x^p\ dy^{\sigma_1}\wedge\cdots\wedge dy^{\sigma_p}\notag\\[8pt]
    &=f(x(y))\sum_{(\sigma_1\cdots\sigma_p)=\sigma\in\mathbb{S}_p}\partial_{y^{\sigma_1}}x^1\cdots\partial_{y^{\sigma_p}}x^p\ \op{sign}\sigma\ dy^1\wedge\cdots \wedge dy^p\notag\\[8pt]
    &=f(x(y))\ \det\partial_yx\ dy^1\wedge\cdots \wedge dy^p\label{dyi}\;,
\end{align}
so that 
$$g(y)=f(x(y))\,\det\partial_yx\;.$$
Then
\begin{align*}
    \int_N\omega&=\int_{\psi(U)}g(y)\ dy^1\cdots dy^p\\
    &=\int_{\psi(U)}f(x(y))\ \det\partial_yx\ dy^1\cdots dy^p\\
    &=\int_{\psi(U)}f(x(y))\ |\det\partial_yx|\ dy^1\cdots dy^p\\
    &=\int_{\varphi(U)}f(x)\ dx^1\cdots dx^p\,,
\end{align*}
where the third equality follows from the orientability assumption and the fourth equality from the coordinate transformation theorem for Lebesgue integrals. This concludes our proof of coordinate-independence for integrals over smooth manifolds.\medskip

Next, we would like to define the integral over a $p$-dimensional smooth manifold $N$ of an arbitrary top-form $\omega\in\Omega^p(N)$\,. This means that we drop the assumption about the support of $\zw\,$, while the assumption that $N$ is orientable and oriented remains valid. Using a partition of unity $(\zeta_\alpha)_\alpha$ subordinate to a locally finite compatible atlas $\mathscr{A}_N=(U_\alpha,\,\varphi_\alpha)_\alpha\,,$ we define the integral of $\omega$ over $N$ by setting
\begin{equation*}
    \int_N\omega=\int_N\left(\sum_\alpha\zeta_\alpha\right)\omega:=\sum_\alpha\int_N\zeta_\alpha\,\omega
\end{equation*}
provided the series on the right-hand side converges in $\mathbb{R}$\,. Note that $\zeta_\alpha\,\omega$ is a top-form whose support is compact and contained in $U_\alpha$\,, so that each of the integrals in the series is defined by \eqref{intN}. It can be verified that $\int_N\omega$ does not depend on the choice of the partition of unity.

\subsection{Integration on $\mathbb{Z}_2^n$-manifolds}

\subsubsection{$\Z_2^n$-Berezinian-sheaf of a $\Z_2^n$-manifold}

Once again let $N$ be a smooth manifold of dimension $p$ and let $(U,\,\varphi=(x^1,...,x^p))$ be a coordinate chart of $N$\,. We denote by $M:=\Omega^1(U)$ the $\mathcal{C}^\infty(U)$-module of differential $1$-forms over $U$\,. Denoting furthermore the real commutative algebra $\mathcal{C}^\infty(U)$ by $\mathcal{A}$ we obtain that $M$ is a free module of rank $p$ over $\mathcal{A}$\,. Considering the exterior algebra $\wedge M$ of $M$ we set
\begin{equation*}
    \op{Det}M:=\wedge^pM=\wedge^p\Gamma(T^*U)=\Gamma(\wedge^pT^*U)=\Omega^p(U)=\mathcal{C}^\infty(U)\Omega\,,
\end{equation*}
where $\Omega$ is the volume form $dx^1\wedge\cdots\wedge dx^p$\,. Of course $\op{Det}M$ is a module of rank $1$ over $\mathcal{A}$\,. Now we make an important observation concerning the relation between $M$ and $\op{Det}M$\,. Namely, a basis transformation in $M$\,, given by $$dy^j=\sum_i\partial_{x^i}y^jdx^i$$ and characterized by $$B=\prescript{t}{}{\partial_xy}\in\op{GL}(p,\,\mathcal{C}^\infty(U))$$ induces a basis transformation in $\op{Det}M$\,, characterized by $\det B$\,. Indeed, looking at \eqref{dxi} and \eqref{dyi} above and exchanging the roles of $x$ and $y$ we find
\begin{equation*}
    dy^1\wedge\cdots\wedge dy^p=\det\partial_xy\ dx^1\wedge\cdots\wedge dx^p
\end{equation*}
with $$\det\partial_xy=\det\prescript{t}{}{\partial_xy}=\det B\,.$$\medskip

Our goal is to generalize $\op{Det}M=\Omega^p(U)$ to the $\mathbb{Z}_2^n$-context, which cannot be done in a straightforward way since there are no $\mathbb{Z}_2^n$-top-forms. As seen in the previous section $\op{Det}M$ is the module of objects that can be integrated over smooth manifolds and by generalizing $\op{Det}M$ to the $\mathbb{Z}_2^n$-setting we intend to find the module of objects that can be integrated over $\mathbb{Z}_2^n$-manifolds.\medskip

We start with a real $\mathbb{Z}_2^n$-algebra $\mathcal{A}$ and a free $\mathbb{Z}_2^n$-module $M$ of total rank $r$ over $\mathcal{A}$\,. The problem we are trying to solve can then be described as finding a free $\mathbb{Z}_2^n$-module $\mathbb{Z}_2^n\op{Ber}M$
of total rank $1$ over $\mathcal{A}$ such that a basis transformation in $M$ characterized by $B\in\mathbb{Z}_2^n\op{GL}_{\gamma_0}(r,\,\mathcal{A})$ induces a basis transformation in $\mathbb{Z}_2^n\op{Ber}M$ characterized by $\mathbb{Z}_2^n\op{Ber}B$.\medskip

Before solving this problem using tools from cohomology theory we briefly recall tensor products of vector spaces and modules.

\begin{Remark}\label{tensorprod}
The tensor product $V\otimes W$ of two real vector spaces is itself a vector space over $\mathbb{R}$\,. If $M$ and $N$ are modules over a commutative ring $\mathcal{R}$ their tensor product $M\otimes_\mathcal{R}N$ is also an $\mathcal{R}$-module. Considering the same situation with $\mathcal{R}$ being an arbitrary not necessarily commutative ring we obtain that $M\otimes_\mathcal{R}N$ is an abelian group or, equivalently, a module over $\mathbb{Z}$\,. Now let $M$ and $N$ be $\mathbb{Z}_2^n$-modules over a real $\mathbb{Z}_2^n$-algebra $\mathcal{A}$\,. The tensor product $M\otimes_\mathcal{A}N$ is a $\mathbb{Z}_2^n$-module over $\mathcal{A}$ as well and taking two copies of $M$ we can define the $\mathbb{Z}_2^n$-symmetric tensor product $M\odot_\mathcal{A}M$\,, which is another $\mathbb{Z}_2^n$-module over $\mathcal{A}$ and we have $$m\odot n=(-1)^{\langle\tilde{m},\tilde{n}\rangle}n\odot m\;.$$
\end{Remark}

Taking the free $\mathbb{Z}_2^n$-module $M$ considered above, we shift the degree of each of its elements by a fixed odd $\mathbb{Z}_2^n$-degree $\gamma$ and obtain a new free $\mathbb{Z}_2^n$-module of total rank r over $\mathcal{A}$\,, which we denote by $M[\gamma]$\,. This shift makes sure that the square of the cohomology operater introduced below vanishes. Taking into account Remark \ref{tensorprod} we obtain that
\begin{equation}
    \mathcal{K}:=\odot_\mathcal{A}M[\gamma]\ \otimes\ \odot_\mathcal{A}M^*
\label{K}\end{equation}
is a $\mathbb{Z}_2^n$-module over $\mathcal{A}$ as tensor product of two such $\mathbb{Z}_2^n$-modules. Furthermore $\mathcal{K}$ can be equipped with a multiplication $\odot$ detailed below and can thus also be seen as a $\mathbb{Z}_2^n$-algebra over $\mathcal{A}$\,. Choosing a basis $(e_i)_i$ of $M$ and denoting the corresponding dual basis of $M^*$ by $(\varepsilon^i)_i$ we define an element
\begin{equation}
    \delta:=\sum_{i=1}^re_i[\gamma]\ \otimes\ \varepsilon^i\in\mathcal{K}\,.
\label{delta}\end{equation}\medskip

Combining the fact that
\begin{equation*}
    \varepsilon^i(e_i)=1_\mathcal{A}
\end{equation*}
with the observation that the identity element $1_\mathcal{A}$ in $\mathcal{A}$ is of degree $\gamma_0$ it becomes clear that $e_i$ and $\varepsilon^i$ must have the same degree for every $i\in\{1,...,r\}$\,. Therefore, the degree of $e_i[\gamma]$ is odd if the degree of $\varepsilon^i$ is even and vice versa, which implies that in each term of $\delta$ there is exactly one odd factor.\medskip

Let $$\sum_{\op{fin}}m[\gamma]\odot n[\gamma]\otimes\alpha^*\in\mathcal{K}$$ be the finite sum of some tensor products of elements in $m[\gamma],\,n[\gamma]\in M[\gamma]$ and $\alpha^*\in M^*$\,. We define the value of $\delta$ on $$\sum_{\op{fin}}m[\gamma]\odot n[\gamma]\otimes\alpha^*$$ by setting
\begin{align*}
    \delta\left(\sum_{\op{fin}}m[\gamma]\odot n[\gamma]\otimes\alpha^*\right)&:=\left(\sum_ie_i[\gamma]\otimes\varepsilon^i\right)\odot\left(\sum_{\op{fin}}m[\gamma]\odot n[\gamma]\otimes\alpha^*\right)\\[8pt]
    &:=\sum_i\sum_{\op{fin}}(-1)^{\langle\tilde{e}_i,\tilde{m}+\gamma+\tilde{n}+\gamma\rangle}(e_i[\gamma]\odot m[\gamma]\odot n[\gamma])\otimes(\varepsilon^i\odot\alpha^*)\,,
\end{align*}
where the term $2\gamma$ in the exponent can be omitted as $2\zg=\gamma_0$\,. If we define the cohomological degree of an element in $\mathcal{K}$ to equal the number of odd factors each of its terms contains, then the cohomological degree of $\delta$ is $1$\,. If an element $\kappa\in\mathcal{K}$ has cohomological degree $l$ then the above definition implies that $\delta(\kappa)$ is of degree $l+1$\,. Hence $\delta$ can be seen as an $\mathcal{A}$-linear map
\begin{equation*}
    \delta:\mathcal{K}^l\rightarrow\mathcal{K}^{l+1}\quad\text{ such that }\quad\delta^2=0\,.
\end{equation*}
Indeed, we have
\begin{align*}
    \delta^2&=\sum_{i,j}(-1)^{\langle\tilde{e}_i,\tilde{e}_j+\gamma\rangle}e_i[\gamma]\odot e_j[\gamma]\ \otimes\ \varepsilon^i\odot\varepsilon^j\\
    &=\sum_{i,j}(-1)^{\langle\tilde{e}_i,\tilde{e}_j+\gamma\rangle}(-1)^{\langle\tilde{e}_i+\gamma,\tilde{e}_j+\gamma\rangle} e_j[\gamma]\odot e_i[\gamma]\ \otimes\ (-1)^{\langle\tilde{e}_i,\tilde{e}_j\rangle}\varepsilon^j\odot\varepsilon^i\\
    &=\sum_{i,j}(-1)^{\langle\tilde{e}_i,\tilde{e}_j+\gamma\rangle+\langle\tilde{e}_i+\gamma,\tilde{e}_j+\gamma\rangle+\langle\tilde{e}_i,\tilde{e}_j\rangle}e_j[\gamma]\odot e_i[\gamma]\ \otimes \ \varepsilon^j\odot\varepsilon^i\\
    &=\sum_{i,j}(-1)^{\langle\gamma,\gamma\rangle}(-1)^{\langle\tilde{e}_i+\gamma,\tilde{e}_j\rangle}e_j[\gamma]\odot e_i[\gamma]\ \otimes \ \varepsilon^j\odot\varepsilon^i\\
    &=-\sum_{i,j}(-1)^{\langle\tilde{e}_i,\tilde{e}_j+\gamma\rangle}e_i[\gamma]\odot e_j[\gamma]\ \otimes \ \varepsilon^i\odot\varepsilon^j\,,
\end{align*}
where the roles of $i$ and $j$ have been interchanged in the last step to show that $\delta^2$ is equal to its opposite and thus vanishes.\medskip

Moreover, it can be shown that the operator $d$ is independent of the choice of the basis $(e_i)_i$ of $M\,.$\medskip

Therefore $(\mathcal{K}^{\boldsymbol{\cdot}},\,\delta)$ is a cochain complex of $\mathbb{Z}_2^n$-modules over $\mathcal{A}$\,. Consequently, its cohomology $\op{H}^{\boldsymbol{\cdot}}(\mathcal{K}^{\boldsymbol{\cdot}},\,\delta)$ is a graded $\mathbb{Z}_2^n$-module over $\mathcal{A}$\,, where graded refers to the cohomology degree. This cohomology can be computed and we state without proof the

\begin{Theorem}\cite{cohomological}
Let $M$ be a free $\Z_2^n$-module of total rank $r$ over a real $\Z_2^n$-algebra $\mathcal{A}$ and let $(\mathcal{K^{\boldsymbol{\cdot}}},\zd)$ be the cochain complex defined by \eqref{K} and \eqref{delta}. For every $k\neq r$ the degree $k$ cohomology $\Z_2^n$-module of $(\mathcal{K}^{\boldsymbol{\cdot}},\,\delta)$ is given by
\begin{equation*}
    \op{H}^k(\mathcal{K}^{\boldsymbol{\cdot}},\,\delta)=0
\end{equation*}
and for $k=r$ we have
\begin{equation*}
    \op{H}^r(\mathcal{K}^{\boldsymbol{\cdot}},\,\delta)=[\Omega]\mathcal{A}\,,
\end{equation*}
which is a free $\mathbb{Z}_2^n$-module over $\mathcal{A}$ of rank $1$ and where $\Omega\in\ker^r\delta\subseteq\mathcal{K}^r$ is the product of all odd vectors among the $e_i[\gamma]$ and the $\varepsilon^i$ associated to a basis $(e_i)_i$ of $M$\,.
\end{Theorem}

Note that $\mathbb{Z}_2^n\op{Ber}M$\,, the free $\mathbb{Z}_2^n$-module over $\mathcal{A}$ of rank $1$ that we are looking for, should be given by $\op{H}^{\boldsymbol{\cdot}}(\mathcal{K}^{\boldsymbol{\cdot}},\,\delta)=\op{H}^r(\mathcal{K}^{\boldsymbol{\cdot}},\,\delta)=[\Omega]\mathcal{A}$\,. It remains to check whether a basis transformation in $M$ characterized by a $\mathbb{Z}_2^n$-matrix $B$ induces a basis transformation in $\op{H}^r(\mathcal{K}^{\boldsymbol{\cdot}},\,\delta)$ characterized by $\mathbb{Z}_2^n\op{Ber}B$\,.\medskip

To this end, we make another small digression on tensor products.

\begin{Remark}
Let $V$ and $W$ be finite dimensional real vector spaces. If $l:V\rightarrow W$ is an isomorphism then $l^{-1}:W\rightarrow V$ and $(l^{-1})^*:V^*\rightarrow W^*$ are isomorphisms as well. Furthermore, we can define an isomorphism $l^\odot:\odot V\rightarrow\odot W$ by setting
\begin{equation*}
    l^\odot(v_1,...,v_p):=l(v_1)\odot\cdots\odot l(v_p),
\end{equation*}
so that $(l^{-1})^{*^\odot}\in\op{Isom}(\odot V^*,\,\odot W^*)$\,. The tensor product of these last two maps yields
\begin{equation*}
    l^\odot\otimes(l^{-1})^{*^\odot}\in\op{Isom}(\odot V\otimes\odot V^*,\,\odot W\otimes\odot W^*)\,.
\end{equation*}
\end{Remark}

If $(e_i)_i$ and $(e'_i)_i$ are two bases in a real vector space $V$ of dimension $p$ then the corresponding basis transformation in $V$ is characterized by some matrix $B\in\op{GL}(p,\,\mathbb{R})$\,, or equivalently by the corresponding automorphism $\beta\in\op{Aut}(V)$\,. Analogously, a basis transformation in a free $\mathbb{Z}_2^n$-module $M$ of rank $r$ over $\mathcal{A}$ is characterized by some $\mathbb{Z}_2^n$-matrix $B\in\mathbb{Z}_2^n\op{GL}_{\gamma_0}(r,\,\mathcal{A})$ that can be identified with an automorphism
\begin{equation*}
    \beta\in\op{Aut}_{\mathcal{A},\gamma_0}(M)\,.
\end{equation*}
The $\mathbb{Z}_2^n$-transpose of the inverse of $B$ corresponds to $(\beta^{-1})^*\in\op{Aut}_{\mathcal{A},\gamma_0}(M^*)$ and we use these automorphisms to construct
\begin{equation*}
    \Phi_B:=\beta^\odot\otimes(\beta^{-1})^{*^\odot}\in\op{Aut}_{\mathcal{A},\gamma_0}(\mathcal{K})\,.
\end{equation*}
Since $\Phi_B$ is actually an invertible cochain map from $(\mathcal{K},\,\delta)$ to itself, by applying the cohomology functor $\op{H}$ to it we obtain
\begin{equation*}
    \op{H}(\Phi_B)\in\op{Aut}_{\mathcal{A},\gamma_0}(\op{H}(\mathcal{K},\,\delta))\,,
\end{equation*}
the map that characterizes the basis transformation in $\op{H}(\mathcal{K},\,\delta)$ which corresponds to the basis transformation in $M$ characterized by $\beta$\,. Observing that $$\op{Aut}_{\mathcal{A},\gamma_0}(\op{H}(\mathcal{K},\,\delta))\cong\mathbb{Z}_2^n\op{GL}_{\gamma_0}(1,\,\mathcal{A})\cong\mathcal{A}^\times_{\gamma_0}$$ we get the map $$\op{H}(\Phi):\mathbb{Z}_2^n\op{GL}_{\gamma_0}(p|\underline{q},\,\mathcal{A})\ni B\mapsto\op{H}(\Phi_B)\in\mathcal{A}^\times_{\gamma_0}\,,$$ which can be shown to satisfy all of the characterizing properties of $\mathbb{Z}_2^n\op{Ber}$\,. Since $\mathbb{Z}_2^n\op{Ber}$ is unique by Theorem \ref{z2nber} we must have $\op{H}(\Phi_B)=\mathbb{Z}_2^n\op{Ber}B$ for all $B\in\mathbb{Z}_2^n\op{GL}_{\gamma_0}(p|\underline{q},\,\mathcal{A})$\,, which implies in particular that a basis transformation in $M$ characterized by $B$ induces a basis tranformation in $H(\mathcal{K},\,\delta)$ characterized by $\mathbb{Z}_2^n\op{Ber}B$ as expected.\medskip

Hence we can finally set
\begin{equation*}
    \mathbb{Z}_2^n\op{Ber}M:=\op{H}(\mathcal{K},\,\delta)=[\Omega]\mathcal{A}\,.
\end{equation*}
Note that $\mathbb{Z}_2^n\op{Ber}M$ can be thought of as the module of algebraic `$\mathbb{Z}_2^n$-top-forms' in view of its similarities with the module of top forms $\op{Det}\tilde{M}=\Omega^p(U)$ in differential geometry, where $\tilde{M}=\Omega^1(U)$ is the module of differential $1$-forms over the algebra $\mathcal{C}^\infty(U)$ of smooth functions on some coordinate domain $U$ of a $p$-dimensional smooth manifold $N$\,. Furthermore, comparing $[\Omega]$ to the volume form $dx^1\wedge\cdots\wedge dx^p$ in differential geometry suggests referring to $[\Omega]$ as algebraic `$\mathbb{Z}_2^n$-Berezinian-volume'. Let us stress once again that if a matrix $B$ represents a basis transformation $$e_j'=e_i\,B^i_j$$ in $M$, then $\Z_2^n\op{Ber}B$ represents the corresponding basis transformation \be[\zW']=[\zW]\,\Z_2^n\op{Ber}B\label{BVT}\ee in $\Z_2^n\op{Ber}M\,.$\medskip

Now consider a $\mathbb{Z}_2^n$-manifold $\mathcal{N}=(N,\,\mathcal{O}_N)$ of dimension $p|\underline{q}$ and a $\mathbb{Z}_2^n$-coordinate-chart $\mathcal{U}=(U,\,\mu)$ of $\mathcal{N}$\,. Then the free $\mathbb{Z}_2^n$-module $M:=\Omega^1\mathcal{N}(U)$ over $\mathcal{A}:=\mathcal{O}_N(U)$ has total rank $$p+\sum_{i=1}^{2^n-1}q_i=:p+q$$ and in the particular case $n=2$ a basis of $M$ is given by
\begin{equation*}
    (e_i)_i=(dx,\,dy,\,d\zx,\,d\eta)\,,
\end{equation*}
where $dx$ stands for the differentials of the $p$ coordinates of degree $(0,\,0)$\,, $dy$ represents the differentials of the $q_1$ coordinates of degree $(1,\,1)$ and similarly for $d\zx$ and $d\eta$\,. Fixing $\gamma=(0,\,1)$ we obtain
\begin{equation*}
    (e_i[\gamma])_i=(dx[\gamma],\,dy[\gamma],\,d\zx[\gamma],\,d\eta[\gamma])\,,
\end{equation*}
where the degrees are given by $((0,\,1),\,(1,\,0),\,(0,\,0),\,(1,\,1))$\,. Furthermore we have the dual basis
\begin{equation*}
    (\varepsilon^i)_i=(\partial_x,\,\partial_y,\,\partial_\zx,\,\partial_\eta)
\end{equation*}
where each $\varepsilon^i$ has the same degree as the corresponding $e_i$. These bases lead to the $\mathbb{Z}_2^2$-Berezinian-volume
\begin{equation*}
    \Omega=dx[\gamma]\odot dy[\gamma]\otimes\partial_\zx\odot\partial_\eta=:\Omega(x,\,y,\,\zx,\,\eta)=\Omega(\mu)
\end{equation*}
and to the module
\begin{equation*}
    (\mathbb{Z}_2^2\op{Ber}\Omega^1\mathcal{N})(U):=\mathbb{Z}_2^2\op{Ber}\Omega^1\mathcal{N}(U)=[\Omega]\mathcal{O}_N(U)=\{[\Omega(\mu)]f(\mu)\}\,,
\end{equation*}
of `$\mathbb{Z}_2^2$-top-forms' of $\mathcal{N}$ over $U$ or local $\mathbb{Z}_2^2$-Berezinian-sections of $\mathcal{N}$ over $U$\,.\medskip

In order to investigate the coordinate transformation law for local Berezinian sections we consider the case $n=1$ and let $\Phi_{\mu\nu}$ be a generic supercoordinate transformation from $\mu=(x,\,\zx)$ to $\nu=(y,\,\eta)$ given by
\begin{equation*}
    \begin{cases}
     y=y(x,\,\zx)\\
     \eta=\eta(x,\,\zx)
    \end{cases},
    \quad\text{and accordingly }\quad
    \begin{cases}
     x=x(y,\,\eta)\\
     \zx=\zx(y,\,\eta)
    \end{cases}.
\end{equation*}
The corresponding basis transformation in $M=\Omega^1\mathcal{N}(U)$ verifies
\begin{equation*}
    dy=dx\,\partial_xy+d\zx\,\partial_\zx y\quad\text{ and }\quad d\eta=dx\,\partial_x\eta+d\zx\,\partial_\zx\eta\
\end{equation*}
or, more precisely,
\begin{equation*}
    dy^j=\sum_idx^i\,\partial_{x^i}y^j+\sum_ad\zx^a\,\partial_{\zx^a}y^j\quad\text{ and }\quad d\eta^b=\sum_idx^i\,\partial_{x^i}\eta^b+\sum_ad\zx^a\,\partial_{\zx^a}\eta^b
\end{equation*}
and is thus characterized by the matrix
\begin{equation*}\renewcommand*{\arraystretch}{1.4}
    B=\left(\begin{array}{c|c}\prescript{t}{}{\partial_xy}&\prescript{t}{}{\partial_x\eta}\\\hline \prescript{t}{}{\partial_\zx y}&\prescript{t}{}{\partial_\zx\eta}\end{array}\right)
    =\prescript{\mathbb{Z}_2t}{}{\left(\begin{array}{c|c}\partial_xy&-\partial_\zx y\\\hline\partial_x\eta&\partial_\zx\eta\end{array}\right)}
    =\prescript{\mathbb{Z}_2t}{}{\mathbb{Z}_2\op{Jac}\Phi_{\mu\nu}}\in\mathbb{Z}_2\op{GL}_0(p|q,\,\mathcal{A})\,.
\end{equation*}
The $\mathbb{Z}_2$-Berezinian of $B$ is then given by
\begin{equation*}
    \mathbb{Z}_2\op{Ber}B=\mathbb{Z}_2\op{Ber}\left(\prescript{\mathbb{Z}_2t}{}{\mathbb{Z}_2\op{Jac}\Phi_{\mu\nu}}\right)=\mathbb{Z}_2\op{\Ber}(\mathbb{Z}_2\op{Jac}\Phi_{\mu\nu})\in\mathcal{A}^\times_0=\mathcal{O}_N(U)^\times_0\,,
\end{equation*}
where the second equality follows from the fact that the $\mathbb{Z}_2^n$-Berezinian, just as the classical determinant, is invariant with respect to taking the transpose of a matrix. This result can actually be generalized to an arbitrary $n\geq1\,,$ so that we have
\begin{equation}
    [\Omega(\nu)]=[\Omega(\mu)]\;\mathbb{Z}_2^n\op{Ber}(\mathbb{Z}_2^n\op{Jac}\Phi_{\mu\nu})\,,
\label{BVT2}\end{equation}
in view of \eqref{BVT}.\medskip

In order to find out which properties the transformation law for local Berezinian sections should have we start considering transformation laws in different contexts.\medskip

For instance, a $(p,q)$-tensor $T\in\otimes^p_qV$
over some real finite-dimensional vector space $V$ can be defined as a tuple $(T^{i_1\cdots i_p}_{j_1\cdots j_q})$ of components in every basis $(e_i)_i$ of $V$ such that the coherent transformation law
\begin{equation*}
    T^{i_1\cdots i_p}_{j_1\cdots j_q}=B^{i_1}_{a_1}\cdots B^{i_p}_{a_p}B'^{b_1}_{j_1}\cdots B'^{b_q}_{j_q}\ T'^{a_1\cdots a_p}_{b_1\cdots b_q}
\end{equation*}
holds. Here $B'=B^{-1}$ and `coherent' means, for instance in the case $(p,q)=(1,0)$, that if
\begin{equation*}
    T^i=B^i_aT'^a\,,\quad T'^a=C^a_bT''^b\,,\quad T^i=D^i_bT''^b
\end{equation*}
characterize basis transformations between $(e_i)_i$ and $(e'_i)_i$\,, between $(e'_i)_i$ and $(e''_i)_i$ and between $(e_i)_i$ and $(e''_i)_i$ respectively, then the matrices $D$ and $BC$ coincide. This is the case since
\begin{equation*}
    D^i_be_i=e''_b=C^a_be'_a=C^a_bB^i_ae_i=(BC)^i_be_i\,.
\end{equation*}

Similarly, a global vector field $X\in\Gamma(TN)$ on a smooth manifold $N$ can be defined in terms of local vector fields $\sum_iX^i\partial_{x^i}$ for some $X^i\in\mathcal{C}^\infty(U)$ on every coordinate chart $(U,\,x)$ of $N$ in conjunction with the coherent transformation law $$X^i=\partial_{y^j}x^iY^j\,.$$ In this case, coherence refers to the fact that if additionally to the above transformation between $x$- and $y$-coordinates we have transformations from $y$- to $z$-coordinates and from $x$- to $z$-coordinates given by
\begin{equation*}
    Y^j=\partial_{z^k}y^jZ^k\quad\text{ and }\quad X^i=\partial_{z^k}x^iZ^k
\end{equation*}
then the matrices $(\partial_zx)$ and $(\partial_yx)(\partial_zy)$ coincide. This is true in view of the theorem of differentiation of composite functions.\medskip

Returning to $\mathbb{Z}_2^n$-geometry we consider a $\mathbb{Z}_2^n$-manifold $\mathcal{N}=(N,\,\mathcal{O}_N)$\,, where the base manifold $N$ is assumed to be orientable and oriented, and an atlas $\mathscr{A}_\mathcal{N}$ of $\mathbb{Z}_2^n$-charts of $\mathcal{N}$\,. Then we define a global $\mathbb{Z}_2^n$-Berezinian-section $$\sigma\in(\mathbb{Z}_2^n\op{Ber}\Omega^1\mathcal{N})(N)$$ of $\mathcal{N}$ as a family $$[\Omega(\mu)]f(\mu),\,[\Omega(\nu)]g(\nu)\,,...$$ of local $\mathbb{Z}_2^n$-Berezinian-sections of $\mathcal{N}$ indexed by the $\mathbb{Z}_2^n$-charts of $\mathscr{A}_\mathcal{N}$ that satisfy the coherent transformation law
\begin{equation}\label{GC}
    f(\mu)=\mathbb{Z}_2^n\op{Ber}(\mathbb{Z}_2^n\op{Jac}\Phi_{\mu\nu})\phi^*(g(\nu))\,,
\end{equation}
which is also referred to as gluing condition and where $\Phi_{\mu\nu}=\Phi=(\phi,\phi^*)$ denotes the transformation from $\mu$- to $\nu$-coordinates.\medskip

Condition \eqref{GC} is natural since if the local sections can be glued they coincide on the coordinate overlaps, i.e., due to \eqref{BVT2}, the section $[\zW(\zm)]f(\zm)$ coincides with the section
$$
[\zW(\zn)]g(\zn)=[\zW(\zm)]\;\Z_2^n\op{Ber}(\Z_2^n\op{Jac}\Phi_{\zm\zn})g(\zn(\zm))=[\zW(\zm)]\;\Z_2^n\op{Ber}(\Z_2^n\op{Jac}\Phi_{\zm\zn})\phi^*(g(\zn))\;.
$$

In order to check whether \eqref{GC} actually defines a {\it coherent} transformation law we consider $\mu$-, $\nu$- and $\omega$-coordinates and denote the coordinate transformations between $\mu$- and $\nu$-coordinates and between $\nu$- and $\omega$-coordinates by $\Phi_{\mu\nu}$ and $\Psi_{\nu\omega}$ respectively. Accordingly, the transformation from $\mu$- to $\omega$-coordinates is given by $\Psi_{\nu\omega}\circ\Phi_{\mu\nu}$\,. Then we have, omitting the prefix $\mathbb{Z}_2^n$\,,
\begin{equation*}
    f(\mu)=\op{Ber}(\op{Jac}\Phi_{\mu\nu})\phi^*(g(\nu))\,
\end{equation*}
and
\begin{equation*}
    \phi^*(g(\nu))=\phi^*(\op{Ber}(\op{Jac}\Psi_{\nu\omega})\phi^*(\psi^*(h(\omega)))\,.
\end{equation*}
Thus $f(\mu)$ can be expressed by
\begin{equation}\label{fmu1}
    f(\mu)=\op{Ber}(\op{Jac}\Phi_{\mu\nu})\,\phi^*(\op{Ber}(\op{Jac}\Psi_{\nu\omega})(\phi^*\circ\psi^*)(h(\omega))
\end{equation}
and by
\begin{equation}\label{fmu2}
    f(\mu)=\op{Ber}(\op{Jac}(\Psi_{\nu\omega}\circ\Phi_{\mu\nu}))(\phi^*\circ\psi^*)(h(\omega))\,.
\end{equation}
Since the $\mathbb{Z}_2^n$-Berezinian is multiplicative we get
\begin{equation*}
    \op{Ber}(\op{Jac}(\Psi_{\nu\omega}\circ\Phi_{\mu\nu}))=\op{Ber}(\phi^*(\op{Jac}\Psi_{\nu\omega})\cdot\op{Jac}\Phi_{\mu\nu})=\op{Ber}(\phi^*(\op{Jac}\Psi_{\nu\omega}))\cdot\op{Ber}(\op{Jac}\Phi_{\mu\nu})\,.
\end{equation*}
Switching the order of $\op{Ber}$ and $\phi^*$ in the expression on the right-hand side and taking into account that $\phi^*(\op{Ber}(\op{Jac}\Psi_{\nu\omega}))$ and $\op{Ber}(\op{Jac}\Phi_{\mu\nu})$ commute as they are of degree $\gamma_0$ we can conclude that \eqref{fmu1} and \eqref{fmu2} are equal and therefore \eqref{GC} is a coherent transformation law.\medskip

In the same fashion as $(\mathbb{Z}_2^n\op{Ber}\Omega^1\mathcal{N})(N)$ we can define $(\mathbb{Z}_2^n\op{Ber}\Omega^1\mathcal{N})(W)$ for any $W\in{\tt Open}(N)$  and since restrictions and the gluing property are included in these definitions we obtain that $\mathbb{Z}_2^n\op{Ber}\Omega^1\mathcal{N}$ is a locally free rank $1$ sheaf of $\mathbb{Z}_2^n$-modules over $\mathcal{O}_N$\,, i.e. a $\mathbb{Z}_2^n$-vector bundle of rank $1$ over $\mathcal{N}$\,. We refer to $\mathbb{Z}_2^n\op{Ber}\Omega^1\mathcal{N}$ as the $\mathbb{Z}_2^n$-Berezinian-sheaf of $\mathcal{N}$\,.

\subsubsection{Integration on $\mathbb{Z}_2$-manifolds}

In Section \ref{intsmooth} we discussed how integration of global top-forms $\Omega^p(M)$ over an oriented smooth manifold $M$ of dimension $p$ works. Similarly, we would now like to integrate global $\mathbb{Z}_2$-Berezinian-sections $(\mathbb{Z}_2\op{Ber}\Omega^1\mathcal{N})(N)$ over a $\mathbb{Z}_2$-manifold $\mathcal{N}$ of dimension $p|q$ whose base manifold is oriented. For this we consider a global $\mathbb{Z}_2$-Berezinian-section $\sigma\in(\mathbb{Z}_2\op{Ber}\Omega^1\mathcal{N})(N)$ that is compactly supported in a $\mathbb{Z}_2$-coordinate-domain $U\subseteq N$\,. The restriction $\mathcal{N}\restr{U}$ can be identified with a $\mathbb{Z}_2$-domain $\mathcal{U}$ equipped with $\mathbb{Z}_2$-coordinates $\mu=(x,\,\zx)$ and $\sigma$ is locally given by
\begin{align*}
    \sigma\restr{U}&=[\Omega(\mu)]f(\mu)\\
    &=[dx[1]\otimes\partial_\zx]f(x,\,\zx)\\
    &=[dx^1[1]\odot\cdots\odot dx^p[1]\otimes\partial_{\zx^q}\odot\cdots\odot\partial_{\zx^1}]f(x,\,\zx)\\
    &=[dx^1\wedge\cdots\wedge dx^p\otimes\partial_{\zx^q}\cdots\partial_{\zx^1}]f(x,\,\zx)\,,
\end{align*}
where the change of notation between the second to last and the last line is motivated by the fact that the differentials $dx^i[1]$ as well as the partial derivatives $\partial_{\zx^a}$ anticommute. The integral of $\sigma$ over $\mathcal{N}$ is then given by
\begin{equation*}
    \int_\mathcal{N}\sigma=\int_\mathcal{U}\sigma\restr{U}=\int_\mathcal{U}[\Omega(\mu)]f(\mu)=\int_\mathcal{U}[dx^1\wedge\cdots\wedge dx^p\otimes\partial_{\zx^q}\cdots\partial_{\zx^1}]f(x,\,\zx)\,.
\end{equation*}
In Section \ref{intsmooth} we defined the integral $\int_Udx^1\wedge\cdots\wedge dx^p\ f(x)$ for $f\in\mathcal{C}^\infty_c(U)$ to be equal to the Lebesgue integral $\int_Udx^1\cdots dx^p\ f(x)$ and verified that this integral is independent of the choice of coordinates in $U$\,. Therefore, we would like to transform $$\int_\mathcal{U}[dx^1\wedge\cdots\wedge dx^p\otimes\partial_{\zx^q}\cdots\partial_{\zx^1}]f(x,\,\zx)$$ into an expression similar to $\int_Udx^1\wedge\cdots\wedge dx^p\ f(x)$ in order to be able to apply the definition from differential geometry. Hence, it is natural to set
\begin{align}
    \int_{\mathcal{N}}\zs&=\int_\mathcal{U}[dx^1\wedge\cdots\wedge dx^p\otimes\partial_{\zx^q}\cdots\partial_{\zx^1}]f(x,\,\zx)\notag\\&:=\int_Udx^1\wedge\cdots\wedge dx^p(\partial_{\zx^q}\cdots\partial_{\zx^1}f(x,\,\zx))\notag\\&=\int_Udx^1\wedge\cdots\wedge dx^p f_{1\ldots q}(x)\notag\\&=\int_Udx^1\cdots dx^p\ f_{1\cdots q}(x)\,,
\end{align}
where $f_{1\ldots q}\in\Ci_c(U)$ is the coefficient of the monomial $\zx^1\zx^2\ldots\zx^q$ in the compactly supported superfunction $f(x,\zx)\,.$

\begin{Remark}
This text differs from most of the literature about integration on supermanifolds as it attempts to approach the idea of differentiating with respect to the odd parameters instead of integrating with respect to them in a natural way instead of providing the definition of a $\mathbb{Z}_2$-integral without any further explanation.
\end{Remark}

Let $V\subseteq N$ be another $\mathbb{Z}_2$-coordinate domain of $\mathcal{N}$ that contains the support of $\sigma$ and denote the $\mathbb{Z}_2$-coordinates of $\mathcal{N}\restr{V}\cong\mathcal{V}$ by  $\nu=(y,\,\eta)$\,. According to the above definition the integral of $\sigma$ over $\mathcal{N}$ can thus be expressed as
\begin{equation*}
    \int_\mathcal{N}\sigma=\int_\mathcal{V}[\Omega(\nu)]g(\nu)=\int_\mathcal{V}[dy^1\wedge\cdots\wedge dy^p\otimes\partial_{\eta^q}\cdots\partial_{\eta^1}]g(y,\,\eta)=\int_Vdy^1\cdots dy^p\ g_{1\cdots q}(y)\,.
\end{equation*}
In order to prove that $\int_\mathcal{N}\sigma$ is coordinate-independent we need to show that
\begin{equation}\label{toshow}
    \int_\mathcal{U}[\Omega(\mu)]f(\mu)=\int_\mathcal{V}[\Omega(\nu)]g(\nu)\,.
\end{equation}
If $\Phi_{\mu\nu}=(\phi,\,\phi^*):\mathcal{U}\rightarrow\mathcal{V}$\,, where restrictions are omitted for the sake of simplicity, denotes the transformation from $\mu$- to $\nu$-coordinates then \eqref{GC} implies that
\begin{equation*}
    f(\mu)=\mathbb{Z}_2\op{Ber}(\mathbb{Z}_2\op{Jac}\Phi_{\mu\nu})\phi^*(g(\nu))\,,
\end{equation*}
so that the statement \eqref{toshow} that has to be proved becomes
\begin{equation}\label{coordtransthm}
    \int_\mathcal{V}[\Omega(\nu)]g(\nu)=\int_\mathcal{U}[\Omega(\mu)]\mathbb{Z}_2\op{Ber}(\mathbb{Z}_2\op{Jac}\Phi_{\mu\nu})\phi^*(g(\nu))\,.
\end{equation}
This result is called coordination transformation theorem in the $\mathbb{Z}_2$-Berezinian-integral and its proof is based on the following fundamental observation: If \eqref{coordtransthm} holds for the coordinate transformations $\Phi_{\mu\nu}:\mathcal{U}\rightarrow\mathcal{V}$ and $\Psi_{\nu\omega}:\mathcal{V}\rightarrow\mathcal{W}$ then it holds for $\Psi_{\nu\omega}\circ\Phi_{\mu\nu}$\,. This is the case since
\begin{align*}
    \int_\mathcal{W}[\Omega(\omega)]h(\omega)&=\int_\mathcal{V}[\Omega(\nu)]\mathbb{Z}_2\op{Ber}(\mathbb{Z}_2\op{Jac}\Psi_{\nu\omega})\psi^*(h(\omega))\\
    &=\int_\mathcal{U}[\Omega(\mu)]\mathbb{Z}_2\op{Ber}(\mathbb{Z}_2\op{Jac}\Phi_{\mu\nu})\cdot\phi^*(\mathbb{Z}_2\op{Ber}(\mathbb{Z}_2\op{Jac}\Psi_{\nu\omega}))\cdot\phi^*(\psi^*(h(\omega))\\
    &=\int_\mathcal{U}[\Omega(\mu)]\mathbb{Z}_2\op{Ber}(\mathbb{Z}_2\op{Jac}(\Psi_{\nu\omega}\circ\Phi_{\mu\nu}))\cdot(\phi^*\circ\psi^*)(h(\omega))\,,
\end{align*}
where the first and second equalities follow from the coordinate transformation theorem for $\Psi_{\nu\omega}$ and for $\Phi_{\mu\nu}$ respectively and the third equality is based on the same cosideration as the equality of \eqref{fmu1} and \eqref{fmu2} above. This observation reduces the proof of \eqref{coordtransthm} to showing that every $\mathbb{Z}_2$-coordinate-transformation $\Phi$ can be decomposed in $n$ types of simple coordinate transformations $\Phi_1,...,\Phi_n$ for some $n\in\mathbb{N}$ and proving that \eqref{coordtransthm} holds for each of the $\Phi_i$\,.

\subsubsection{Integration on $\mathbb{Z}_2^2$-manifolds}\label{IZ22}

Let $\mathcal{N}=(N,\,\mathcal{O}_N)$ be a $\mathbb{Z}_2^2$-manifold of dimension $1|(1,\,1,\,1)$ with oriented base, consider a $\mathbb{Z}_2^2$-Berezinian-section $$\sigma\in(\mathbb{Z}_2^2\op{Ber}\Omega^1\mathcal{N})(N)$$ that is compactly supported in a $\mathbb{Z}_2^2$-coordinate-domain $U\subseteq N$ and assume that $\mathcal{N}\restr{U}$ is isomorphic to the $\mathbb{Z}_2^2$-domain $\mathcal{U}$ with $\mathbb{Z}_2^2$-coordinates $\mu=(x,\,y,\,\zx,\,\eta)$\,. Then $\sigma$ locally reads as
\begin{align*}
    \sigma\restr{U}&=[\Omega(\mu)]f(\mu)\\
    &=[dx[\gamma]\odot dy[\gamma]\otimes\partial_\zx\odot\partial_\eta]f(x,\,y,\,\zx,\,\eta)\\
    &=[dx\odot dy\otimes\partial_\eta\partial_\zx]f(x,\,y,\,\zx,\,\eta)\,,
\end{align*}
where the change of notation between the second to last and the last line is due to the fact that the partial derivatives $\partial_\zx$ and $\partial_\eta$ commute with each other and the differentials $dx$ and $dy$ commute with each other whether we shift their degree by one of the two possible values of gamma or not. The integral of $\sigma$ over $\mathcal{N}$ is given by
\begin{equation*}
    \int_\mathcal{N}\sigma=\int_\mathcal{U}\sigma\restr{U}=\int_\mathcal{U}[\Omega(\mu)]f(\mu)=\int_\mathcal{U}[dx\odot dy\otimes\partial_\eta\partial_\zx]f(x,\,y,\,\zx,\,\eta)
\end{equation*}
and we need an idea for the definition of the integral on the right-hand side. The above discussion of $\mathbb{Z}_2$-integrals suggests differentiating $f(x,\,y,\,\zx,\,\eta)$ with respect to the odd parameters $\zx$ and $\eta$\,, which leads to the following integral with respect to the standard variable $x$ and with respect to the formal parameter $y$\,:
\begin{equation*}
    \int_\mathcal{U}[dx\odot dy\otimes\partial_\eta\partial_\zx]f(x,\,y,\,\zx,\,\eta):=\int dx\int dy\ \partial_\eta\partial_\zx f(x,\,y,\,\zx,\,\eta)=\int dx\int dy\ \sum_{k=0}^\infty f_{11k}(x)y^k\,.
\end{equation*}
From this expression we would like to obtain an integral of a smooth compactly supported function in $x$ with respect to $x$ that we can define as in standard differential geometry. For any $\ell\in[0,\,\infty)$ we have $f_{11\ell}\in\mathcal{C}^\infty_c(U)$ and therefore, for any $\ell\in[0,\,\infty)$\,, setting
\begin{equation*}
    \int dy\ \sum_{k=0}^\infty f_{11k}(x)y^k:=f_{11\ell}(x)
\end{equation*}
allows us to define a Lebesgue integral as in the $\mathbb{Z}_2$-case. One could argue that since $dy$ is in the space that is dual to the space $\partial_\zx$ and $\partial_\eta$ belong to and we chose the coefficient of the highest degree term in $\zx\eta$ we should now choose the coefficient of the lowest degree term in $y$\,. This means we set
\begin{equation*}
    \int dx\int dy\ \sum_{k=0}^\infty f_{11k}(x)y^k:=\int_Udx\ f_{110}(x)\,,
\end{equation*}
where the integral on the right-hand side is the Lebesge integral over the subset of $\mathbb{R}^p$ that is isomorphic to $U$\,.\medskip

To validate this idea for the integral of a $\mathbb{Z}_2^2$-Berezinian-section over a $\mathbb{Z}_2^2$-manifold we have to prove coordinate-independence, i.e. the $\mathbb{Z}_2^2$-analogue to \eqref{coordtransthm}. However, there is a fundamental problem that impedes a straightforward implementation of our idea and in the following we will illustrate this problem by means of an example.\medskip

Let $$\mathcal{N}=\mathcal{U}^{1|(1,1,1)}=(\ ]0,\,1[\ ,\,\mathcal{C}^\infty_{1|(1,1,1)})$$ be a $\mathbb{Z}_2^2$-manifold equipped with global coordinate systems $\mu=(x,\,y,\,\zx,\,\eta)$ and $\nu=(X,\,Y,\,\Xi,\,\Eta)$ and consider the coordinate transformation $\Phi_{\mu\nu}$ given by
\begin{equation}
    \begin{cases}
     X=x\\
     Y=y+\zx\eta\\
     \Xi=\zx\\
     \Eta=\eta\,.
    \end{cases}
\label{CT1}\end{equation}
Furthermore, pick a function $\alpha\in\mathcal{C}^\infty_c(\ ]0,\,1[\ )$ that verifies $\int_0^1dx\ \alpha(x)=1$ and define a $\mathbb{Z}_2^2$-Berezinian-section $$\sigma\in(\mathbb{Z}_2^2\op{Ber}\Omega^1\mathcal{N})(\ ]0,\,1[\ )\,,$$ compactly supported in $]0,\,1[$\,, by setting
\begin{equation*}
    \sigma=[\Omega(\nu)]g(\nu)=[\Omega(X,\,Y,\,\Xi,\,\Eta)]\alpha(X)Y\,.
\end{equation*}
Assuming that the coordinate-independence theorem holds for the integral of $\sigma$ over $\mathcal{N}$ we compute
\begin{equation*}
    \int_\mathcal{N}\sigma=\int_\mathcal{U}[\Omega(\nu)]g(\nu)=\int_{]0,1[}dx\ 0=0
\end{equation*}
and
\begin{align*}
    \int_\mathcal{N}\sigma&=\int_\mathcal{U}[\Omega(\mu)]\mathbb{Z}_2^2\op{Ber}(\mathbb{Z}_2^2\op{Jac}\Phi_{\mu\nu})\phi^*(g(\nu))\\
    &=\int_\mathcal{U}[\Omega(\mu)]\mathbb{Z}_2^2\op{Ber}\left(\begin{array}{cc||cc}1&0&0&0\\0&1&\eta&\zx\\\hline\hline0&0&1&0\\0&0&0&1\end{array}\right)(\alpha(x)y+\alpha(x)\zx\eta)\\
    &=\int_{]0,1[}dx\ \alpha(x)=1\,,
\end{align*}
which is a contradiction and thus means that we cannot integrate compactly supported $\mathbb{Z}_2^2$-Berezinian-sections over $\mathbb{Z}_2^2$-manifolds in a straightforward way.
More information on the modification of signs that is used in the $\mathbb{Z}_2^2$-Jacobian can be found for instance in \cite{local}, page 9.\medskip

This problem also appears in $\mathbb{Z}_2$-geometry, both in the approach described in this text and in the alternative deWitt-Rogers approach. For example, using our approach to integration on $\mathbb{Z}_2$-manifolds we can create a problematic situation that is similar to the one in $\mathbb{Z}_2^2$-geometry described above as follows.\medskip

Consider the $\mathbb{Z}_2$-manifold
\begin{equation*}
    \mathcal{N}=\mathcal{U}^{1|2}=(\ ]0,\,1[\ ,\,\mathcal{C}^\infty_{1|2})
\end{equation*}
with global coordinate systems $\mu=(x,\,\zx^1,\,\zx^2)$ and $\nu=(y,\,\eta^1,\,\eta^2)$ and a coordinate transformation $\Phi_{\mu\nu}$ given by
\begin{equation}
    \begin{cases}
    y=x+\zx^1\zx^2\\\eta^1=\zx^1\\\eta^2=\zx^2\,.
    \end{cases}
\label{CT2}\end{equation}
Define $\sigma\in(\mathbb{Z}_2\op{Ber}\Omega^1\mathcal{N})(\ ]0,\,1[\ )$ by setting
\begin{equation*}
    \sigma=[\Omega(\nu)]g(\nu)=[\Omega(y,\,\eta^1,\,\eta^2)]y\,.
\end{equation*}
Then we have
\begin{equation*}
    \int_\mathcal{N}\sigma=\int_\mathcal{U}[\Omega(\nu)]g(\nu)=\int_{]0,1[}dx\ 0=0
\end{equation*}
and
\begin{align*}
    \int_\mathcal{N}\sigma&=\int_\mathcal{U}[\Omega(\mu)]\mathbb{Z}_2\op{Ber}(\mathbb{Z}_2\op{Jac}\Phi_{\mu\nu})\phi^*(g(\nu))\\
    &=\int_\mathcal{U}[\Omega(\mu)]\mathbb{Z}_2\op{Ber}\left(\begin{array}{c|cc}1&-\zx^2&\zx^1\\\hline0&1&0\\0&0&1\end{array}\right)(x+\zx^1\zx^2)\\
    &=\int_{]0,1[}dx\ 1=1\,,
\end{align*}
which means that the integral $\int_\mathcal{N}\sigma$ is not coordinate-independent. Note that in this case $\sigma$ is not compactly supported in $]0,\,1[$ and as stated above we can ensure coordinate-independence when requiring the $\mathbb{Z}_2$-Berezinian-sections that are integrated to be compactly supported in some coordinate domain. In $\mathbb{Z}_2^2$-geometry it does not suffice to assume $\sigma$ to be compactly supported in order to avoid the problem generated by transformations of the type \eqref{CT1}, \eqref{CT2}. However, there are other strategies to avoid this problem in $\mathbb{Z}_2^2$-geometry, two of which will be discussed in the following.\medskip

The first strategy comprises a reduction of the set of integrable objects. More precisely, one can prove that if the coefficient $g(\nu)$ of a $\mathbb{Z}_2^2$-Berezinian-section $[\Omega(\nu)]g(\nu)$ does not contain the term $g_{100}(X)Y$ then the coefficient $f(\mu)$ of this section in any other coordinate system $\mu$ does not contain the term $f_{100}(x)y$ and refer to sections with such coefficients as compactly supported with respect to the degree $(1,\,1)$ parameter $y$\,. It can be shown that the integral of $\mathbb{Z}_2^2$-Berezinian-sections which are compactly supported with respect to $x$ and with respect to $y$ is well-defined, see \cite{towards}, page 15.\medskip

The second strategy is new and involves changing the nature of the integrable objects. This idea comes from complex analysis.

\begin{Remark}
Let $a_1,...,a_N$ be elements in a simply connected open subset $U\subseteq\mathbb{C}$ and consider a function $f:U\rightarrow\mathbb{C}$ that is holomorphic in $V:=U\backslash\{a_1,...,a_N\}$\,, i.e. that is complex differentiable in $V$. This also means that $f$ is complex analytic in $V$\,, i.e. for each $z_0\in V$ there is a power series at $z_0$ that converges to $f(z)$ at every point $z$ that is close enough to $z_0$\,. If $\gamma$ is a positively oriented simple closed rectifiable curve in $V$ the residue theorem states that the integral of $f$ around $\gamma$ is given by
\begin{equation*}
    \oint_\gamma dz\ f(z)=2\pi i\sum_k\op{R}(f,\,a_k)\,,
\end{equation*}
where the sum is taken over all $k$ such that $a_k$ is inside $\gamma$ and $\op{R}(f,\,a_k)$ denotes the residue of $f$ at $a_k$\,, which can be computed by differentiating and taking limits. The residue of $f$ at $a_k$ can be seen as
\begin{equation*}
    \frac{1}{2\pi i}\oint_\mathscr{C}dz\ f(z)\,,
\end{equation*}
where $\mathscr{C}$ denotes a positively oriented simple closed rectifiable curve in $V$ that contains $a_k$ and none of the other singularities. Moreover, for $f$ defined as a Laurent series about $a_i\,$, i.e. defined as $$f(z)=\sum_{k=-\infty}^{+\infty}c_k(z-a_i)^k$$ its residue at $a_i$ is given by $\op{R}(f,\,a_i)=c_{-1}$\,. In particular, the integral of a Laurent series about $0$ that is holomorphic in $\mathbb{C}\backslash\{0\}$ around a positively oriented simple closed rectifiable curve $\gamma$ that contains $0$ is given by
\begin{equation*}
    \oint_\gamma dz\ \sum_{k=-\infty}^{+\infty}c_kz^k=2\pi i\,c_{-1}\,.
\end{equation*}
\end{Remark}

Our idea is to proceed similarly in $\mathbb{Z}_2^2$-geometry and set
\begin{equation*}
    \int dy\ \sum_{k=-m}^{+\infty}f_{k11}(x)y^k:=f_{-111}(x)\,.
\end{equation*}
To implement this idea we consider a $\mathbb{Z}_2^2$-domain $\mathcal{N}=\mathcal{U}^{1|(1,1,1)}=(U,\,\mathcal{C}^\infty_{1|(1,1,1)})$ with global coordinates $\mu=(x,\,y,\,\zx,\,\eta)$\,, where $U\in{\tt Open}(\mathbb{R})$\,. Denoting $\mathcal{C}^\infty_{1|(1,1,1)}(U)$ by $\mathcal{C}^\infty(\mu)$\,, a generic superfunction $f\in\mathcal{C}^\infty(\mu)$ is given by
\begin{equation*}
    f(\mu)=\sum_{k=0}^{+\infty}\left(\sum_{a,b\in\{0,1\}}f_{kab}(x)\zx^a\eta^b\right)y^k
\end{equation*}
and we now define a generic Laurent series $L\in\mathcal{L}^\infty(\mu)$ by setting
\begin{equation*}
    L(\mu)=\sum_{k=-m}^\infty\left(\sum_{a,b\in\{0,1\}}f_{kab}(x)\zx^a\eta^b\right)y^k\,,
\end{equation*}
where the lower bound defined by $m\in\N$ is finite but not fixed. It can be verified that $\mathcal{L}^\infty(\mu)$ is a $\mathbb{Z}_2^2$-commutative associative unital $\mathbb{R}$-algebra. Note that dividing a superfunction by a non-negative power of $y$ yields a Laurent series:
\begin{equation*}
    \frac{\sum_{k=0}^{+\infty}\left(\sum_{a,b\in\{0,1\}}f_{kab}(x)\zx^a\eta^b\right)y^k}{y^m}=\sum_{\kappa=-m}^\infty\left(\sum_{a,b\in\{0,1\}}f_{\kappa+m\;a\,b}(x)\zx^a\eta^b\right)y^\kappa\in\mathcal{L}^\infty(\mu)\,.
\end{equation*}
This indicates that $\mathcal{L}^\infty(\mu)$ is the localization of $\mathcal{C}^\infty(\mu)$ at the multiplicative subset $\mathcal{P}(\mu)=\{y^m\,|\,m\in\mathbb{N}\}\subseteq\mathcal{C}^\infty(\mu)$\,, where multiplicative subset refers to a multiplicatively closed subset that contains $1$\,. Since localizations of $\mathbb{Z}_2^2$-commutative rings such as $\mathcal{C}^\infty(\mu)$ are similar to localizations at commutative rings we recall the concept of localization in the commutative context.

\begin{Remark}
A localization of a commutative ring $R$ at a multiplicative subset $S\subseteq R$ can be seen as a method to add inverses to $R$\,. More precisely, a localization of $R$ at $S$ is defined as a commutative ring $\mathcal{L}$ together with a ring morphism $L:R\rightarrow\mathcal{L}$ such that the image $L(s)$ of any element $s\in S$ is invertible in $\mathcal{L}$\,.\\

The construction of a localization $(\mathcal{L},\,L)$ can be done by generalizing the construction of the rational numbers $\mathbb{Q}$\,. First we introduce an equivalence relation $\sim$ in $R\times S$ by setting
\begin{equation*}
    (r,\,s)\sim(r',\,s')\Leftrightarrow(rs'-r's)\sigma=0
\end{equation*}
for some $\sigma\in S$\,. Denoting the equivalence class of $(r,\,s)\in R\times S$ under $\sim$ by $\frac{r}{s}$ we define the commutative ring
\begin{equation*}
    \mathcal{L}:=RS^{-1}:=\left\{\frac{r}{s}\,|\,r\in R,\,s\in S\right\}
\end{equation*}
and the ring morphism
\begin{equation*}
    L:R\ni r\mapsto\frac{r}{1}\in RS^{-1}\,.
\end{equation*}
Since $L(s)=\frac{s}{1}$ has inverse $\frac{1}{s}\in RS^{-1}$ for all $s\in S$ we can confirm that $(RS^{-1},\,L)$ is a localization of $R$ at $S$\,.\\

It can be observed that $(RS^{-1},\,L)$ is universal in the sense that for any ring morphism $r:R\rightarrow\mathcal{R}$ that sends every element $s\in S$ to an invertible element in the commutative ring $\mathcal{R}$ there exists a unique ring morphism $u:RS^{-1}\rightarrow\mathcal{R}$ such that the following diagram commutes:
\begin{equation*}
    \begin{tikzcd}
    R\arrow[rightarrow]{r}{L}\arrow{dr}{r}&RS^{-1}\arrow{d}{u}\\&\mathcal{R}\,.
    \end{tikzcd}
\end{equation*}
If $L$ is injective this universal property means that for any ring morphism $r:R\rightarrow\mathcal{R}$ valued in a commutative ring that sends every element in $S$ to a unit in $\mathcal{R}$ there exists a unique ring morphism $u$ that coincides with $r$ on $\mathcal{R}$\,.
\end{Remark}

Continuing the implementation of the above idea from complex analysis in $\mathbb{Z}_2^2$-geometry we consider a $\mathbb{Z}_2^2$-manifold $\mathcal{N}=(N,\,\mathcal{O}_N)$ of dimension $1|(1,\,1,\,1)$ with oriented base manifold and an atlas $\mathscr{A}_\mathcal{N}$ of $\mathbb{Z}_2^2$-coordinate-charts of $\mathcal{N}$\,.
\begin{Definition}
    A generalized $\mathbb{Z}_2^2$-Berezinian-section of $\mathcal{N}$ over $N$ is a family
    \begin{equation*}
        [\Omega(\mu)]L(\mu),\,[\Omega(\nu)]\Lambda(\nu),...
    \end{equation*}
    indexed by the $\mathbb{Z}_2^2$-charts in $\mathscr{A}_\mathcal{N}$ of local generalized $\mathbb{Z}_2^2$-Berezinian-sections whose coefficients are Laurent series and satisfy the coherent transformation law
    \begin{equation}\label{TL}
        L(\mu)=\mathbb{Z}_2^2\op{Ber}(\mathbb{Z}_2^2\op{Jac}\Phi_{\mu\nu})\phi^{*^\sim}(\Lambda(\nu))\,,
    \end{equation}
    where $\Phi_{\mu\nu}:\mu=(x,\,y,\,\zx,\,\eta)\rightarrow\nu=(X,\,Y,\Xi,\,\Eta)$ is the coordinate transformation from $\zm$ to $\zn$ and
    \begin{equation}\label{phitilde}
        \phi^{*^\sim}(\Lambda(\nu)):=\sum_{k=-m}^{+\infty}\sum_{a,b}f_{kab}(\phi^*X)(\phi^*\Xi)^a(\phi^*\Eta)^b(\phi^*Y)^k\,.
    \end{equation}
\end{Definition}
To make sure the right-hand side of \eqref{phitilde} is an element in $\mathcal{L}^\infty(\mu)\,,$ is suffices to show that $(\zvf^*Y)^{-1}\in\mathcal{L}^\infty(\zm)\,,$ which can be done, but we will not repeat the proof here. Indeed, then $(\zvf^*Y)^k\in\mathcal{L}^\infty(\zm)$ for all negative $k$ and the whole term indexed by $k$ in the series over $k$ belongs to $\mathcal{L}^\infty(\zm)\,,$ as the sum over $a,b$ is a superfunction. It follows that the finite sum over all negative $k$ is in $\mathcal{L}^\infty(\zm)$ just as the series over all $k\,,$ since the pullback of a superfunction is a superfunction. Thus we actually have $\phi^{*^\sim}(\Lambda(\nu))\in\mathcal{L}^\infty(\mu)$ and obtain that $\phi^{*^\sim}$ is a ring morphism from $\mathcal{L}^\infty(\nu)$ to $\mathcal{L}^\infty(\mu)$ that coincides with $\phi^*$ on $\mathcal{C}^\infty(\nu)$\,.\medskip

In view of the universal property of the localization $(\mathcal{L}^\infty(\nu),\,L_\nu)$ of $\mathcal{C}^\infty(\nu)$ at $\mathcal{P}(\nu)$ we make the following observation. Denoting the localization map of the localization $\mathcal{L}^\infty(\mu)$ of $\mathcal{C}^\infty(\mu)$ at $\mathcal{P}(\mu)$ by $L_\mu$ and noting that $$L_\mu\circ\phi^*:\mathcal{C}^\infty(\nu)\rightarrow\mathcal{L}^\infty(\mu)$$ is a ring morphism that sends every $Y^k\in\mathcal{P}(\nu)$ to $$L_\zm(\zvf^*\,Y^k)=\frac{(\zvf^*Y)^k}{1}\,,$$ which is invertible in $\mathcal{L}^\infty(\zm)$ since $(\zvf^*Y)^{-k}\in\mathcal{L}^\infty(\zm)\,.$ Hence, in view of universality, there exists a unique ring morphism $u$ such that the following diagram commutes:
\begin{equation*}
    \begin{tikzcd}
    \mathcal{C}^\infty(\nu)\arrow[hookrightarrow]{rr}{L_\nu}\arrow{rd}{\phi^*}&&\mathcal{L}^\infty(\nu)\arrow{dd}{u}\\
    &\mathcal{C^\infty(\mu)}\arrow[hookrightarrow]{rd}{L_\mu}&\\
    &&\mathcal{L}^\infty(\mu)\quad.
    \end{tikzcd}
\end{equation*}
Since in the case of Laurent series the multiplicative subset at which we localize does not contain any zero divisor, the localization maps are injective and we can rephrase our preceding statement saying that there exists a unique ring morphism $u:\mathcal{L}^\infty(\nu)\rightarrow\mathcal{L}^\infty(\mu)$ that coincides with $\phi^*$ on $\mathcal{C}^\infty(\nu)$\,. Hence $\phi^{*^\sim}$ is the unique ring morphism from $\mathcal{L}^\infty(\nu)$ to $\mathcal{L}^\infty(\mu)$ that coincides with $\phi^*$ on $\mathcal{C}^\infty(\nu)$\,.\medskip

We are now prepared to check that the transformation law \eqref{TL} is indeed coherent. This means that if $\Phi_{\zm\zn}:\mu\rightarrow\nu$ and $\Psi_{\zn\zw}:\nu\rightarrow\omega$ are coordinate transformations, we must have
\begin{align*}
    L(\mu)&=\mathbb{Z}_2^2\op{Ber}(\mathbb{Z}_2^2\op{Jac}\Phi_{\mu\nu})\phi^*(\mathbb{Z}_2^2\op{Ber}(\mathbb{Z}_2^2\op{Jac}\Psi_{\nu\omega}))(\phi^{*^\sim}\circ\psi^{*^\sim})(l(\omega))\\
    &=\mathbb{Z}_2^2\op{Ber}(\mathbb{Z}_2^2\op{Jac}(\Psi_{\nu\omega}\circ\Phi_{\mu\nu}))(\psi\circ\phi)^{*^\sim}(l(\omega))\,.
\end{align*}
As we already know that
\begin{equation*}
    \mathbb{Z}_2^2\op{Ber}(\mathbb{Z}_2^2\op{Jac}\Phi_{\mu\nu})\phi^*(\mathbb{Z}_2^2\op{Ber}(\mathbb{Z}_2^2\op{Jac}\Psi_{\nu\omega}))=\mathbb{Z}_2^2\op{Ber}(\mathbb{Z}_2^2\op{Jac}(\Psi_{\nu\omega}\circ\Phi_{\mu\nu}))\;,
\end{equation*}
the above equality boils down to the coherence condition
\begin{equation}\label{CC}
    (\psi\circ\phi)^{*^\sim}=\phi^{*^\sim}\circ\psi^{*^\sim}\,.
\end{equation}
Although \eqref{CC} is trivial when considering the pullbacks without extending them to Laurent series, its direct verification in the case involving extensions is not obvious at all. However, we can argue that $(\psi\circ\phi)^{*^\sim}$ is the unique ring morphism from $\mathcal{L}^\infty(\omega)$ to $\mathcal{L}^\infty(\mu)$ that coincides with $\phi^*\circ\psi^*$ on $\mathcal{C}^\infty(\omega)$ and since $\phi^{*^\sim}\circ\psi^{*^\sim}$ is a ring morphism from $\mathcal{L}^\infty(\omega)$ to $\mathcal{L}^\infty(\mu)$ that coincides with $\phi^*\circ\psi^*$ on $\mathcal{C}^\infty(\omega)$ both morphims must be equal.\medskip

Finally, if $\mathcal{N}=(N,\,\mathcal{O}_N)$ is a $\mathbb{Z}_2^2$-manifold of dimension $1|(1,\,1,\,1)$ with oriented base, we define the integral over $\mathcal N$ of a generalized $\mathbb{Z}_2^2$-Berezinian-section $\mathfrak s$ that is compactly supported in a $\mathbb{Z}_2^2$-coordinate-domain $U\subseteq N$ such that $\mathcal{N}\restr{U}$ is isomorphic to the $\mathbb{Z}_2^2$-domain $\mathcal{U}$ with coordinates $\mu=(x,\,y,\,\zx,\,\eta)$\,, by setting
\begin{equation*}
    \int_\mathcal{N}\mathfrak{s}=\int_\mathcal{U}[\Omega(\mu)]L(\mu)=\int_\mathcal{U}[dx\odot dy\otimes\partial_\eta\partial_\zx]L(x,\,y,\zx,\,\eta):=\int dx\int dy\ \sum_{k=-m}^{+\infty}f_{k11}(x)y^k
\end{equation*}
as before and setting
\begin{equation*}
    \int dy\ \sum_{k=-m}^{+\infty}f_{k11}(x)y^k:=f_{-111}(x)
\end{equation*}
motivated by the development from complex analysis discussed above so that we finally obtain the definition
\begin{equation*}
    \int_\mathcal{N}\mathfrak{s}:=\int_Udx\ f_{-111}(x)\,,
\end{equation*}
where the right-hand side denotes the Lebesgue integral of the coefficient $f_{-111}\in\mathcal{C}_c^\infty(U)$ with respect to the standard coordinate $x$\,. It can be shown that this definition is coordinate-independent as desired.

\subsubsection{Outlook}

Having discussed integration of compactly supported generalized $\mathbb{Z}_2^2$-Berezinian-sections over $\mathbb{Z}_2^2$-manifolds of dimension $1|(1,\,1,\,1)$,\, the question arises whether this integration theory can be extended to `higher' settings. If $\mathcal{N}=(N,\mathcal{O})$ is a $\mathbb{Z}_2^n$-manifold of dimension $p|\underline{q}$ whose ideal sheaf is denoted as usual by $\mathcal{J}$ and which locally has $\mathbb{Z}_2^n$-coordinates $$\mu=(x,\,y,\zx)=(x^1,...,x^p,\,y^1,...,y^{q_0},\,\zx^1,...,\zx^{q_1})\,,$$ where $x$ denotes the coordinates of degree $\gamma_0$\,, the tuple $y$ the coordinates of even degree different from $\gamma_0$ and $\zx$ the coordinates of odd degree, we generalize Laurent series and end up with generalized fractions in the sense of algebraic topology. They appear as an explicit description of the $q_0$-th $\mathcal{O}(U)$-module $\op{H}^{q_0}_\mathcal{J}(U,\,\mathcal{O})$ of the $\mathcal{J}$-local cohomology of $\mathcal{O}$ over $U\in{\tt Open}(N)$ and we can integrate the compactly supported vectors of this module. This $\mathbb{Z}_2^n$-integration-theory is related to Grothendieck duality and requires the use of an appropriate group of admissible coordinate transformations that allows to work around the problematic monomials of the type \eqref{CT1} and \eqref{CT2} discussed in Subsection \ref{IZ22}.

\phantomsection 

\end{document}